%% file: main.tex
\pdfoutput=1
\documentclass[12pt,a4paper]{article}
\usepackage{ifthen} % for conditional statements
\newboolean{pdflatex}
\setboolean{pdflatex}{true} % False for eps figures 
\usepackage{lscape}
\usepackage{adjustbox}

\newboolean{articletitles}
\setboolean{articletitles}{true} % False removes titles in references

\newboolean{uprightparticles}
\setboolean{uprightparticles}{false} %True for upright particle symbols

\def\paperauthors{LHCb collaboration} % Leave as is for PAPER, CONF and FIGURE
\def\paperasciititle{Progress in measurements of the CKM angle gamma and charm mixing at LHCb} % Set ASCII title here !! MAKE sure it's only ASCII characters !! 
\def\papertitle{Progress in measurements of the CKM angle $\gamma$ and charm mixing \\ at LHCb} % Latex formatted title
\def\paperkeywords{{High Energy Physics}, {LHCb}} % Comma separated list
\def\papercopyright{\the\year\ CERN for the benefit of the LHCb collaboration}
\def\paperlicence{CC BY 4.0 licence}
\def\paperlicenceurl{https://creativecommons.org/licenses/by/4.0/}

\newif\ifEnableSectionTOCLinks
\EnableSectionTOCLinksfalse % deactivated

\input{LHCb/preamble}
\usepackage{bm}  % Bold math in titles
\usepackage{booktabs} % for input tables
\usepackage{siunitx}
\usepackage[capitalise]{cleveref}
\newcommand{\cmmnt}[1]{\ignorespaces}

\usepackage{multirow}

\newcommand{\omcl}{\ensuremath{1-{\rm CL}}\xspace}
\newcommand\suthreef{{\ensuremath{\mathrm{SU}(3)_{\rm F}}}\xspace}
\newcommand\ptwiddle[1]{\mathord{\mathop{#1}\limits^{\scriptscriptstyle(\sim)}}}

\newcommand{\rb}{\ensuremath{r_{B}}\xspace}
\newcommand{\db}{\ensuremath{\delta_{B}}\xspace}
\newcommand{\rd}{\ensuremath{r_{D}}\xspace}
\newcommand{\dd}{\ensuremath{\delta_{D}}\xspace}
\newcommand{\xD}{\ensuremath{x}\xspace}
\newcommand{\yD}{\ensuremath{y}\xspace}
\newcommand{\qopD}{\ensuremath{|q/p|}\xspace}
\newcommand{\xth}{\ensuremath{x_{12}}\xspace}
\newcommand{\yth}{\ensuremath{y_{12}}\xspace}
\newcommand{\phiD}{\ensuremath{\phi_2}\xspace}
\newcommand{\phiM}{\ensuremath{\phi^M_2}\xspace}
\newcommand{\phiG}{\ensuremath{\phi^{\Gamma}_2}\xspace}
\newcommand{\rdkpi}{\ensuremath{r_{D}^{K\pi}}\xspace}
\newcommand{\ddkpi}{\ensuremath{\delta_{D}^{K\pi}}\xspace}
\newcommand{\ycpHH}{\ensuremath{y_{\CP}^{\hh}}\xspace}
\newcommand{\ycpRS}{\ensuremath{y_{\CP}^{\Km\pip}}\xspace}
\newcommand{\etaCP}{\ensuremath{\eta_{\CP}}\xspace}
\newcommand{\rbdk} {\ensuremath{r_{\Bpm}^{D\Kpm}}\xspace}
\newcommand{\dbdk} {\ensuremath{\delta_{\Bpm}^{D\Kpm}}\xspace}
\newcommand{\rbdpi} {\ensuremath{r_{\Bpm}^{D\pipm}}\xspace}
\newcommand{\dbdpi} {\ensuremath{\delta_{\Bpm}^{D\pipm}}\xspace}
\newcommand{\rbdstk} {\ensuremath{r_{\Bpm}^{\Dstar\Kpm}}\xspace}
\newcommand{\dbdstk} {\ensuremath{\delta_{\Bpm}^{\Dstar\Kpm}}\xspace}
\newcommand{\rbdstpi} {\ensuremath{r_{\Bpm}^{\Dstar\pipm}}\xspace}
\newcommand{\dbdstpi} {\ensuremath{\delta_{\Bpm}^{\Dstar\pipm}}\xspace}
\newcommand{\rbdkst} {\ensuremath{r_{\Bpm}^{D\Kstarpm}}\xspace}
\newcommand{\dbdkst} {\ensuremath{\delta_{\Bpm}^{D\Kstarpm}}\xspace}
\newcommand{\kbdkst} {\ensuremath{\kappa_{\Bpm}^{D\Kstarpm}}\xspace}
\newcommand{\rbdkstz} {\ensuremath{r_{\Bd}^{D\Kstarz}}\xspace}
\newcommand{\dbdkstz} {\ensuremath{\delta_{\Bd}^{D\Kstarz}}\xspace}
\newcommand{\kbdkstz} {\ensuremath{\kappa_{\Bd}^{D\Kstarz}}\xspace}
\newcommand{\rbdkpipi} {\ensuremath{r_{\Bpm}^{D\Kpm\pip\pim}}\xspace}
\newcommand{\rbdpipipi} {\ensuremath{r_{\Bpm}^{D\pipm\pip\pim}}\xspace}
\newcommand{\rbdsk} {\ensuremath{r_{\Bs}^{\Dsmp\Kpm}}\xspace}
\newcommand{\dbdsk} {\ensuremath{\delta_{\Bs}^{\Dsmp\Kpm}}\xspace}
\newcommand{\rbdskpipi} {\ensuremath{r_{\Bs}^{\Dsmp\Kpm\pip\pim}}\xspace}
\newcommand{\dbdskpipi} {\ensuremath{\delta_{\Bs}^{\Dsmp\Kpm\pip\pim}}\xspace}
\newcommand{\rbdmpi} {\ensuremath{r_{\Bd}^{\Dmp\pipm}}\xspace}
\newcommand{\dbdmpi} {\ensuremath{\delta_{\Bd}^{\Dmp\pipm}}\xspace}

\newcommand{\Fppp}    {\ensuremath{F^{+}_{\pi\pi\piz}}\xspace}
\newcommand{\Fkkp}    {\ensuremath{F^{+}_{KK\piz}}\xspace}
\newcommand{\rdkpp}   {\ensuremath{r_{\D}^{K\pi\piz}}\xspace}
\newcommand{\ddkpp}   {\ensuremath{\delta_{\D}^{K\pi\piz}}\xspace}
\newcommand{\kdkpp}   {\ensuremath{\kappa_{\D}^{K\pi\piz}}\xspace}
\newcommand{\Fpppp}   {\ensuremath{F^{+}_{4\pi}}\xspace}
\newcommand{\Fkkpp}   {\ensuremath{F^{+}_{KK\pi\pi}}\xspace}
\newcommand{\ddkppp}  {\ensuremath{\delta_{\D}^{K3\pi}}\xspace}
\newcommand{\kdkppp}  {\ensuremath{\kappa_{\D}^{K3\pi}}\xspace}
\newcommand{\rdkskpi}  {\ensuremath{r_{\D}^{\KS K\pi}}\xspace}
\newcommand{\ddkskpi}  {\ensuremath{\delta_{\D}^{\KS K\pi}}\xspace}
\newcommand{\kdkskpi}  {\ensuremath{\kappa_{\D}^{\KS K\pi}}\xspace}

\def\RKpi      {\ensuremath{R_{\kaon\pion}}\xspace}
\def\AKpi      {\ensuremath{A_{\kaon\pion}}\xspace}
\def\AKpitilde {\ensuremath{\ptwiddle{A}_{\kaon\pion}}\xspace}
\def\AKpit     {\ensuremath{\tilde{A}_{_{\kaon\pion}}}\xspace}
\def\cKpi      {\ensuremath{c_{\kaon\pion}}\xspace}
\def\cKpipr    {\ensuremath{c_{\kaon\pion}^{\prime}}\xspace}
\def\DcKpi     {\ensuremath{\Delta c_{\kaon\pion}}\xspace}
\def\DcKpipr   {\ensuremath{\Delta c_{\kaon\pion}^{\prime}}\xspace}
\def\DcKpitilde  {\ensuremath{\Delta \ptwiddle{c}_{\kaon\pion}}\xspace}
\def\rmix{\ensuremath{   r_\textnormal{mix}}\xspace}
\def\pmix{\ensuremath{\phi_\textnormal{mix}}\xspace}

\newcommand{\hp}{\ensuremath{h^+}\xspace}
\renewcommand{\hm}{\ensuremath{h^-}\xspace}
\newcommand{\hpm}{\ensuremath{h^\pm}\xspace}
\newcommand{\hmp}{\ensuremath{h^\mp}\xspace}

\newcommand{\BuDh}{\ensuremath{\Bm\to \D\hm}\xspace}
\newcommand{\BuDK}{\ensuremath{\Bm\to \D\Km}\xspace}
\newcommand{\BuDpi}{\ensuremath{\Bm\to \D\pim}\xspace}
\newcommand{\BuDsth}{\ensuremath{\Bm\to\Dstar\hm}\xspace}
\newcommand{\BuDKst}{\ensuremath{\Bm\to\D\Kstarm}\xspace}
\newcommand{\BuDhpipi}{\ensuremath{\Bpm\to\D\hpm\pip\pim}\xspace}
\newcommand{\BdDKst}{\ensuremath{\Bd\to\D\Kstarz}\xspace}
\newcommand{\BdDpi}{\ensuremath{\Bd\to\Dmp\pipm}\xspace}
\newcommand{\BsDsK}{\ensuremath{\Bs\to\Dsmp\Kpm}\xspace}
\newcommand{\BsDsKpipi}{\ensuremath{\Bs\to\Dsmp\Kpm\pip\pim}\xspace}

\newcommand{\hh}{\ensuremath{\hp\hm}\xspace}
\newcommand{\KK}{\ensuremath{\Kp\Km}\xspace}
\newcommand{\pipi}{\ensuremath{\pip\pim}\xspace}

\newcommand{\Dzhh}     {\ensuremath{\Dz\to \hh}\xspace}
\newcommand{\DzKK}     {\ensuremath{\Dz\to \KK}\xspace}
\newcommand{\Dzpipi}   {\ensuremath{\Dz\to \pipi}\xspace}
\newcommand{\Dhhpiz}   {\ensuremath{D\to \hp\hm\piz}\xspace}
\newcommand{\Dhpipipi} {\ensuremath{D\to \hpm\pimp\pip\pim}\xspace}
\newcommand{\Dpipipipi}{\ensuremath{D\to \pip\pim\pip\pim}\xspace}
\newcommand{\DKpi}      {\ensuremath{D\to \Kp\pim}\xspace}
\newcommand{\DzKpi}     {\ensuremath{\Dz\to \Kpm\pimp}\xspace}
\newcommand{\DzKpPim}   {\ensuremath{\Dz\to \Kp\pim}\xspace}
\newcommand{\DzKmPip}   {\ensuremath{\Dz\to \Km\pip}\xspace}
\newcommand{\DKpipiz}   {\ensuremath{D\to \Kpm\pimp\piz}\xspace}
\newcommand{\DKpipipi}  {\ensuremath{D\to \Kpm\pimp\pip\pim}\xspace}
\newcommand{\DzKpipipi} {\ensuremath{\Dz\to \Kpm\pimp\pip\pim}\xspace}
\newcommand{\DKSpipi}   {\ensuremath{D\to\KS\pip\pim}\xspace}
\newcommand{\DzKSpipi}  {\ensuremath{\Dz\to\KS\pip\pim}\xspace}
\newcommand{\DKShh}     {\ensuremath{D\to\KS\hp\hm}\xspace}
\newcommand{\DKSKpi}   {\ensuremath{D\to \KS\Kpm\pimp}\xspace}
\newcommand{\DKpipi}   {\ensuremath{\Dp\to\Km\pip\pip}\xspace}
\newcommand{\Dshhh}    {\ensuremath{\Ds\to\hp\hm\pip}\xspace}
\newcommand{\Dhhpr}     {\ensuremath{D\to \hpm h^{\prime \mp}}\xspace}
\newcommand{\Dhhprpiz}  {\ensuremath{D\to \hpm h^{\prime \mp}\piz}\xspace}
\newcommand{\Dhhpipi}   {\ensuremath{D\to \hp\hm\pip\pim}\xspace}
\newcommand{\DKKpipi}   {\ensuremath{D\to \Kp\Km\pip\pim}\xspace}
\newcommand{\Dzpipipiz} {\ensuremath{\Dz\to \pip\pim\piz}\xspace}

\newcommand{\DY    }{\ensuremath{\Delta Y        }\xspace}
\newcommand{\DYhh  }{\ensuremath{\Delta Y_{\hh}  }\xspace}
\newcommand{\DYKK  }{\ensuremath{\Delta Y_{\KK}  }\xspace}
\newcommand{\DYpipi}{\ensuremath{\Delta Y_{\pipi}}\xspace}
\newcommand{\DYpppeff}{\ensuremath{\Delta Y^\textnormal{eff}_{\pipi \piz}}\xspace}

\newcommand{\acp}[1]{\ensuremath{a^{\textnormal{d}}_{{#1}}}\xspace}
\newcommand{\AcpDhh  }{\ensuremath{\acp{\hh}  }\xspace}
\newcommand{\AcpDKK  }{\ensuremath{\acp{\KK}  }\xspace}
\newcommand{\AcpDpipi}{\ensuremath{\acp{\pipi}}\xspace}
\newcommand{\AcpDKpi }{\ensuremath{\acp{\Kp\pim}}\xspace}
\newcommand{\Ahh}[1]{\ensuremath{A_{\hh}^{#1}}\xspace}
\newcommand{\AKK}[1]{\ensuremath{A_{\KK}^{#1}}\xspace}
\newcommand{\Apipi}[1]{\ensuremath{A_{\pipi}^{#1}}\xspace}

\def\rQCD{{\ensuremath{r_{\hh}}}\xspace}

\newcommand{\ddhhQCD}{\ensuremath{\delta_{\hh}}\xspace}
\newcommand{\ddKKQCD}{\ensuremath{\delta_{\KK}}\xspace}
\newcommand{\ddpipiQCD}{\ensuremath{\delta_{\pipi}}\xspace}

\newcommand{\rdk}{\rbdk}
\newcommand{\ddk}{\dbdk}
\newcommand{\rdpi}{\rbdpi}
\newcommand{\ddpi}{\dbdpi}
\newcommand{\rdstk}{\rbdstk}
\newcommand{\ddstk}{\dbdstk}
\newcommand{\rdstpi}{\rbdstpi}
\newcommand{\ddstpi}{\dbdstpi}
\newcommand{\rdkst}{\rbdkst}
\newcommand{\ddkst}{\dbdkst}
\newcommand{\rdkstz}{\rbdkstz}
\newcommand{\ddkstz}{\dbdkstz}
\newcommand{\rdkpipi}{\rbdkpipi}
\newcommand{\rdpipipi}{\rbdpipipi}
\newcommand{\ldsk}{\rbdsk}
\newcommand{\ddsk}{\dbdsk}
\newcommand{\ldskpipi}{\rbdskpipi}
\newcommand{\ddskpipi}{\dbdskpipi}
\newcommand{\ldmpi}{\rbdmpi}
\newcommand{\ddmpi}{\dbdmpi}
\newcommand{\rDkpi}{\rdkpi}
\newcommand{\dDkpi}{\ddkpi}

\newcommand{\runOneTwo}{Runs 1--2}

\providecommand{\AD}{\ensuremath{a_{\Kp\pim}^{\textnormal{d}}}\xspace}

\newcommand{\gCKMfitter}{\ensuremath{(66.2^{\,+\,0.8}_{\,-\,3.2})\degrees}\xspace}
\newcommand{\gUTfit}{\ensuremath{(65.7\pm 1.3)\degrees}\xspace}

\newcommand{\ninputsRunThree}{218\xspace}
\newcommand{\nparamsRunThree}{52\xspace}
\newcommand{\chisqminRunThree}{191.1\xspace}
\newcommand{\goodnessOfFitRunThree}{8.9\%\xspace}

\newcommand{\resultGRunThree}{\ensuremath{\left(63.2^{\,+\,2.3}_{\,-\,2.2}\right)\degrees}\xspace}
\newcommand{\resultLdmpiRunThree}{\ensuremath{0.030^{\,+\,0.014}_{\,-\,0.012}}\xspace}
\newcommand{\resultXtwelveRunThree}{\ensuremath{\left(4.2^{\,+\,0.4}_{\,-\,0.5}\right)\times10^{-3}}\xspace}
\newcommand{\resultYtwelveRunThree}{\ensuremath{\left(6.18^{\,+\,0.17}_{\,-\,0.18}\right)\times10^{-3}}\xspace}
\newcommand{\resultPhiMRunThree}{\ensuremath{\left(4^{\,+\,11}_{\,-\,10}\right)\mrad}\xspace}
\newcommand{\resultPhiGRunThree}{\ensuremath{\left(19^{\,+\,23}_{\,-\,24}\right)\mrad}\xspace}
\newcommand{\resultQPRunThree}{\ensuremath{-0.007^{\,+\,0.013}_{\,-\,0.012}}\xspace}
\newcommand{\resultPhiRunThree}{\ensuremath{(-14 \pm 16)\mrad}\xspace}
\newcommand{\resultAcpDKKRunThree}{\ensuremath{(0.9 \pm 0.5)\times10^{-3}}\xspace}
\newcommand{\resultAcpDpipiRunThree}{\ensuremath{(2.4 \pm 0.6)\times10^{-3}}\xspace}
\newcommand{\resultDdKpiRunThree}{\ensuremath{(192.9^{\,+\,2.3}_{\,-\,2.4})\degrees}\xspace}

\newcommand{\resultGTIRunThree}{\ensuremath{(62.5^{\,+\,2.3}_{\,-\,2.4})\degrees}\xspace}
\newcommand{\resultGTDRunThree}{\ensuremath{(76 \pm 10)\degrees}\xspace}
\newcommand{\resultGBuDsthRunThree}{\ensuremath{(76^{\,+\,7}_{\,-\,8})\degrees}\xspace}
\newcommand{\resultGphisRunThree}{\ensuremath{(73 \pm 10)\degrees}\xspace}

\newcommand{\resultPhiMUnbiased}    {\ensuremath{(0 \pm 12)\mrad}\xspace}
\newcommand{\resultAcpDKKUnbiased}  {\ensuremath{(0.9 \pm 0.5)\times10^{-3}}\xspace}
\newcommand{\resultAcpDpipiUnbiased}{\ensuremath{(2.8 \pm 0.7)\times10^{-3}}\xspace}

\newcommand{\gammaPrevious}{\ensuremath{\g = (65.4^{\,+\,3.8}_{\,-\,4.2})\degrees}\xspace}  % LHCb-PAPER-2021-033
\newcolumntype{L}{>{$}l<{$}}
\newcolumntype{R}{>{$}r<{$}}
\newcolumntype{C}{>{$}c<{$}}

\begin{document}

%%%%%%%%%%%%%%%%%%%%%%%%%
%%%%% Title     %%%%%%%%%
%%%%%%%%%%%%%%%%%%%%%%%%%
\renewcommand{\thefootnote}{\fnsymbol{footnote}}
\setcounter{footnote}{1}

% %%%%%%% CHOOSE TITLE PAGE--------
\input{title-LHCb-PAPER}
% %%%%%%%%%%%%% ---------

\renewcommand{\thefootnote}{\arabic{footnote}}
\setcounter{footnote}{0}

\cleardoublepage

%%%%%%%%%%%%%%%%%%%%%%%%%
%%%%% Main text %%%%%%%%%
%%%%%%%%%%%%%%%%%%%%%%%%%

\pagestyle{plain} % restore page numbers for the main text
\setcounter{page}{1}
\pagenumbering{arabic}

%% Comment before a final submission.
% \linenumbers

%% This is the main body
%% It is useful to have a single file so comments are not missed in overleaf.
\input{body}
\clearpage

\addcontentsline{toc}{section}{References}
%\setboolean{inbibliography}{true}
\bibliographystyle{LHCb/LHCb}
\bibliography{main,LHCb/standard,LHCb/LHCb-PAPER,LHCb/LHCb-CONF,LHCb/LHCb-DP,LHCb/LHCb-TDR,LHCb/LHCb-PUB}
 
\newpage
\input{LHCb/Authorship_LHCb-PAPER-2026-058}

\end{document}

%% file: LHCb/preamble.tex
\usepackage[top=1in, bottom=1.25in, left=1in, right=1in]{geometry}

\usepackage{microtype}
\usepackage{lineno}  % for line numbering during review
\usepackage{xspace} % To avoid problems with missing or double spaces after
\usepackage{caption} %these three command get the figure and table captions automatically small

\usepackage{graphicx}  % to include figures (can also use other packages)
\usepackage{color}
\usepackage{colortbl}
\graphicspath{{./figs/}} % Make Latex search fig subdir for figures
\usepackage{amsmath} % Adds a large collection of math symbols
\usepackage{amssymb}
\usepackage{amsfonts}
\usepackage{upgreek} % Adds in support for greek letters in roman typeset

\newcommand*\patchAmsMathEnvironmentForLineno[1]{%
\expandafter\let\csname old#1\expandafter\endcsname\csname #1\endcsname
\expandafter\let\csname oldend#1\expandafter\endcsname\csname
end#1\endcsname
 \renewenvironment{#1}%
   {\linenomath\csname old#1\endcsname}%
   {\csname oldend#1\endcsname\endlinenomath}%
}
\newcommand*\patchBothAmsMathEnvironmentsForLineno[1]{%
  \patchAmsMathEnvironmentForLineno{#1}%
  \patchAmsMathEnvironmentForLineno{#1*}%
}
\AtBeginDocument{%
\patchBothAmsMathEnvironmentsForLineno{equation}%
\patchBothAmsMathEnvironmentsForLineno{align}%
\patchBothAmsMathEnvironmentsForLineno{flalign}%
\patchBothAmsMathEnvironmentsForLineno{alignat}%
\patchBothAmsMathEnvironmentsForLineno{gather}%
\patchBothAmsMathEnvironmentsForLineno{multline}%
\patchBothAmsMathEnvironmentsForLineno{eqnarray}%
}

\usepackage[pdftex,
            pdfauthor={\paperauthors},
            pdftitle={\paperasciititle},
            pdfkeywords={\paperkeywords}]{hyperref}
\usepackage{hyperxmp}
\hypersetup{
    pdfcopyright={Copyright (C) \papercopyright},
    pdflicenseurl={\paperlicenceurl}
}
\usepackage[colorinlistoftodos,textsize=scriptsize]{todonotes}

\usepackage[bottom,flushmargin,hang,multiple]{footmisc}

\usepackage[all]{hypcap} % Internal hyperlinks to floats.

\input{LHCb/lhcb-symbols-def} % Add in the predefined LHCb symbols

\hypersetup{
  colorlinks   = true, %Colours links instead of ugly boxes
  urlcolor     = blue, %Colour for external hyperlinks
  linkcolor    = blue, %Colour of internal links
  citecolor    = red   %Colour of citations
}

\ifEnableSectionTOCLinks
    \usepackage[explicit]{titlesec} % to change headings
    
    \let\oldcontentsline\contentsline
    \renewcommand

    \titleformat{\section}{\normalfont\Large\bf}{\hyperlink{tocsection.\thesection}{{\thesection} \parbox[t]{\dimexpr\textwidth-1pc}{#1}}}{1pc}{}

    \titleformat{\subsection}{\normalfont\bf}{\hyperlink{tocsubsection.\thesubsection}{{\thesubsection} \parbox[t]{\dimexpr\textwidth-1pc}{#1}}}{1pc}{}

    \titleformat{name=\section,numberless}[display]{}{}{0pt}{\normalfont\Huge\bfseries #1}
\fi

\usepackage{cite} % Allows for ranges in citations
\usepackage{LHCb/mciteplus}

%% file: LHCb/lhcb-symbols-def.tex
\usepackage{xspace} 
\usepackage{upgreek}

\def\lhcb   {\mbox{LHCb}\xspace}

\def\besiii {\mbox{BESIII}\xspace}

\def\lhc    {\mbox{LHC}\xspace}

\def\MagUp {\mbox{\em Mag\kern -0.05em Up}\xspace}

\ifthenelse{\boolean{uprightparticles}}%
{
 
 \def\Pgamma      {\ensuremath{\upgamma}\xspace}

 \def\Ppi         {\ensuremath{\uppi}\xspace}

 \def\Ppsi        {\ensuremath{\uppsi}\xspace}

 \def\PDelta      {\ensuremath{\Delta}\xspace}                 
 \def\PXi         {\ensuremath{\Xi}\xspace}                 
 \def\PLambda     {\ensuremath{\Lambda}\xspace}                 
 \def\PSigma      {\ensuremath{\Sigma}\xspace}                 
 \def\POmega      {\ensuremath{\Omega}\xspace}                 
 \def\PUpsilon    {\ensuremath{\Upsilon}\xspace}
 \let\oldPi\Pi
 \def\PPi         {\ensuremath{\oldPi}\xspace}

 \def\PB      {\ensuremath{\mathrm{B}}\xspace}                 
 \def\PD      {\ensuremath{\mathrm{D}}\xspace}                 
 \def\PJ      {\ensuremath{\mathrm{J}}\xspace}                 
 \def\PK      {\ensuremath{\mathrm{K}}\xspace}                 
 \def\Pb      {\ensuremath{\mathrm{b}}\xspace}                 
 \def\Pc      {\ensuremath{\mathrm{c}}\xspace}                 
 \def\Pd      {\ensuremath{\mathrm{d}}\xspace}                 

 \def\Ps      {\ensuremath{\mathrm{s}}\xspace}                 
                  
 \def\Pu      {\ensuremath{\mathrm{u}}\xspace}                 
 \def\thebaroffset{0.0em}
}
{
 
 \def\Pgamma      {\ensuremath{\gamma}\xspace}

 \def\Ppi         {\ensuremath{\pi}\xspace}

 \def\Ppsi        {\ensuremath{\psi}\xspace}                 
                  
 \mathchardef\PDelta="7101
 \mathchardef\PXi="7104
 \mathchardef\PLambda="7103
 \mathchardef\PSigma="7106
 \mathchardef\POmega="710A
 \mathchardef\PUpsilon="7107
 \mathchardef\PPi="7105
 \def\PB      {\ensuremath{B}\xspace}                 
 \def\PD      {\ensuremath{D}\xspace}                 
 \def\PJ      {\ensuremath{J}\xspace}                 
 \def\PK      {\ensuremath{K}\xspace}                 
 \def\Pb      {\ensuremath{b}\xspace}                 
 \def\Pc      {\ensuremath{c}\xspace}                 
 \def\Pd      {\ensuremath{d}\xspace}                 

 \def\Ps      {\ensuremath{s}\xspace}                 
                  
 \def\Pu      {\ensuremath{u}\xspace}                 
 \def\thebaroffset{0.18em}
}
\newcommand{\offsetoverline}[2][\thebaroffset]{\kern #1\overline{\kern -#1 #2}}%

\makeatletter
\ifcase \@ptsize \relax% 10pt
  \newcommand{\miniscule}{\@setfontsize\miniscule{4}{5}}% \tiny: 5/6
\or% 11pt
  \newcommand{\miniscule}{\@setfontsize\miniscule{5}{6}}% \tiny: 6/7
\or% 12pt
  \newcommand{\miniscule}{\@setfontsize\miniscule{5}{6}}% \tiny: 6/7
\fi
\makeatother

\DeclareRobustCommand{\optbar}[1]{\shortstack{{\miniscule (\rule[.5ex]{1.25em}{.18mm})}
  \\ [-.7ex] $#1$}}

\def\g      {{\ensuremath{\Pgamma}}\xspace}

\def\uquark    {{\ensuremath{\Pu}}\xspace}
\def\uquarkbar {{\ensuremath{\overline \uquark}}\xspace}

\def\dquark    {{\ensuremath{\Pd}}\xspace}
\def\dquarkbar {{\ensuremath{\overline \dquark}}\xspace}

\def\squark    {{\ensuremath{\Ps}}\xspace}
\def\squarkbar {{\ensuremath{\overline \squark}}\xspace}

\def\cquark    {{\ensuremath{\Pc}}\xspace}
\def\cquarkbar {{\ensuremath{\overline \cquark}}\xspace}

\def\bquark    {{\ensuremath{\Pb}}\xspace}
\def\bquarkbar {{\ensuremath{\overline \bquark}}\xspace}

\def\pion   {{\ensuremath{\Ppi}}\xspace}
\def\piz    {{\ensuremath{\pion^0}}\xspace}
\def\pip    {{\ensuremath{\pion^+}}\xspace}
\def\pim    {{\ensuremath{\pion^-}}\xspace}
\def\pipm   {{\ensuremath{\pion^\pm}}\xspace}
\def\pimp   {{\ensuremath{\pion^\mp}}\xspace}

\def\kaon    {{\ensuremath{\PK}}\xspace}
\def\Kbar    {{\ensuremath{\offsetoverline{\PK}}}\xspace}

\def\KorKbar {\kern \thebaroffset\optbar{\kern -\thebaroffset \PK}{}\xspace}
\def\Kz      {{\ensuremath{\kaon^0}}\xspace}
\def\Kzb     {{\ensuremath{\Kbar{}^0}}\xspace}
\def\Kp      {{\ensuremath{\kaon^+}}\xspace}
\def\Km      {{\ensuremath{\kaon^-}}\xspace}
\def\Kpm     {{\ensuremath{\kaon^\pm}}\xspace}
\def\Kmp     {{\ensuremath{\kaon^\mp}}\xspace}
\def\KS      {{\ensuremath{\kaon^0_{\mathrm{S}}}}\xspace}

\def\Kstarz  {{\ensuremath{\kaon^{*0}}}\xspace}

\def\Kstar   {{\ensuremath{\kaon^*}}\xspace}

\def\Kstarm  {{\ensuremath{\kaon^{*-}}}\xspace}
\def\Kstarpm {{\ensuremath{\kaon^{*\pm}}}\xspace}

\def\Dbar    {{\ensuremath{\offsetoverline{\PD}}}\xspace}
\def\D       {{\ensuremath{\PD}}\xspace}

\def\DorDbar {\kern \thebaroffset\optbar{\kern -\thebaroffset \PD}\xspace}
\def\Dz      {{\ensuremath{\D^0}}\xspace}
\def\Dzb     {{\ensuremath{\Dbar{}^0}}\xspace}
\def\Dp      {{\ensuremath{\D^+}}\xspace}
\def\Dm      {{\ensuremath{\D^-}}\xspace}

\def\Dmp     {{\ensuremath{\D^\mp}}\xspace}
\def\DpDm    {\ensuremath{\Dp {\kern -0.16em \Dm}}\xspace}
\def\Dstar   {{\ensuremath{\D^*}}\xspace}

\def\theDstarz{{\ensuremath{\D^{*}(2007)^{0}}}\xspace}

\def\Ds      {{\ensuremath{\D^+_\squark}}\xspace}
\def\Dsp     {{\ensuremath{\D^+_\squark}}\xspace}
\def\Dsm     {{\ensuremath{\D^-_\squark}}\xspace}

\def\Dsmp    {{\ensuremath{\D^{\mp}_\squark}}\xspace}

\def\B       {{\ensuremath{\PB}}\xspace}
\def\Bbar    {{\ensuremath{\offsetoverline{\PB}}}\xspace}

\def\BorBbar {\kern \thebaroffset\optbar{\kern -\thebaroffset \PB}\xspace}
\def\Bz      {{\ensuremath{\B^0}}\xspace}
\def\Bzb     {{\ensuremath{\Bbar{}^0}}\xspace}
\def\Bd      {{\ensuremath{\B^0}}\xspace}

\def\BdorBdbar {\kern \thebaroffset\optbar{\kern -\thebaroffset \Bd}\xspace}
\def\Bu      {{\ensuremath{\B^+}}\xspace}
\def\Bub     {{\ensuremath{\B^-}}\xspace}
\def\Bp      {{\ensuremath{\Bu}}\xspace}
\def\Bm      {{\ensuremath{\Bub}}\xspace}
\def\Bpm     {{\ensuremath{\B^\pm}}\xspace}
\def\Bmp     {{\ensuremath{\B^\mp}}\xspace}
\def\Bs      {{\ensuremath{\B^0_\squark}}\xspace}
\def\Bsb     {{\ensuremath{\Bbar{}^0_\squark}}\xspace}
\def\BsorBsbar {\kern \thebaroffset\optbar{\kern -\thebaroffset \Bs}\xspace}

\def\jpsi     {{\ensuremath{{\PJ\mskip -3mu/\mskip -2mu\Ppsi}}}\xspace}

\def\Y#1S{\ensuremath{\PUpsilon{(#1S)}}\xspace}

\def\LorLbar     {\kern \thebaroffset\optbar{\kern -\thebaroffset \PLambda}\xspace}

\newcommand{\decay}[2]{\ensuremath{\mathinner{#1\!\to #2}}\xspace}

\def\to                 {\ensuremath{\rightarrow}\xspace}

\def\grpsutw    {{\ensuremath{\mathrm{SU}(2)}}\xspace}

\def\order   {{\ensuremath{\mathcal{O}}}\xspace}

\def\CP                {{\ensuremath{C\!P}}\xspace}
\def\CPT               {{\ensuremath{C\!PT}}\xspace}

\newcommand{\phis}{{\ensuremath{\phi_{\squark}}}\xspace}
\newcommand{\betas}{{\ensuremath{\beta_{\squark}}}\xspace}

\def\AT#1     {\ensuremath{A_{\mathrm{T}}^{#1}}\xspace}           % 2

\def\C#1      {\ensuremath{\mathcal{C}_{#1}}\xspace}                       % 9
\def\Cp#1     {\ensuremath{\mathcal{C}_{#1}^{'}}\xspace}                    % 7
\def\Ceff#1   {\ensuremath{\mathcal{C}_{#1}^{\mathrm{(eff)}}}\xspace}        % 9  
\def\Cpeff#1  {\ensuremath{\mathcal{C}_{#1}^{'\mathrm{(eff)}}}\xspace}       % 7
\def\Ope#1    {\ensuremath{\mathcal{O}_{#1}}\xspace}                       % 2
\def\Opep#1   {\ensuremath{\mathcal{O}_{#1}^{'}}\xspace}                    % 7

\def\ycp        {\ensuremath{y_{\CP}}\xspace}

\newcommand{\ket}[1]{\ensuremath{|#1\rangle}}              % {b}
\newcommand{\aunit}[1]{\ensuremath{\text{\,#1}}}       
\newcommand{\tev}{\aunit{Te\kern -0.1em V}\xspace}
\newcommand{\gev}{\aunit{Ge\kern -0.1em V}\xspace}
\newcommand{\mev}{\aunit{Me\kern -0.1em V}\xspace}
\newcommand{\kev}{\aunit{ke\kern -0.1em V}\xspace}
\newcommand{\ev}{\aunit{e\kern -0.1em V}\xspace}
 
\newcommand{\mevc}{\ensuremath{\aunit{Me\kern -0.1em V\!/}c}\xspace}
\newcommand{\gevc}{\ensuremath{\aunit{Ge\kern -0.1em V\!/}c}\xspace}
\newcommand{\mevcc}{\ensuremath{\aunit{Me\kern -0.1em V\!/}c^2}\xspace}
\newcommand{\gevcc}{\ensuremath{\aunit{Ge\kern -0.1em V\!/}c^2}\xspace}
\def\fb   {\ensuremath{\aunit{fb}}\xspace}
\def\invfb   {\ensuremath{\fb^{-1}}\xspace}

\def\order{{\ensuremath{\mathcal{O}}}\xspace}
\newcommand{\chisq}{\ensuremath{\chi^2}\xspace}
\newcommand{\chisqndf}{\ensuremath{\chi^2/\mathrm{ndf}}\xspace}

\def\gsim{{~\raise.15em\hbox{$>$}\kern-.85em
          \lower.35em\hbox{$\sim$}~}\xspace}
\def\lsim{{~\raise.15em\hbox{$<$}\kern-.85em
          \lower.35em\hbox{$\sim$}~}\xspace}

\newcommand{\Imag}{\ensuremath{\mathcal{I}m}\xspace}

\def\degrees{\ensuremath{^{\circ}}\xspace}

\def\mrad{\aunit{mrad}\xspace}

\def\tell1  {TELL1\xspace}
\def\ukl1   {UKL1\xspace}

\newcommand{\eg}{\mbox{\itshape e.g.}\xspace}
\newcommand{\ie}{\mbox{\itshape i.e.}\xspace}

\newcommand{\cf}{\mbox{\itshape cf.}\xspace}

\newcommand{\vs}{\mbox{\itshape vs.}\xspace}
\newcommand{\lhcborcid}[1]{\href{https://orcid.org/#1}{\hspace*{0.1em}\raisebox{-0.45ex}{\includegraphics[width=1em]{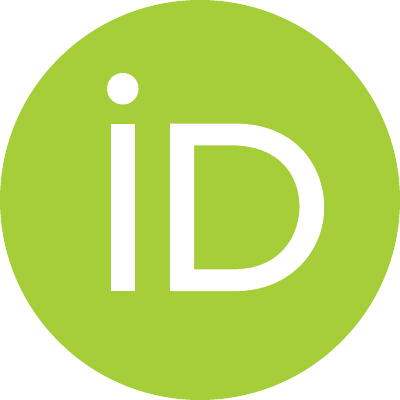}}}}

%% file: title-LHCb-PAPER.tex
% ===============================================================================
% Purpose: LHCb-PAPER journal paper title page template
% Author: 
% Created on: 2010-09-25
% ===============================================================================

%%%%%%%%%%%%%%%%%%%%%%%%%
%%%%%  TITLE PAGE  %%%%%%
%%%%%%%%%%%%%%%%%%%%%%%%%
\begin{titlepage}
\pagenumbering{roman}

% Header ---------------------------------------------------
\vspace*{-1.5cm}
\centerline{\large EUROPEAN ORGANIZATION FOR NUCLEAR RESEARCH (CERN)}
\vspace*{1.5cm}
\noindent
\begin{tabular*}{\linewidth}{lc@{\extracolsep{\fill}}r@{\extracolsep{0pt}}}
\ifthenelse{\boolean{pdflatex}}% Logo format choice
{\vspace*{-1.5cm}\mbox{\!\!\!\includegraphics[width=.14\textwidth]{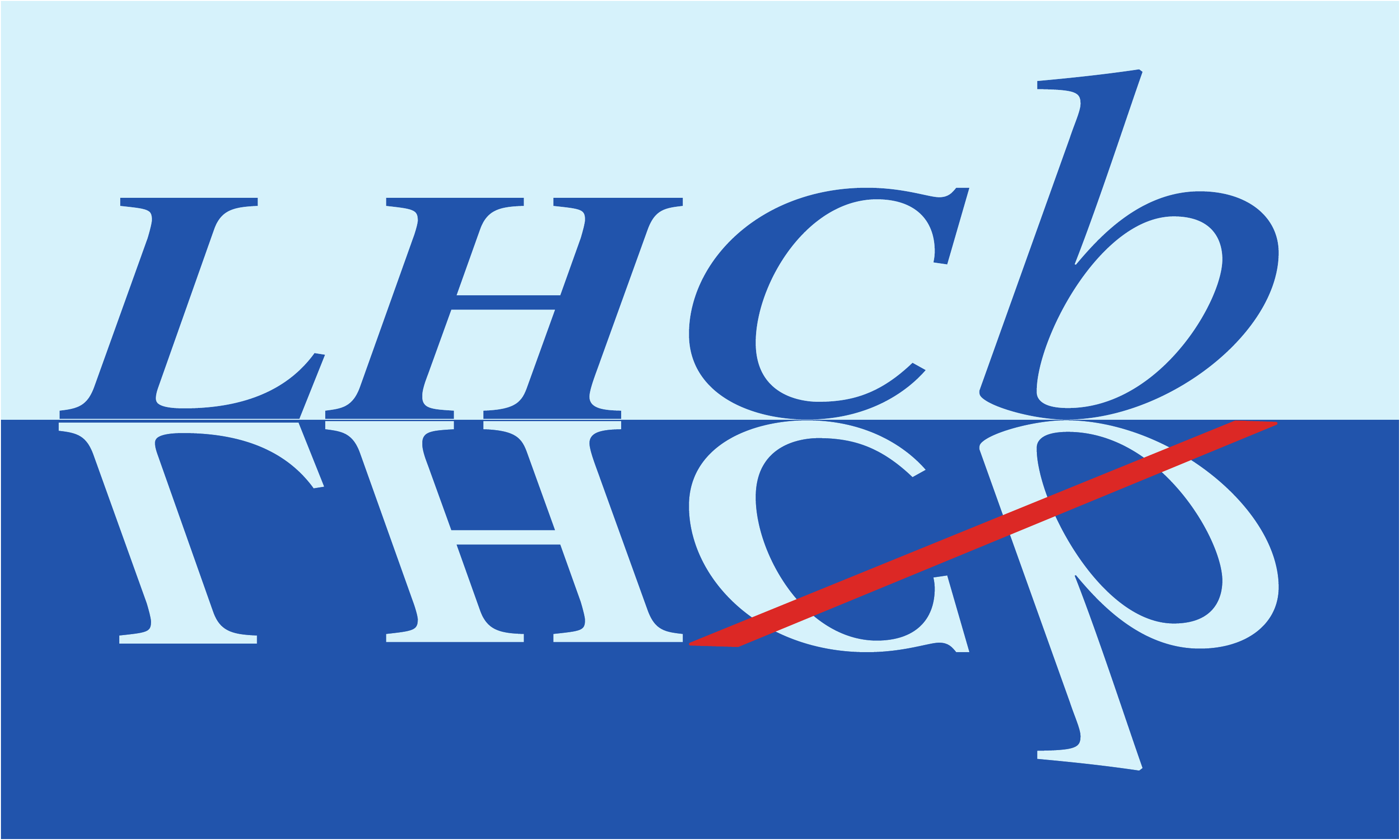}} & &}%
{\vspace*{-1.2cm}\mbox{\!\!\!\includegraphics[width=.12\textwidth]{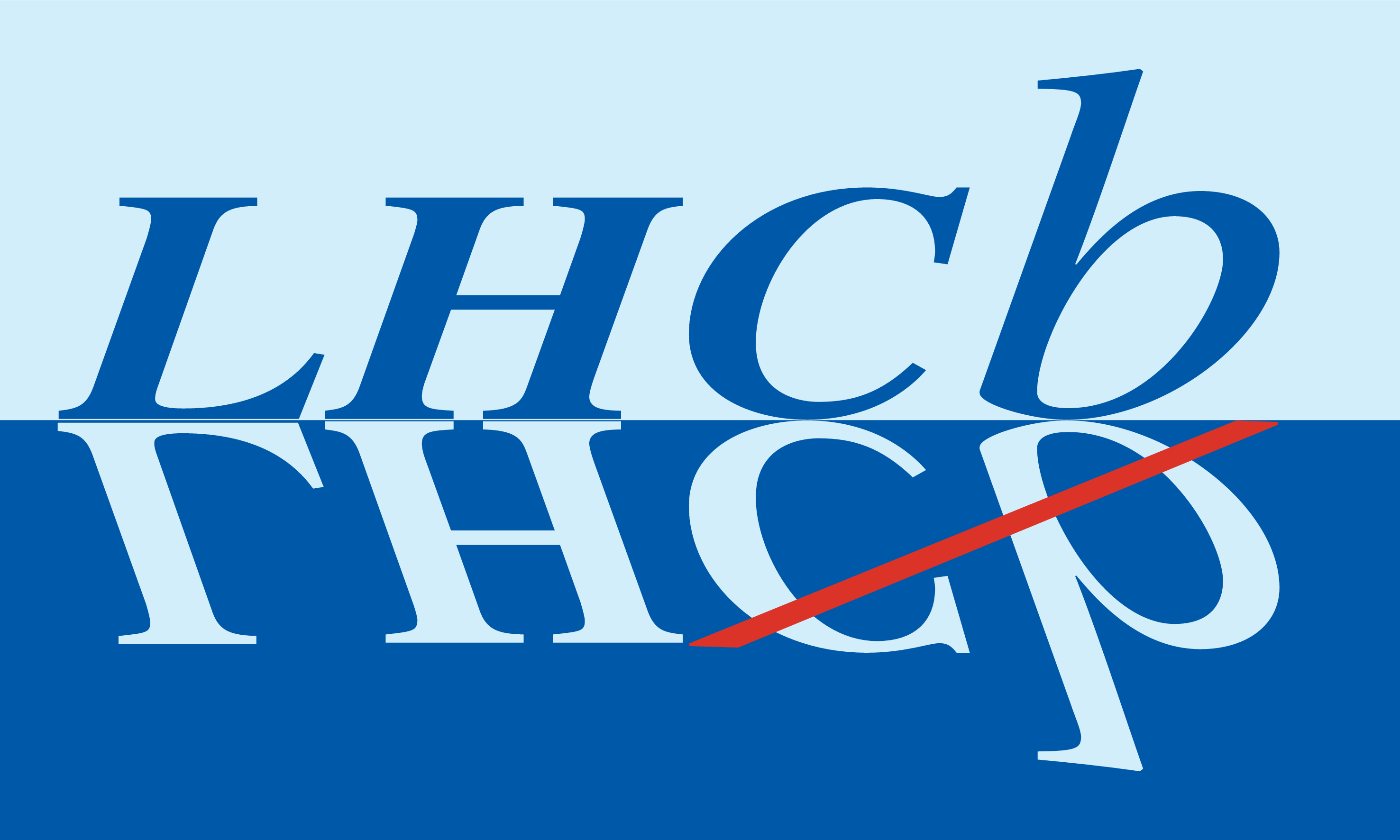}} & &}%
\\
 & & CERN-EP-2026-298 \\  % ID 
 & & LHCb-PAPER-2026-058 \\  % ID 
 & & 19 November 2026 \\
 & & \\
\end{tabular*}

\vspace*{3.0cm}

% Title --------------------------------------------------
{\normalfont\bfseries\boldmath\huge
\begin{center}
% DO NOT EDIT HERE. Instead edit macro in main.tex to keep metadata correct
  \papertitle 
\end{center}
}

\vspace*{2.0cm}

% Authors -------------------------------------------------
\begin{center}
% Edit macro in main.tex to keep metadata correct
\paperauthors\footnote{Authors are listed at the end of this paper.}
\end{center}

\vspace{\fill}

% Abstract -----------------------------------------------
\begin{abstract}
  \noindent
  Measurements of the CKM angle $\gamma$ and of the charm mixing and $C\!P$-violation parameters enable stringent and complementary tests of the weak sector of the Standard Model.
  The two categories of measurements are intimately related, since they both depend on hadronic parameters of $D^0$ meson decays, and knowledge of charm mixing is required to determine $\gamma$.
  The progress made in these sectors at the LHCb experiment is reviewed, and a new simultaneous combination of all relevant LHCb measurements is presented.
  The combination includes measurements based on data collected during 2011--2018, as well as the first measurement of $\gamma$ with data collected in 2024.
  The results,
  \mbox{$\gamma = \left(63.2^{\,+\,2.3}_{\,-\,2.2}\right)^{\circ}$} modulo $180^{\circ}$,
  \mbox{$x_{12} = \left(4.2^{\,+\,0.4}_{\,-\,0.5}\right)\times10^{-3}$},
  \mbox{$y_{12} = \left(6.18^{\,+\,0.17}_{\,-\,0.18}\right)\times10^{-3}$},
  \mbox{$\phi^M_2 = \left(4^{\,+\,11}_{\,-\,10}\right)\,\mathrm{mrad}$} and
  \mbox{$\phi^{\Gamma}_2 = \left(19^{\,+\,23}_{\,-\,24}\right)\,\mathrm{mrad}$},
  are the most precise to date.
  The $C\!P$ asymmetries in $D^0\to K^+K^-$ and $D^0\to\pi^+\pi^-$ decays are also determined under different theoretical assumptions.
\end{abstract}

\vspace*{2.0cm}

\begin{center}
  Submitted to Rep.~Prog.~Phys.
  % Citation style from https://iopscience.iop.org/journal/0034-4885
\end{center}

\vspace{\fill}

{\footnotesize 
% Edit macro in main.tex to keep metadata correct
\centerline{\copyright~\papercopyright. \href{\paperlicenceurl}{\paperlicence}.}}
\vspace*{2mm}

\end{titlepage}

%%%%%%%%%%%%%%%%%%%%%%%%%%%%%%%%
%%%%%  EOD OF TITLE PAGE  %%%%%%
%%%%%%%%%%%%%%%%%%%%%%%%%%%%%%%%

%  empty page follows the title page ----
\newpage
\setcounter{page}{2}
\mbox{~}

%% file: body.tex
\section{Introduction}
\label{sec:introduction}

The Cabibbo--Kobayashi--Maskawa (CKM) matrix describes the misalignment between the weak-interaction and mass eigenstates of quarks, the latter of which are determined by their Yukawa couplings to the Higgs field~\cite{Cabibbo:1963yz,Kobayashi:1973fv}.
It is an approximately diagonal $3 \times 3$ unitary matrix with complex elements, $V_{ij}$, that describe the coupling strength between the up-type quark $i$ and the down-type quark $j$ under the charged-current weak interaction.
A single irreducible complex phase in this matrix is the only experimentally established source of charge--parity (\CP) violation in the Standard Model (SM).
This \CP violation, however, is far too small to generate the baryon asymmetry observed in the Universe, so additional sources of \CP violation must exist~\cite{vandeVis:2025efm}.
These additional interactions beyond the SM (BSM) may also be responsible for the hierarchical structure of the CKM matrix and of quark masses~\cite{Arkani-Hamed:2026uos}, which in the SM is accommodated by a large number of free parameters.
Establishing whether the CKM picture provides a complete description of quark-flavour dynamics at the energies currently accessible at particle colliders, and quantifying the room that remains for BSM physics, is therefore one of the central goals of flavour physics.

Unitarity implies the relations $\sum_k V^{\phantom{*}}_{ki} V^*_{kj} = 0$ and $\sum_k V^{\phantom{*}}_{ik} V^*_{jk} = 0$ for any $i \neq j$, each of which can be represented by a triangle in the complex plane.
Of the six triangles so obtained, one is conventionally singled out and referred to as \emph{the} Unitarity Triangle.
It corresponds to the columns $i=d$, $j=b$, has three sides of comparable length, and is defined by the relation
\begin{equation*}
	V^{\phantom{*}}_{ud}V^*_{ub} + V^{\phantom{*}}_{cd}V^*_{cb} + V^{\phantom{*}}_{td}V^*_{tb} = 0 \, .
\end{equation*}
The SM can be stringently tested by overconstraining this triangle. 
Each side and angle can be independently determined experimentally and compared with the indirect value inferred from a global fit to all remaining observables.

The internal angles of the Unitarity Triangle, $\alpha$, $\beta$ and \g (also denoted as $\phi_2$, $\phi_1$ and $\phi_3$, respectively), are \CP-violating weak phases, of which only
\mbox{$\g \equiv \arg{(-V^{\phantom{*}}_{ud}V^*_{ub}/V^{\phantom{*}}_{cd}V^*_{cb} )}$}
can be determined using solely measurements of tree-level decays~\cite{Gronau:1991dp,Gronau:1990ra,Atwood:1996ci,Atwood:2000ck,Bondar,Giri:2003ty,Belle:2004bbr,Grossman:2002aq}.
The most precise of these determinations exploit the interference between $b\to c\bar{u}s$ and $b\to u\bar{c}s$ amplitudes in decays such as \BuDK, in which the \D meson is reconstructed in a final state common to \Dz and \Dzb decays.
The theoretical uncertainty on the interpretation of experimental observables related to these decays in terms of the angle $\gamma$ is due to second-order electroweak corrections and is negligible~\cite{Brod:2013sga}, under the assumption that BSM effects are not sizeable at tree level~\cite{Brod:2014bfa,Lenz:2019lvd}.
Conversely, all hadronic parameters can be directly determined from data.
This makes \g unique among the parameters of the Unitarity Triangle:
it is a standard candle whose direct determination is limited by experimental precision alone, with no hadronic theoretical ambiguity.
Discrepancies between direct measurements of \g and the indirect value derived from global fits, which assume the validity of the SM and hence unitarity of the CKM matrix, would be an unambiguous sign of BSM physics.
Furthermore, comparisons between values of \g extracted from decays of different beauty-meson (\B) species provide sensitivity to possible BSM effects at tree level, given the different topologies of the Feynman diagrams involved~\cite{Brod:2014bfa,Lenz:2019lvd}.

For much of the history of the field, this comparison was of limited value because the direct determination was far less precise than the indirect one.
Combinations of the full \mbox{BaBar} and Belle datasets, both reported in 2013, had uncertainties in excess of $10\degrees$, giving \mbox{$\g = (69^{\,+\,17}_{\,-\,16})\degrees$}~\cite{Lees:2013nha} and \mbox{$\g = (68^{\,+\,15}_{\,-\,14})\degrees$}~\cite{Trabelsi:2013uj}, respectively. 
The first LHCb combination, published in the same year, was of comparable precision, \mbox{$\g = (72^{\,+\,15}_{\,-\,16})\degrees$}~\cite{LHCb-PAPER-2013-020}.
The decade since has seen a dramatic change in the situation, driven by the vast beauty datasets predominantly accumulated by the LHCb experiment.
The LHCb combinations reported \mbox{$\g = (72.2^{\,+\,6.8}_{\,-\,7.3})\degrees$} in 2016~\cite{LHCb-PAPER-2016-032} and \mbox{$\g = (65.4^{\,+\,3.8}_{\,-\,4.2})\degrees$} in 2021~\cite{LHCb-PAPER-2021-033}.
Progress has not been confined to LHCb: a combination of Belle and Belle~II results gives \mbox{$\gamma = (78.6^{\,+\,7.2}_{\,-\,7.3})\degrees$}~\cite{Belle:2024knt}, an important independent check in a complementary experimental environment.
The uncertainty on the direct determination thus fell by a factor of approximately three between the 2013 and 2021 LHCb combinations, and now approaches that of the indirect determinations, \mbox{$\g = \gCKMfitter$}~\cite{CKMfitter2015} in a frequentist framework and \mbox{$\g = \gUTfit$}~\cite{UTfit-UT} in a Bayesian one.
Closing the remaining gap in precision between the direct and indirect determinations is a key goal of the LHCb experiment, and their comparison is an increasingly incisive test of the SM.
The determination of \g is dependent on \D hadronic parameters, and moreover is affected by \Dz--\Dzb mixing. 
Therefore, its determination is intimately connected to the charm sector.

Neutral charm mesons provide a complementary test of the CKM mechanism.
Both \Dz--\Dzb oscillation and \CP violation in \Dz decays arise due to flavour-changing neutral currents (FCNCs) that do not occur at tree level in the SM, and can therefore be significantly affected by BSM interactions.
The \Dz meson is the only neutral meson composed of up-type quarks in which these processes occur, thus it provides information that is complementary to that from the \Kz and $B_{(s)}^0$ mesons composed of down-type quarks.
Finally, sensitivity to BSM interactions is enhanced by the fact that FCNCs in the charm system are accidentally much more suppressed by the Glashow--Iliopoulos--Maiani mechanism~\cite{Glashow:1970gm,Buras:2020_GIM} than in \Kz and $B^0_{(s)}$ mesons,
due to the smallness of the CKM matrix elements connecting the first two quark generations to the third, and to the much smaller mass of the down-type quarks running in the Feynman-diagram loops with respect to the top quark that enters the loops in \Kz and $B_{(s)}^0$ mesons.
While uncertainties on nonperturbative quantum-chromodynamics (QCD) effects limit the theoretical interpretation of available measurements of mixing and \CP violation~\cite{Lenz:2020awd,Pajero:2022vev,Friday:2025gpj}, even assuming that the observed values are entirely due to BSM interactions sets limits on the scale of these interactions that are more stringent than those for $B_{(s)}^0$ mesons, and comparable to those for \Kz mesons~\cite{EuropeanStrategyforParticlePhysicsPreparatoryGroup:2020qin}.

Progress in the charm sector over the past decade has been as rapid as for beauty decays.
The experimental precision is dominated by LHCb measurements~\cite{HFLAV:2024ctg}.
In particular, \CP violation in charm decays was first observed in 2019 by the LHCb experiment~\cite{LHCb-PAPER-2019-006}, which also established a nonzero mass difference between the neutral charm-meson eigenstates in 2021~\cite{LHCb-PAPER-2021-009}, almost fifteen years after the first observation of \Dz--\Dzb mixing~\cite{BaBar:2007kib,BELLE:2007dgh,Petersen:2007vf}.
Conversely, \CP violation in charm mixing continues to elude experimental detection.

This article reviews the progress made in the determination of \g and of the charm mixing and \CP-violation parameters at LHCb during Runs~1 and 2 of the \lhc, together with the first result from Run~3, and presents an updated simultaneous combination of all relevant LHCb measurements.
It is a continuation of the combinations reported in Refs.~\cite{LHCb-PAPER-2013-020,LHCb-PAPER-2016-032,LHCb-PAPER-2021-033}, and supersedes those results as well as the preliminary combinations of Refs.~\cite{LHCb-CONF-2022-003-adapted,LHCb-CONF-2024-004-adapted,LHCb-CONF-2025-003}.
Relative to Ref.~\cite{LHCb-PAPER-2021-033}, the scope has broadened in multiple respects.
First, a substantial number of new or updated beauty measurements are included, among them measurements of \g using \BuDsth~\cite{LHCb-PAPER-2023-012,LHCb-PAPER-2023-029},
\BuDKst~\cite{LHCb-PAPER-2024-023}, \BdDKst~\cite{LHCb-PAPER-2023-009,LHCb-PAPER-2023-040}, and \BuDh decays with multibody \D final states~\cite{LHCb-PAPER-2022-017,LHCb-PAPER-2025-019},
together with updated decay time dependent measurements of \BsDsK decays~\cite{LHCb-PAPER-2024-020}.\footnote{Throughout the article, $D^{(*)}$ denotes an arbitrary superposition of $\D^{(*)0}$ and $\Dbar{}^{(*)0}$ mesons, \hpm denotes either a \pipm or a \Kpm meson, and
charge conjugation is implied unless stated otherwise.
Natural units, with $c=\hbar=1$, are used.}
The first \g measurement using data collected during Run~3 with the upgraded LHCb detector, recorded in 2024 alone and corresponding to an integrated luminosity of $5.8\invfb$, already reaches a precision of $6.7\degrees$~\cite{LHCb-PAPER-2026-010} and is also incorporated into the combination.
Second, the charm inputs are extended and updated~\cite{LHCb-PAPER-2022-020,LHCb-PAPER-2022-024,LHCb-PAPER-2024-008,LHCb-PAPER-2025-029},
as is the parametrisation of charm mixing, with results reported in both the theoretical and the phenomenological parametrisations~\cite{Kagan:2020vri}.
Finally, the \CP asymmetries in \DzKK and \Dzpipi decays are now determined within the combination itself rather than treated as external constraints, and alternative theory assumptions used to determine their values from time-integrated measurements are discussed.

Throughout, the statistical procedure follows the frequentist treatment described in detail in Ref.~\cite{LHCb-PAPER-2013-020} and used in all subsequent LHCb combinations.
External constraints, principally from measurements of strong-phase parameters and coherence factors with quantum-correlated \D\Dbar pairs at the CLEO and \besiii experiments, and from the world averages of $\beta$ and $\phi_s$, are treated as auxiliary inputs.

The remainder of this article is organised as follows.
The formalism of charm mixing and of \CP violation in charm decays is set out in \cref{sec:charm}, along with definitions of the charm observables.
\Cref{sec:beauty} fixes the notation used for beauty decays and illustrates the two ingredients that make a joint treatment with charm necessary and desirable: the corrections to the \B decay rates induced by charm mixing, and the dependence of these rates on the same strong \D parameters that charm observables are sensitive to.
The methods used to determine \g in the beauty sector, both time-integrated and time-dependent, are also described.
The full set of experimental inputs, and the external constraints applied to them, are given in \cref{sec:inputs}, as is a description of the statistical treatment.
The results of the combination are presented in \cref{sec:results}, and conclusions and an outlook towards the precision achievable with future LHCb data are provided in \cref{sec:conclusions}.

\section{Charm sector}
\label{sec:charm}

The theoretical and phenomenological parametrisations of charm mixing are introduced in \cref{sec:charm-mixing}, followed by the parametrisation of \CP violation in \Dz decays in \cref{sec:charm-cpv}.
The observables included in the combination are then described in \cref{sec:charm-observables}, along with their expressions in terms of the theoretical parameters and a discussion of the assumptions that their interpretation is based on.
Expressions for the same observables in the phenomenological parametrisation are given in \cref{app:phenomenological}.

\subsection{Mixing formalism}
\label{sec:charm-mixing}

Quantum oscillations of \Dz into \Dzb mesons and vice versa can be conveniently described through the two dimensionless mixing parameters
\mbox{$\xth \equiv 2 \lvert M_{12} \rvert / \Gamma$} and
\mbox{$\yth \equiv \lvert \Gamma_{12} \rvert / \Gamma$},
where \mbox{$\bm{H}\equiv \bm{M} - \tfrac{i}{2}\bm{\Gamma}$} is the effective Hamiltonian of the Hilbert space spanned by the \ket{\Dz} and \ket{\Dzb} flavour eigenstates, $\bm{M}$ and $\bm{\Gamma}$ are Hermitian matrices, and $\Gamma$ is the \Dz decay width (\CPT symmetry is assumed)~\cite{Grossman:2009mn,Kagan:2009gb,Kagan:2020vri}.
In this process, \CP violation is described by two additional \CP-odd weak phases, for which a convention must be chosen, since $\bm{H}$ is not invariant under rephasing of the fields of the \cquark and \uquark quarks and of \Dz mesons.
A natural choice is to rely on the U-spin decomposition of the matrix elements $M_{12}$ and $\Gamma_{12}$.\footnote{
    U-spin is the \grpsutw group that describes the approximate symmetry of QCD interactions under the exchange of \dquark and \squark quarks.
}  
In the SM, $M_{12}$ can be written as
\begin{equation}
\label{eq:M12}
    M_{12}^{\rm SM} =   \frac{(\lambda_s - \lambda_d)^2}{4} M_2
                      + \frac{(\lambda_s - \lambda_d)\lambda_b}{2} M_1
                      + \frac{(\lambda_b)^2}{4} M_0,
\end{equation}
see Sec.~IV.A of Ref.~\cite{Kagan:2020vri}, where $\lambda_i \equiv V_{ci}V^\ast_{ui}$ and $M_{0,1,2}$ are the $\Delta U_3 = 0$ matrix elements of the $\Delta U = 0, 1, 2$ multiplets, respectively.
The subscript of $M_i$ also denotes its order in U-spin breaking, implying that $M_0$ is the only non-vanishing term in the U-spin limit.
While the hadronic matrix element $M_2$ is the most suppressed by approximate U-spin symmetry, the first term in \cref{eq:M12} is the expected dominant contribution to $M_{12}^{\rm SM}$ due to the strong hierarchy of the CKM elements, $\left\lvert\lambda_b\right\rvert \ll \left\lvert\lambda_s - \lambda_d\right\rvert$.
An analogous expansion holds also for $\Gamma_{12}$.
The two weak phases governing \Dz mixing are then defined as the phases of $M_{12}$ and $\Gamma_{12}$ with respect to their $\Delta U = 2$ SM components,
\begin{equation}
    \phiM \equiv \arg\left(\frac{M_{12}}{\left(\lambda_s - \lambda_d\right)^2 \lvert M_{2}\rvert}\right),\qquad
    \phiG \equiv \arg\left(\frac{\Gamma_{12}}{\left(\lambda_s - \lambda_d\right)^2 \lvert\Gamma_{2}\rvert}\right).
\end{equation}
Each of the phases \phiM and \phiG equals zero or $\pi$ in the limit of \CP symmetry.
As the matrices $\bm{M}$ and $\bm{\Gamma}$ describe dispersive and absorptive amplitudes, respectively, the \xth and \phiM (\yth and \phiG) parameters are often referred to as dispersive (absorptive).
This parametrisation of charm mixing is referred to as \emph{theoretical}, since it ensures a direct connection between the experimental observables and the matrix elements $M_{12}$ and $\Gamma_{12}$ that are the subject of theory predictions.
It also provides more compact expressions for the experimental observables, with most observables depending on only two rather than four parameters, and highlights their dependence on dispersive or absorptive processes.

An alternative \emph{phenomenological} parametrisation, previously used in the literature, is based on the definition of the eigenstates of $\bm{H}$, \mbox{$\ket{D_{1,2}} \equiv p\ket{\Dz}\pm q\ket{\Dzb}$}, where $p$ and $q$ are complex numbers such that $|p|^2 + |q|^2=1$,
and the $D_1$ ($D_2$) state corresponds to the $+$ ($-$) sign.
The mixing parameters are defined as $x \equiv (m_2 - m_1) / \Gamma$ and $y \equiv (\Gamma_2 - \Gamma_1) / 2\Gamma$, where $m_i$ ($\Gamma_i$) is the mass (decay width) of $\ket{D_{i}}$.
Violation of the \CP symmetry in the mixing amplitudes is then parametrised by \qopD and \phiD, where \phiD is the relative phase of $q/p$ with respect to the complex conjugate of the $\Delta U = 2$ SM component of $\Gamma_{12}$~\cite{Kagan:2020vri},
\begin{equation}
    \phiD \equiv \arg\left(\frac{q}{p}\left(\lambda_s - \lambda_d\right)^2 \Gamma_{2}\right).
\end{equation}
In the limit of \CP symmetry, $\qopD = 1$ and $\phiD \in \{0, \pi\}$.
In contrast, \CP violation in mixing and in interference between the decays with and without mixing corresponds to the conditions $\qopD \neq 1$ and $\phiD \notin \{0, \pi\}$, respectively, each of which depends on a single parameter, while in the theoretical parametrisation \CP violation in mixing corresponds to the condition $\sin\phi_{12} \neq 0$, where $\phi_{12} \equiv \phiM - \phiG$.

The two parametrisations are equivalent; expressions to change from (\xth, \yth, \phiM, \phiG) to (\xD, \yD, \qopD, \phiD) can be found in Ref.~\cite{Kagan:2020vri}.
In the limit of small \CP violation in charm mixing, $\xth \approx \lvert x \rvert$ and $\yth \approx \lvert y \rvert$ up to relative corrections that are second order in $\sin\phi_{12}$.
Whereas all the theoretical parameters are observable, in the phenomenological parametrisation the overall sign of \xD and \yD is not, and \phiD is observable only modulo $\pi$:
the transformation $q \to -q$ swaps the definitions of $D_1$ and $D_2$, which flips the sign of both \xD and \yD, while also shifting the value of \phiD by $\pi$.
In the following, the conventions that $\CP\ket{\Dz} = - \ket{\Dzb}$ and that $D_1$ ($D_2$) is approximately \CP odd (even) are adopted, whereby \phiD is approximately 0 rather than $\pi$.

Both mixing parameters arise only at second order in U-spin breaking~\cite{Falk:2001hx}, and are proportional to the second power of $\lambda$ in the Wolfenstein parametrisation of the CKM matrix~\cite{Wolfenstein:1983yz,Buras:1994ec,Charles:2004jd}, where $\lambda = \sin\theta_\textnormal{C} \approx 0.225$~\cite{CKMfitter2015} and $\theta_\textnormal{C}$ is the Cabibbo angle.
As a result, they are both of order $10^{-2}$, such that mixing is two orders of magnitude smaller than in any other neutral-meson system.
Their size is dominated by long-distance contributions that have so far resisted reliable theoretical prediction~\cite{Lenz:2020awd,Pajero:2022vev,Friday:2025gpj,Xie:2025hzx,DiCarlo:2025mnm,Dulibic:2025emg}.
Providing precise predictions for the \CP-violating phases \phiM and \phiG is equally challenging; available estimates lie in the range 1--5\mrad~\cite{Kagan:2020vri}.

\subsection{\texorpdfstring{{\boldmath \CP} violation in {\boldmath \Dz} decays}{CP violation in D0 decays}}
\label{sec:charm-cpv}

The decays of \Dz mesons can be classified according to the order of their decay amplitudes in $\lambda$.
Amplitudes proportional to the zeroth, first and second power of $\lambda$ are referred to as Cabibbo-favoured (CF), singly Cabibbo-suppressed (SCS) and doubly Cabibbo-suppressed (DCS), respectively.

Violation of the \CP symmetry is negligible in most CF and DCS decays, such as \DzKpi and \DzKpipipi, since these decays
correspond to quark-level transitions involving quarks of four different flavours, \decay{\cquark}{\uquark\squark\dquarkbar} or \decay{\cquark}{\uquark\dquark\squarkbar}.
Therefore, they are insensitive to QCD penguin and chromomagnetic dipole operators.
Building explicit models that introduce sizeable \CP violation in these decays is challenging~\cite{Bergmann:1999pm}.
An exception is given by decays that result from a superposition of amplitudes containing \Kz and \Kzb mesons, such as \DzKSpipi, where the interference of the DCS and CF amplitudes can introduce asymmetries up to the $10^{-4}$ level~\cite{Bigi:1994aw,Pajero:2022vev}.
These effects are undetectable within current experimental sensitivity, and in \lhcb measurements they are accounted for by systematic uncertainties~\cite{LHCb-PAPER-2015-042,LHCb-PAPER-2019-001,LHCb-PAPER-2021-009,LHCb-PAPER-2022-020}.
Therefore, the present combination neglects \CP violation in all CF and DCS decays.

Conversely, \CP violation is expected at a detectable level in SCS \decay{\cquark}{\uquark\squark\squarkbar} and \decay{\cquark}{\uquark\dquark\dquarkbar} transitions, and indeed \CP violation has been observed in \Dzhh decays~\cite{LHCb-PAPER-2019-006,LHCb-PAPER-2022-024}.
Assuming unitarity of the CKM matrix, the amplitudes of these decays can be parametrised as
\begin{equation}
    \label{eq:d0-to-hh-amplitude}
    A(\Dzhh) \equiv \pm \Ahh{1} \, \frac{\lambda_s^\ast - \lambda_d^\ast}{2} + \Ahh{0} \, \lambda_b^\ast,
\end{equation}
where the plus (minus) sign refers to the \KK (\pipi) final state and the superscript of \Ahh{i} indicates the U-spin difference between the initial and final states.
The amplitudes \Ahh{1} and \Ahh{0} are often referred to as ``tree'' and ``penguin'', even though this does not directly correspond to the topologies of the Feynman diagrams involved, \eg, \Ahh{0} also receives contributions from (long-distance) penguin contractions of tree operators, and \Ahh{1} contributions from the U-spin breaking difference of the penguin contributions (including penguin contractions) with inner \squark and \dquark quarks~\cite{Grossman:2019xcj}.
It follows from \cref{eq:d0-to-hh-amplitude} that the \CP asymmetries in the decay are 
\begin{equation}
    \label{eq:acphh-param}
    \AcpDhh \approx \pm 4 \, \Imag\left(\frac{\lambda_b}{\lambda_s - \lambda_d}\right) \,
    \rQCD
    \sin\ddhhQCD,
\end{equation}
where $\rQCD \equiv \lvert \Ahh{0} / \Ahh{1}\rvert$, $\ddhhQCD \equiv \arg(\Ahh{0} / \Ahh{1})$, and terms of higher order in $\left\lvert\lambda_b/(\lambda_s-\lambda_d)\right\rvert$ are neglected.
The CKM factor
$4 \, \Imag(\lambda_b / (\lambda_s - \lambda_d)) \approx 1.3 \times 10^{-3}$~\cite{CKMfitter2015}
suppresses these asymmetries in the SM, making them sensitive probes of BSM physics.
However, the hadronic factors $\rQCD \sin\ddhhQCD$ are governed by nonperturbative QCD dynamics that cannot yet be computed reliably.
The measured asymmetries~\cite{LHCb-PAPER-2019-006,LHCb-PAPER-2022-024} require $\rQCD \sin\ddhhQCD$ to be of order unity~\cite{Grossman:2019xcj}, and whether this is consistent with the SM remains under discussion~\cite{Lenz:2020awd,Pajero:2022vev,Friday:2025gpj,Pich:2023kim,Pich:2026zux,Mohamed:2026hab}.
Sum rules relating the \CP asymmetries of different decays,
derived from approximate symmetries of QCD such as isospin and U-spin,
have been proposed as a more reliable, if not conclusive, way of testing the consistency of the measured asymmetries with the SM~\cite{Grossman:2012ry,Hiller:2012xm,Muller:2015lua,Dery:2021mll,Gavrilova:2022hbx,Gavrilova:2024npn,Fleischer:2025zhl,Gavrilova:2026ryc,Grossman:2026qew}.
For \Dzhh decays, the U-spin limit implies $\AKK{i} = \Apipi{i}$ for $i = 0,1$ and $\ddKKQCD = \ddpipiQCD$, which lead to
\begin{equation}
    \label{eq:u-spin}
    \frac{\AcpDKK}{\AcpDpipi} = -1.
\end{equation}
However, U-spin symmetry is known to be broken in the branching fractions of $\mbox{\DzKK}$ and \Dzpipi decays, possibly due to large rescattering and hence broken-penguin contributions to the amplitudes~\cite{Grossman:2019xcj}, although this interpretation is not universally accepted~\cite{Pich:2023kim,Pich:2026zux}.
An improved prediction based on U-spin symmetry can be made by allowing for the known U-spin breaking in \Ahh{1}, but not in \Ahh{0},
\begin{equation}
    \label{eq:u-spin-improved}
    \frac{\AcpDKK}{\AcpDpipi}\left\lvert\frac{\AKK{1}}{\Apipi{1}}\right\rvert = -1,
\end{equation}
where $\lvert\AKK{1}/\Apipi{1}\rvert$ can be calculated from the experimental value of the branching fractions~\cite{PDG2026} according to \namecref{eq:u-spin}~(2.12) of Ref.~\cite{Grossman:2019xcj}.\footnote{
  The contributions of \Ahh{0} to the branching fractions are negligible due to the smallness of the CKM matrix elements multiplying them in \cref{eq:d0-to-hh-amplitude}.}
This relation still relies on the assumption $\sin\ddKKQCD = \sin\ddpipiQCD$.
If these strong phases are small, even a mild U-spin breaking in them can produce large deviations from \cref{eq:u-spin-improved}, and the reliability of this relation has been questioned~\cite{Franco:2012ck}.
The predictions of \cref{eq:u-spin,eq:u-spin-improved} are nevertheless shown for reference in the figures reporting confidence regions for \AcpDpipi \vs \AcpDKK.

\subsection{Observables}
\label{sec:charm-observables}
Two classes of charm observables enter the combination: those sensitive to the mixing parameters, obtained from time-dependent analyses of \Dz decays; and time-integrated measurements of \Dzhh decays, which are sensitive to \CP violation in the corresponding decay amplitudes.
These two classes are described in \cref{sec:charm-observables-mix,sec:charm-observables-cpv-decay}, respectively.
In the following, $\tau$ is used to designate the \Dz proper decay time in units of \Dz mean lifetimes, $\tau \equiv t(\Dz) / \tau_{\Dz}$.

\subsubsection{\texorpdfstring{Mixing and {\boldmath \CP} violation in mixing}{Mixing and CP violation in mixing}}
\label{sec:charm-observables-mix}
The cleanest access to the phases \phiM and \phiG is provided by decays proceeding via a superposition of DCS and CF amplitudes, such as \DzKpPim and \DzKSpipi.
This is because in the SM final-state-dependent corrections to the mixing observables are negligible at the precision of current and planned experiments~\cite{Kagan:2020vri} or, in the case of final states involving neutral kaons, can be precisely calculated~\cite{Stahl:2014xlz,LHCb-PAPER-2026-011}.
These decays also provided some of the earliest evidence for \Dz mixing~\cite{BaBar:2007kib} and the first separate determinations of the \xth and \yth parameters~\cite{LHCb-PAPER-2021-009}.

The first analysis of this kind measured the time-dependent ratio of the decay rates of wrong-sign (WS) \DzKpPim and right-sign (RS) \DzKmPip decays,
\begin{equation}
    \label{eq:ws-to-rs-def}
	\RKpi^+(\tau) \equiv \frac{\Gamma(\decay{\Dz (\tau)}{\Kp\pim})}{\Gamma(\decay{\Dz (\tau)}{\Km\pip})},   \qquad
	\RKpi^-(\tau) \equiv \frac{\Gamma(\decay{\Dzb(\tau)}{\Km\pip})}{\Gamma(\decay{\Dzb(\tau)}{\Kp\pim})},
\end{equation}
where $\Dz(\tau)$ denotes the time evolution at time $\tau$ of
a meson created at time zero in its \Dz flavour eigenstate.
These ratios are parametrised, up to second order in the mixing parameters, as
\begin{equation}
    \label{eq:ws-to-rs}
    \RKpi^\pm(\tau) = (\rdkpi)^2 (1 \pm \AKpi)
                    + \rdkpi \sqrt{1 \pm \AKpi} (\cKpi \pm \DcKpi) \tau
                    + (\cKpipr \pm \DcKpipr) \tau^2,
\end{equation}
where
$(\rdkpi)^2$ is the \CP-averaged ratio between the DCS and CF decay widths,
\begin{equation}
    \label{eq:rdkpi}
    (\rdkpi)^2 \equiv \frac{1}{2}\left(\left\lvert\frac{A(\DzKpPim)}{A(\DzKmPip)}\right\rvert^2 + \left\lvert\frac{A(\decay{\Dzb}{\Km\pip})}{A(\decay{\Dzb}{\Kp\pim})}\right\rvert^2\right),
\end{equation}
and, neglecting \CP violation in CF decays, $\AKpi \approx \AcpDKpi$.
The terms linear and quadratic in decay time in \cref{eq:ws-to-rs} arise from the decay rates in the numerators of \cref{eq:ws-to-rs-def}, and correspond to the interference between DCS decays and CF decays following \Dz oscillation, and to CF decays following oscillation, respectively.
Conversely, the RS decays in the denominators are used as normalisation channels, to reduce the impact of efficiency variations as a function of $\tau$ on the determination of the WS observables.
They follow an approximately exponential decay in $\tau$ with negligible corrections from \Dz mixing~\cite{Pajero:2021jev}, since the tree-level decay is CF, while the decay amplitude following \Dz oscillation is DCS.
The coefficients of the terms proportional to $\tau$ and $\tau^2$ in \cref{eq:ws-to-rs} are related to the theoretical parameters by
\begin{equation}
\label{eq:ws-to-rs-ccpr}
\begin{aligned}
    \cKpi    &\approx - \yth \cos\phiG\cos\ddkpi + \xth\cos\phiM\sin\ddkpi\,,\\
    \DcKpi   &\approx - \xth \sin\phiM\cos\ddkpi - \yth\sin\phiG\sin\ddkpi\,,\\
    \cKpipr  &\approx \dfrac{1}{4} \left(\xth^2 + \yth^2\right)\,, \\
    \DcKpipr &\approx \dfrac{1}{2}\xth\yth\sin(\phiM-\phiG)\,,
\end{aligned}
\end{equation}
where \ddkpi is the strong-phase difference between CF and DCS decays,\footnote{
There are multiple conventions in the literature for the strong phase \ddkpi.
Here, the convention typically employed in $\gamma$ studies is used, in which $\ddkpi\to\pi$ in the U-spin limit, \ie the relative minus sign between the CKM matrix elements of the DCS and CF decay amplitudes is treated as a strong rather than a weak phase.
This convention is shifted by $\pi$ with respect to that employed by the HFLAV Charm group, $\delta_{K\pi}$, which is in turn the opposite of the phase $\Delta_{\Km\pip}$ defined in Ref.~\cite{Kagan:2020vri}: $\ddkpi = \pi + \delta_{K\pi} = \pi - \Delta_{\Km\pip}$.}
\begin{equation}
    \label{eq:phase-definition}
    \begin{aligned}
     \phiM + \ddkpi &\equiv \arg\left(M_{12}\frac{A(\decay{\Dz}{\Km\pip})}{A(\decay{\Dzb}{\Km\pip})}\right)\,,\\
     \phiM - \ddkpi &\equiv \arg\left(M_{12}\frac{A(\decay{\Dz}{\Kp\pim})}{A(\decay{\Dzb}{\Kp\pim})}\right)\,.
    \end{aligned}
\end{equation}
The linear term in $\tau$ in \cref{eq:ws-to-rs} is measured with much better relative precision than the quadratic term~\cite{LHCb-PAPER-2016-033,LHCb-PAPER-2017-046,LHCb-PAPER-2024-008,LHCb-PAPER-2024-044}.
As \phiM and \phiG are close to zero, the measurement of the \CP-averaged WS-to-RS ratio is mostly sensitive to the combination $- \yth \cos\ddkpi + \xth\sin\ddkpi$.
Since \ddkpi is close to $\pi$, as expected in the U-spin limit, and \xth and \yth have approximately the same size,
this is in turn mostly sensitive to \yth, with the final precision being limited by how well \ddkpi is known.
Correspondingly, the measurement of \DcKpi through the asymmetry of the WS-to-RS ratios of \Dz and \Dzb decays is mostly sensitive to \phiM and provides only limited sensitivity to \phiG.

An analogous measurement of the decay-time-dependent ratio of the WS-to-RS rates can also be performed using \DzKpipipi decays.
In this case, the strong-phase difference varies across the phase space, and thus measuring the ratio in disjoint regions of the phase space provides significant sensitivity to \xth and \phiG~\cite{LHCb-PAPER-2025-029}.
However, the sensitivity to the mixing parameters is significantly limited by the knowledge of the hadronic parameters of the \Dz decay, which are measured at charm factories~\cite{Evans:2019wza,BESIII:2026oav}.

Finally, a method similar to the WS-to-RS ratio, the so-called bin-flip method~\cite{DiCanto:2018tsd}, can be applied to the self-conjugate decay \DzKSpipi in regions of the phase space.
For this decay, the \Dz hadronic parameters are known more precisely than for \DzKpipipi decays.
The results are expressed in terms of the following four observables,
\begin{equation}
\label{eq:binflip}
\begin{aligned}
      x_{\CP}  &=  \xth \cos\phiM,
    & \Delta x &= -\yth \sin\phiG, \\
      y_{\CP}  &=  \yth \cos\phiG,
    & \Delta y &=  \phantom{-}\xth \sin\phiM,
\end{aligned}
\end{equation}
where $x_{\CP}$ and $y_{\CP}$ are \CP even and $\Delta x$ and $\Delta y$ are \CP odd.
Measurements relying on this method currently provide the most precise determination of \xth~\cite{LHCb-PAPER-2019-001,LHCb-PAPER-2021-009,LHCb-PAPER-2022-020}.
Furthermore, together with analogous evidence from \DzKpipipi decays, these measurements determine the phase \phiM to be approximately zero rather than $\pi$,
\ie that the shorter-lived mass eigenstate of the \Dz--\Dzb system is also the heavier.

Singly Cabibbo-suppressed decays can also be used to constrain \Dz mixing.
In particular, the most sensitive observable is related to the time-dependent analysis of \Dzhh decays, using the RS \DzKmPip decay as a normalisation channel.
The ratio of the \CP-averaged decay rates of these channels is given by~\cite{Pajero:2021jev}
\begin{equation}
    \label{eq:ycp-observable}
    \frac{\Gamma(\decay{\Dz(\tau)}{\hh    }) + \Gamma(\decay{\Dzb(\tau)}{\hh    })}
         {\Gamma(\decay{\Dz(\tau)}{\Km\pip}) + \Gamma(\decay{\Dzb(\tau)}{\Kp\pim})}
        \propto 1 - (\ycpHH - \ycpRS) \tau,
\end{equation}
where second-order terms in the mixing parameters are omitted, but are accounted for in the systematic uncertainties of the measurements~\cite{LHCb-PAPER-2018-038,LHCb-PAPER-2021-041},
the subleading correction \ycpRS quantifies the deviation of the RS decay rate from an exponential function in $\tau$,
and the two terms are equal to
\begin{equation}
    \label{eq:ycp-rs}
    \begin{aligned}
    \ycpHH &= \yth \cos\phi^\Gamma_{\hp\hm}, \\
    \ycpRS &= \rdkpi (\xth\cos\phiM\sin\ddkpi + \yth\cos\phiG\cos\ddkpi).
    \end{aligned}
\end{equation}
Contrary to DCS and CF decays, the phase $\phi^\Gamma_{\hh} \equiv \phiG + \delta\phi_{\hh}$ in the first expression contains a final-state-dependent contribution, $\delta\phi_{\hh}$.
However, $\delta\phi_{\hh}$ is of the order of $10^{-3}$~\cite{Kagan:2020vri}, such that relative final-state-dependent corrections to \ycpHH are of the order of $\sin\phiG\sin\delta\phi_{\hh} = \order(10^{-6})$ and can be neglected.
Therefore, in the following $\ycpHH = \ycp$ is assumed, with \ycp defined in \cref{eq:binflip}.
Note that, unlike all observables defined in the previous paragraphs, the observable in \cref{eq:ycp-observable} is sensitive only to \yth and \phiG, up to subleading corrections due to \ycpRS that are suppressed by $\rdkpi \approx 6\%$.
This is a consequence of the \hh final state being \CP-even, corresponding to a zero strong-phase difference between \Dz decays with and without mixing.
This observable currently provides the most precise determination of the mixing parameter \yth.
Moreover, it was the first to establish that \phiG is approximately zero rather than $\pi$~\cite{BELLE:2007dgh}, \ie that the nearly \CP-even \Dz eigenstate is also the shorter lived, which was later confirmed by measurements in \DzKpi, \DzKSpipi and \DzKpipipi decays.

\CP violation in mixing can also be investigated using the final state \Dzhh,
measuring the time dependence of the asymmetry between \Dz and \Dzb decays~\cite{LHCb-PAPER-2014-069,LHCb-PAPER-2016-063,LHCb-PAPER-2019-032,LHCb-PAPER-2020-045},
\begin{equation}
    \label{eq:acp-hh-time-dep}
    \frac{\Gamma(\decay{\Dz(\tau)}{\hh}) - \Gamma(\decay{\Dzb(\tau)}{\hh    })}
         {\Gamma(\decay{\Dz(\tau)}{\hh}) + \Gamma(\decay{\Dzb(\tau)}{\hh    })}
    = \AcpDhh + \DYhh \tau,
\end{equation}
where \AcpDhh is the \CP asymmetry in the decay, terms of second order in the mixing parameters are neglected, and the observable of interest \DYhh is equal to~\cite{LHCb-PAPER-2020-045}
\begin{equation}
    \label{eq:DeltaY}
    \begin{aligned}
    \DYhh &\approx - \xth \sin\phiM + \yth \cos\phiG\, \AcpDhh + \xth\cos\phiM\AcpDhh\cot\ddhhQCD, \\
          &\approx - \xth \sin\phiM + \yth \, \AcpDhh \Big(1 + \frac{\xth}{\yth}\cot\ddhhQCD\Big),
    \end{aligned}
\end{equation}
where the second line exploits the experimental observation $\phiM \approx 0$, $\phiG \approx 0$, and \ddhhQCD was defined in \cref{eq:acphh-param}.
Since the final state is \CP even, this observable depends on \xth and \phiM only, up to final-state-dependent corrections from \CP violation in the decay.
These corrections imply that external determinations of \AcpDhh and assumptions on the size of \ddhhQCD are required to use \DYhh measurements to constrain the value of \phiM, as discussed in \cref{sec:charm-observables-cpv-decay}.

A similar measurement has been performed using quasi-\CP-even \Dzpipipiz decays \cite{LHCb-PAPER-2024-003}.
The measurement is integrated over the phase space of the final state,
which dilutes the sensitivity by a factor $2\Fppp - 1$,
where \Fppp is the \CP-even fraction of \Dzpipipiz decays.
Neglecting \CP violation in the decay,
the first-order coefficient of the Taylor expansion of the time-dependent asymmetry in $\tau$ is equal to $\DYpppeff = (2 \Fppp - 1) \DY$,
where $\DY \equiv - \xth \sin\phiM = - \Delta y$.

\subsubsection{\texorpdfstring{{\boldmath \CP} violation in decay}{CP violation in decay}}
\label{sec:charm-observables-cpv-decay}
The \CP asymmetries in \Dzhh decays, \acp{\hh}, are typically determined from measurements of time-integrated \CP asymmetries~\cite{LHCb-PAPER-2014-013,LHCb-PAPER-2015-055,LHCb-PAPER-2016-035,LHCb-PAPER-2019-006,LHCb-PAPER-2022-024},
\begin{equation}
    \label{eq:acp-dz-to-hh}
    A_{\CP}(\Dzhh) \approx \acp{\hh} + \DYhh \langle \tau_{\hh} \rangle,
\end{equation}
\cf \cref{eq:acp-hh-time-dep},
where $\langle \tau_{\hh} \rangle$ is the average decay time in units of \Dz mean lifetimes in the analysed sample, and the expression for \DYhh is given in \cref{eq:DeltaY}.
While the first term on the right-hand side of \cref{eq:DeltaY} is universal, the others depend on the final state.
Due to the smallness of $\yth \, \AcpDhh$, which is of the order of $10^{-5}$, approximately one order of magnitude below the current experimental uncertainty of the \DYhh measurements, final-state-dependent contributions to \DYhh are likely negligible at the current level of experimental precision.
However, they could be sizeable if $\cot\ddhhQCD$ is enhanced to $\order(10)$.
While this scenario is unlikely, since it would require fine tuning of \ddhhQCD and would similarly enhance \rQCD in \cref{eq:acphh-param} by $\order(10)$, making a SM explanation for the size of \acp{\hh} even more challenging, it cannot be excluded a priori.
Neglecting the final-state-dependent contributions is referred to as the assumption of \emph{universality} for the charm-mixing weak phases, which is adopted in the baseline combination and is implemented by setting $\AcpDhh = 0$ on the right-hand side of \cref{eq:DeltaY}.

\section{Beauty sector}
\label{sec:beauty}

\begin{figure}[b]
  \centering
  \includegraphics[width=0.495\textwidth]{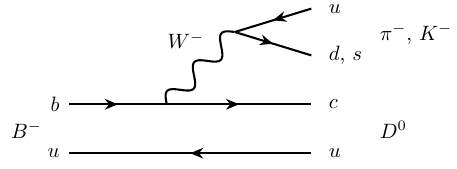}
  \includegraphics[width=0.495\textwidth]{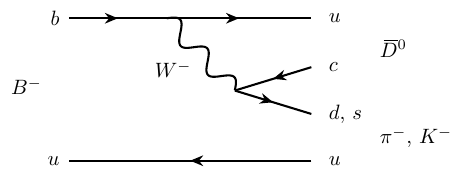}
  \caption{
    Leading-order Feynman diagrams for $B^-\to Dh^-$ decays with a (left) favoured $b \to c$ and (right) suppressed $b \to u$ transition.
    The relative suppression between the two diagrams arises both from the smaller magnitude of the CKM matrix elements involved in the right diagram,
    and from the fact that the left diagram is colour-favoured while the right diagram is colour-suppressed.
  }
  \label{fig:feynman}
\end{figure}

The CKM angle \g can be directly measured by analysing decays that are sensitive to the interference between favoured $\decay{\bquark}{\cquark\uquarkbar q}$ and suppressed $\decay{\bquark}{\uquark\cquarkbar q}$ transition amplitudes, where $q$ denotes an \squark or \dquark quark.
The ratio of the suppressed to favoured amplitudes is \mbox{$\rb e ^{i(\db\mp\g)}$}, where the $-$ or $+$ sign indicates whether the initial state contains a \bquark or \bquarkbar quark, \rb is the ratio of the amplitude magnitudes, and \db their \CP-even strong-phase difference.\footnote{
  A definition of these quantities that is invariant under rephasing of the quark and meson fields requires the subsequent \D decay to be considered simultaneously.
}
Sensitivity to \g arises only from the interference between these two amplitudes.
This is typically studied in \B decays such as \BuDh, illustrated in~\cref{fig:feynman}, where the \D meson decays to a final state, $f$, accessible to both \Dz and \Dzb mesons.
The \BuDh decay rate for two-body \B and \D decays is given by
\begin{equation}
	\Gamma(\decay{\Bmp}{\D\hmp}) \propto \left\lvert \rd e^{-i\dd} + \rb e ^{i(\db\mp\g)} \right\rvert ^2 
    = \rd^2 + \rb^2 + 2 \, \rd \rb \cos(\db+\dd\mp\g),
	\label{eq:no_mix_rate}
\end{equation}
where \rd and \dd are the magnitude ratio and the opposite of the strong-phase difference between the $\Dz\to f$ and $\Dzb\to f$ amplitudes, see \eg \cref{eq:rdkpi,eq:phase-definition}.
Here, \CP violation and mixing in charm are neglected for brevity. 
The mixing contributions to some observables are significant and are discussed later in this section. 
At the current level of precision, \CP violation in charm decays is negligible, and not incorporated in the combination.

The hadronic parameters $(\rb, \db)$ and $(\rd,\dd)$ are specific to each \B decay and subsequent \D decay, respectively, whereas the \CP-odd weak phase difference between \Bm and \Bp amplitudes, $-2\g$, is shared by all such decays.
In the case of multibody \B and \D decays, two coherence factors $\kappa_B$ and $\kappa_D$ additionally multiply the last term in the equation, to account for a dilution of the interference term due to variations in the strong-phase difference across the phase space of the decays.
Typically, the best sensitivity to \g is achieved with final states $f$ such that $\rd \leq 1$, which ensures that the size of the last term in the equation is of similar size to those that do not depend on \g.
The decay rate for the final state $\bar{f}$ is obtained from \cref{eq:no_mix_rate}, with the same proportionality coefficient, by substituting $\rd \to 1 / \rd$ and $\dd \to - \dd$ and multiplying the right-hand side by $\rd^2$.
If $\rd \ll 1$, it constitutes a useful normalisation channel for which interference effects are small.

\Cref{eq:no_mix_rate} is invariant under the transformation $(\gamma, \db) \to (\gamma + \pi, \db + \pi)$.
Therefore, the value of the angle $\gamma$ is determined only up to an ambiguity of $\pi$.
The equation contains at least three unknown parameters, more when \B coherence factors or the \D hadronic parameters are not trivial (\ie, $\kappa_B \neq1$, $\kappa_D \neq 1 $, $\rd \neq 1$ or $\dd\neq0, \,\pi$), such that the value of \g cannot be determined from a single pair of \Bmp decay rates.
The strategy that has been pursued instead, and which underpins this article, is to combine results from many different \B- and \D-decay modes to overconstrain the parameters of each \B decay, provided that the corresponding \rd, \dd and $\kappa_D$ parameters are constrained by other measurements.
In combinations prior to 2021, these charm parameters were taken exclusively as external inputs from dedicated charm-factory measurements.
The \B samples now available are large enough that \ddkpi, the strong-phase difference between the \DzKmPip and \decay{\Dzb}{\Km\pip} amplitudes defined in \cref{eq:phase-definition}, can be determined by the beauty inputs themselves with a precision comparable to that on \g and \db.
This has consequences beyond the beauty sector: since the precision attainable on \yth depends on how well \ddkpi is known, through the measurement of the WS-to-RS ratio set out in \cref{sec:charm-observables-mix},
a simultaneous combination of the beauty and charm sectors improves the precision on \yth.
Exploiting the combined constraining power of the beauty and charm sectors has become even more relevant after the recent \lhcb measurement of charm mixing in \DzKpipipi decays~\cite{LHCb-PAPER-2025-029}.
In this case, the uncertainty on the \DKpipipi hadronic parameters limits the precision not only of both charm-mixing parameters,
but also of \g as determined from \BuDh decays with \DKpipipi~\cite{LHCb-PAPER-2022-017}.

Charm mixing affects the determination of \g also in a different way, by giving rise to additional terms in \cref{eq:no_mix_rate}.
Incorporating them up to first order in \xth and \yth, the decay rate becomes~\cite{Rama:2013voa}
\begin{align}
	\Gamma(\decay{\Bmp}{\D\hmp}) \propto & \phantom{+} \rd^2 + \rb^2 + 2\kappa_D\kappa_B \rd \rb \cos(\db+\dd\mp\g) \nonumber \\
		& - \left[ (1+\rb^2)\kappa_D\rd\cos\dd + (1+\rd^2)\kappa_B\rb\cos(\db\mp\g)\right] \yth \nonumber \\
		& + \left[ (1-\rb^2)\kappa_D\rd\,\sin\dd - (1-\rd^2)\kappa_B\rb\sin(\db\mp\g)\right] \xth,
	\label{eq:with_mix_rate}
\end{align}
where the decay rate is integrated over the phase space of the \B and \D decays.
These terms have the largest impact on final states where CF and DCS \D decay amplitudes interfere.
Neglecting charm mixing would introduce a bias in the determination of $\gamma$ of the same order as the current precision. 
A simultaneous determination of \g, \xth and \yth therefore not only improves the precision on the parameters of both sectors,
but is also required for a consistent account of the effect of charm mixing on the beauty observables.

Having set out the general method employed to measure \g in \B decays,
\cref{sec:timeintg,sec:timedepg} describe the specific approaches used in time-integrated and time-dependent measurements, respectively.

\subsection{\texorpdfstring{Time-integrated measurements of {\boldmath $\gamma$}}{Time-integrated measurements of gamma}}
\label{sec:timeintg}

Time-integrated methods are classified mainly according to the final state of the neutral charm meson, as described in the following paragraphs.
All of the observables described below follow from \cref{eq:no_mix_rate}, specialised to a particular \B decay and \D final state.
It is convenient to make the amplitudes explicit, writing the two interfering \B amplitudes as
\mbox{$A(\Bm\to\Dz \hm) \equiv A_c e^{i\delta_c     }$} and
\mbox{$A(\Bm\to\Dzb\hm) \equiv A_u e^{i(\delta_u-\g)}$},
and the two \D amplitudes as
\mbox{$A(\Dz\to f     ) \equiv A_f e^{i\delta_f}$} and
\mbox{$A(\Dz\to\bar{f}) \equiv A_{\bar{f}} e^{i\delta_{\bar{f}}}$},
where the subscripts $c$ and $u$ refer to the $\bquark\to\cquark\uquarkbar q$ and $\bquark\to\uquark\cquarkbar q$ transitions.
The hadronic parameters appearing in \cref{eq:no_mix_rate} are then \mbox{$\rb = A_u/A_c$}, \mbox{$\db = \delta_u - \delta_c$},
\mbox{$\rd = A_{f} / A_{\bar{f}}$} and \mbox{$\dd = \delta_{\bar{f}} - \delta_f$}.\footnote{
    See \cref{eq:phase-definition} for a rigorous definition of \dd that is invariant under rephasing of the quark and meson fields.}
For multibody \D final states, $A_f$, $A_{\bar{f}}$, $\delta_f$ and $\delta_{\bar{f}}$ are functions of the decay phase space,
and the coherence factor $\kappa_D$ arises when that dependence is integrated over.
The rate for the \CP-conjugate process is obtained by substituting $\g \rightarrow -\g$.
The following discussion considers \BuDK decays, in which the \D mesons are directly produced in the \Bmp decays.
Analogous equations are obtained for \BuDpi decays, where sensitivity to \g is however diluted due to the smaller value of $\rb \approx 0.5\%$, to be compared with $\rb \approx 10\%$ for \BuDK decays.
The \BuDpi decays are therefore usually employed as normalisation channels to correct for experimental efficiencies that are approximately the same in \BuDK and \BuDpi decays, to constrain production asymmetries between $\Bm$ and $\Bp$ mesons, and to constrain \D hadronic parameters.
In the following, the notation $\decay{\Bm}{[f]_\D \hm}$ is used to indicate a $\decay{\Bm}{D \hm}$ decay followed by a $\decay{\D}{f}$ decay.

\subsubsection{\texorpdfstring{{\boldmath \CP} eigenstate charm meson decays}{CP eigenstate charm meson decays}}

The analysis of the decay of neutral \D mesons into two-body \CP-eigenstate final states $f_{\CP}$,
for which $\rd=1$ and $\dd=0$ ($\pi$) for \CP-even (-odd) states,
was first proposed by Gronau, London and Wyler~\cite{Gronau:1990ra,Gronau:1991dp}.
In this case, \cref{eq:no_mix_rate} becomes
\begin{equation}
  \label{eq:glw1}
  \Gamma(\Bmp\rightarrow [f_{\CP}]_\D\Kmp)\propto A_c^2(1+r_B^2 + 2 \, \etaCP\, r_B\cos(\delta_B\mp\gamma))\,,
\end{equation}
where \etaCP is the \CP eigenvalue of the final state.
To cancel most systematic uncertainties, these rates are normalised to those of quasi-flavour-specific charm decays such as \decay{\D}{\Km\pip},
where
\begin{equation}
  \label{eq:CA_finalstate}
  \Gamma(\Bmp\rightarrow [ \Kmp\pipm]_\D \Kmp)
  \propto A_c^2 (1 + \order(\rdkpi \rb))
\end{equation}
and the relative size of the interference term sensitive to \CP violation with respect to the leading one is suppressed by $\rdkpi = \order(\lambda^2)$ with respect to \cref{eq:glw1}.
From \cref{eq:glw1,eq:CA_finalstate}, and accounting for the branching fractions of the \D decays, the observables are defined as
\begin{align}
  A_{\CP\pm} & \equiv   \frac{\Gamma(B^-\rightarrow D_{\CP\pm}K^-)-\Gamma(B^+\rightarrow D_{\CP\pm}K^+)}
                             {\Gamma(B^-\rightarrow D_{\CP\pm}K^-)+\Gamma(B^+\rightarrow D_{\CP\pm}K^+)}
                             \label{eq:glw3def}\,, \\
      R_{\CP\pm} & \equiv \frac{\Gamma(\Bpm\rightarrow D_{\CP\pm}\Kpm)}{\Gamma(\Bpm\rightarrow [\Kpm\pimp]_\D\Kpm)} \times \frac{\Gamma(\Bpm\rightarrow [\Kpm\pimp]_\D \pipm)}{\Gamma(\Bpm\rightarrow  D_{\CP\pm} \pipm)} \,,
                           \label{eq:glw2def}
\end{align}
where $D_{\CP\pm}$ denote the \CP-even (-odd) eigenstates of the neutral \D meson, and the rates in \cref{eq:glw2def} denote the \CP averages of these processes.
These four observables are connected to the three unknowns \g, \rb and \db through
\begin{align}
  A_{\CP\pm} & = \pm 2 \,\rb \sin\db \sin\gamma / R_{\CP\pm} \,, \label{eq:glw3} \\
  R_{\CP\pm} & = 1+\rb^2 \pm 2 \,\rb\cos\db\cos\gamma        \,, \label{eq:glw2}
\end{align}
where terms of order $\mathcal O(\rdkpi\rb)$ and $\mathcal O(\rbdpi)$ are neglected for simplicity of presentation.
However, only three of these equations are independent since, from \cref{eq:glw3},
\mbox{$A_{\CP+} R_{\CP+} = - A_{\CP-} R_{\CP-}$}.
On top of the ambiguity $(\gamma, \db) \to (\gamma + \pi, \db + \pi)$ described in \cref{sec:beauty},
these equations are also invariant under the transformations
$(\gamma, \db) \to (\pi - \gamma, \pi - \db)$ and
$(\gamma, \db) \to (\db, \gamma)$,
leading to an eightfold ambiguity in the determination of \g. 

\subsubsection{Charm meson decays with Cabibbo-favoured and doubly Cabibbo-suppressed contributions}

In this type of analysis, first proposed by Atwood, Dunietz and Soni~\cite{Atwood:1996ci,Atwood:2000ck}, the neutral \D mesons are reconstructed in non-self-conjugate final states that receive contributions from CF and DCS amplitudes, such as $\Kp\pim$.
The \B decay rate is the result of interference between the favoured \decay{\Bm}{\Dz\Km} decay followed by the DCS $\Dz\to\Kp\pim$ decay and the suppressed $B^-\rightarrow \Dzb K^-$ decay followed by the CF $\Dzb\to\Kp\pim$ decay.
As a consequence, the interfering amplitudes are of similar magnitude, so that large interference effects that provide sensitivity to \g can occur.
\Cref{eq:no_mix_rate} becomes
\begin{equation}
  \Gamma(B^\mp\rightarrow [K^\pm\pi^\mp]_\D K^\mp) \propto r_B^2+(\rdkpi)^2 + 2\,r_B\,\rdkpi\cos(\delta_B+\ddkpi\mp\gamma)\, ,
  \label{eq:ads_rate}
\end{equation}
where both \rdkpi and \ddkpi can be measured independently in charm decays or determined directly in the combination.
The statistically independent observables $R_\mp$ are defined as
\begin{equation}
  \label{eq:Rpm_ADS}
  R_\mp \equiv \frac{\Gamma(B^\mp\rightarrow [\Kpm\pimp]_D \Kmp)}
                    {\Gamma(B^\mp\rightarrow [\Kmp\pipm]_D \Kmp)}
                    = \frac{r_B^2 + (\rdkpi)^2 + 2\, \rb\, \rdkpi\cos(\db+\ddkpi\mp\g)}
                           {1+r_B^2 (\rdkpi)^2 + 2\, \rb\, \rdkpi\cos(\db-\ddkpi\mp\g)}\,.
\end{equation}
An additional observable is based on the \CP asymmetry for the favoured
\D meson decay,
\begin{align}
A_{\rm fav} &= \frac{\Gamma(B^- \to [K^-\pi^+]_\D K^-) - \Gamma(B^+\to [K^+\pi^-]_\D K^+)}
                    {\Gamma(B^- \to [K^-\pi^+]_\D K^-) + \Gamma(B^+\to [K^+\pi^-]_\D K^+)} \nonumber \\
            &= \frac{2 \, \rb \, \rdkpi\sin(\db - \ddkpi)\sin\gamma}
                    {1 + r_B^2 (\rdkpi)^2 + 2 \rb \, \rdkpi \cos(\db - \ddkpi)\cos\gamma}\,.
\label{eq:afav}
\end{align}
The presence of the phase \ddkpi formally resolves the $(\gamma, \db) \to (\pi - \gamma, \pi - \db)$ ambiguity affecting \CP-eigenstate \D final states.
However, since \ddkpi is close to $\pi$ due to approximate U-spin symmetry, the ambiguity is not resolved in practice.
The same problem is present in multibody decays such as \DKpipipi and \DKpipiz, where the same equations apply up to the introduction of a coherence factor $\kappa_\D$.
However, their analysis can be performed separately for different regions of the phase space of the \D meson, for which the $\kappa_\D$, \rd and \dd parameters differ, lifting the degeneracy and providing an unambiguous determination of \g.

\subsubsection{Self-conjugate multibody charm meson decays}

In this approach, proposed by Bondar, Poluektov, Giri, Grossman, Soffer and Zupan~\cite{Bondar,Giri:2003ty,Belle:2004bbr}, the \D mesons are selected in multibody self-conjugate final states.
Here, self conjugacy implies that the phase-space integrated strong-phase difference between the two decays is trivial (zero or $\pi$).
Thus, if the measurement is performed in a phase-space dependent manner, no external input on strong phase differences is needed to determine $\gamma$ and all associated parameters, although in practice these inputs turn out to be advantageous. 
This is in contrast with the analysis of \DKpipipi decays described above, where the \B strong phase difference cannot be determined without the constraint on the phase-space integrated charm strong-phase difference provided by the charm factories. 

The most notable example is that of \DKSpipi decays, where the phase-space-dependent $\Dz$ and $\Dzb$ decay amplitudes can be expressed in terms of the squared masses of the $\KS\pi^-$  and $\KS\pi^+$ systems, denoted $m^{2}_-$ and $m^{2}_+$ respectively. As \CP violation in the $D$-meson decay can be neglected, the amplitudes satisfy $A_fe^{i\delta_f}=A(m_-^2,m_+^2)$ and ${A_{\bar{f}}e^{i\delta_{\bar{f}}}=A(m_+^2,m_-^2)}$, with a relative strong phase $\delta_D(m_-^2,m_+^2)$.
The rate in \cref{eq:no_mix_rate} becomes
\begin{equation}
  \label{eq:dalitzrate1}
  \begin{aligned}
  &\frac{d^2 \Gamma(B^\mp\to [\KS\pi^+\pi^-]_\D K^\mp)}{d m_-^2 \, d m_+^2} \\
  &\qquad\propto |A_\mp|^2+r_B^2|A_\pm|^2 + 2|A_\mp||A_\pm|\rb\cos(\db + \dd(m_\mp^2,m_\pm^2)\mp\gamma)\\
  &\qquad=       |A_{\mp}|^2+r_B^2|A_{\pm}|^2+2|A_\mp||A_\pm|\left(x_\mp \cos \delta_D(m_\mp^2,m_\pm^2)-y_\mp \sin \delta_D(m_\mp^2,m_\pm^2)\right)\,,
\end{aligned}
\end{equation}
where the notation is simplified using $A_{\pm}= A(m_\pm^2,m_\mp^2)$, and the \CP-violating observables $x_\pm$  and $y_\pm$ are defined as
\begin{equation}
  x_\pm=r_B \cos\left(\db\pm\g\right), \quad
  y_\pm=r_B \sin\left(\db\pm\g\right).
\end{equation}
This Dalitz-plot-based method can be implemented in a model-dependent way, by parametrising the \D-decay amplitude across the Dalitz plot, or in a model-independent way, by analysing the decays separately in disjoint regions of the Dalitz plot.
The same method can be applied to other multibody decays, such as \decay{\D}{\KS\Kp\Km} and \Dhhpipi.
However, the resulting precision is lower since the former \D decay is phase-space suppressed, and the latter proceeds through SCS amplitudes only, whereas \DKSpipi decays receive contributions from both CF and DCS amplitudes, leading to a larger branching fraction and larger interference effects.

\subsubsection{Non-self-conjugate multibody singly Cabibbo-suppressed charm decays}

In these analyses, first proposed by Grossman, Ligeti and Soffer~\cite{Grossman:2002aq}, the neutral \D mesons are selected in multibody SCS decays, such as \DKSKpi.
Two classes of decays are distinguished, the so-called same-sign (SS) $\Bm\to [\KS\Km\pip]_D \Km$ and opposite-sign (OS) $\Bm\to[\KS\Kp\pim]_D \Km$ decays,
where the nomenclature refers to the relative charge of the charged kaons from the \Bm and \D decays.
As for the \DKpipipi decays described above, \cref{eq:no_mix_rate} gives the rates of \Bmp decays
\begin{equation}
\begin{aligned}
  \Gamma(B^\mp\rightarrow [\KS\Kmp\pipm]_\D K^\mp) &\propto  1 + r_B^2 r_D^2 + 2\rb r_D \kappa_D \cos (\db \mp \g - \delta_D )\, ,  \nonumber \\
  \Gamma(B^\mp\rightarrow [\KS\Kpm\pimp]_\D K^\mp) &\propto  r_B^2 + r_D^2 + 2\rb r_D \kappa_D \cos ( \db \mp \g + \delta_D )\, ,
\end{aligned}
\end{equation}
where $\kappa_D$ is the coherence factor describing the interference in the \D decay.
Since both the \Dz and \Dzb decays are SCS, the value of \rd is much larger than \rb, unlike for the \DKpi decays described above, such that the interference effects, and hence the sensitivity to \g, are smaller.

\subsubsection{\texorpdfstring{Additional {\boldmath \B} decay modes}
                               {Additional B decay modes}
}
All the methods described above can be extended to additional \Bm decay modes, where \D mesons are produced through decays of resonances like \theDstarz, with a different set of \rb and \db parameters~\cite{Bondar:2004bi}.
Another interesting option is to apply the methods to \B decays where the \D meson is produced in association with multiple light hadrons,
such as \decay{\Bm}{\D\KS\pim} or \decay{\Bzb}{\D\Km\pip}~\cite{Gronau:2002mu,Gershon:2009qr}.
These measurements can be performed in a limited region around the \Kstar resonance, or across the entire \B Dalitz plot, to gain sensitivity similarly to what is done with multibody \D decays.

\subsection{\texorpdfstring{Time-dependent measurements of {\boldmath $\gamma$}}{Time-dependent measurements of gamma}}
\label{sec:timedepg}

It is also possible to measure \g using time-dependent analyses
of neutral \Bd and \Bs decays~\cite{Dunietz:1987bv,Aleksan:1991nh,Fleischer:2003yb}.
These methods still rely on the interference of $\decay{b}{u\cquarkbar q}$ and $\decay{b}{c\uquarkbar q}$ tree-level amplitudes; however, interference is achieved through $B^0_{(s)}$ mixing.
Therefore, the sum of \g and the mixing phase is measured, namely $\gamma+2\beta$ ($\gamma - 2\beta_s$) in the \Bz (\Bs) systems.
These measurements require external measurements of $\beta$ and $\beta_s$ to provide sensitivity to \g~\cite{HFLAV:2024ctg,HFLAV:2022esi:zenodo.19371140}.
For the $\BsDsK$ and $\Bs\to D_s^\mp K^\pm\pip\pim$ measurements, phase shifts from penguin or BSM contributions in $\phis$ determinations are assumed to be negligible, implying that $-2\betas=\phis$.
Comparisons of time-integrated determinations of \g with those from $\Bs\to D_s^\mp K^\pm(\pip\pim)$ decays therefore probe possible new \CP-violating contributions that would emerge only in the interference of the mixing amplitudes with the $b\to c\bar{u}s$ and $b\to u\bar{c}s$ amplitudes~\cite{FleischerMalami:2022}.

In $\Bs \to D_s^\mp K^\pm$ decays, both $b\to c \uquarkbar\squark$ and $b\to u\cquarkbar\squark$ amplitudes are colour-allowed and of the same order in the Wolfenstein parameter $\lambda$, $\mathcal{O}(\lambda^3)$, so that the interference effects are expected to be large.
For $f=\Dsm\Kp$, the time-dependent decay rates of the initially produced flavour eigenstates are given by
\begin{align}
  \frac{\textnormal{d} \Gamma(\Bs(t) \to f)}{\textnormal{d} t}
  \propto e^{-\Gamma_s t} \bigg[
    & \cosh\left(\frac{\Delta\Gamma_s t}{2}\right) + A^{\Delta\Gamma}_f \sinh\left(\frac{\Delta\Gamma_s t}{2}\right) \nonumber\\
    & + \ C_f \cos\left(\Delta m_s t\right) - S_f \sin\left(\Delta m_s t\right) \bigg]\,,
  \label{eq:decay_rates}
\end{align}
where $\Gamma_s=(\Gamma_{\rm L}+\Gamma_{\rm H})/2$, $\Delta\Gamma_s=\Gamma_{\rm L}-\Gamma_{\rm H}$ and $\Delta m_s=m_{\rm H}-m_{\rm L}$ are the mixing parameters
of the \Bs system.
The $C_f$ and $S_f$ terms have the opposite sign for \Bsb decays,
while the rates to the final state $\bar f = \Dsp\Km$ follow by replacing $f$ with $\bar f$ in \cref{eq:decay_rates}.
Finally, \CP violation in mixing is neglected since external experimental constraints on this phenomenon~\cite{HFLAV:2024ctg} are much tighter than the precision of the measurements described here, whose sensitivity to \CP violation therefore comes from the interference between decays with and without mixing.

Neglecting \CP violation in the decay, the six coefficients $C_f$, $C_{\bar f}$, $S_f$, $S_{\bar f}$, $A_f^{\Delta\Gamma}$ and $A_{\bar f}^{\Delta\Gamma}$ depend on $r_B$, $\delta_B$ and $\gamma-2\beta_s$ as
\begin{equation}
\begin{gathered}
C_f=-C_{\bar f} = \frac{1-r_B^2}{1+r_B^2}\,,\\[1ex]
\begin{alignedat}{2}
A_f^{\Delta\Gamma} &= -\frac{2\,r_B\cos(\delta_B-(\gamma-2\beta_s))}{1+r_B^2}\,,
&\qquad A_{\bar f}^{\Delta\Gamma} &= -\frac{2\,r_B\cos(\delta_B+(\gamma-2\beta_s))}{1+r_B^2}\,,\\[1ex]
S_f &= \frac{2\,r_B\sin(\delta_B-(\gamma-2\beta_s))}{1+r_B^2}\,,
&\qquad S_{\bar f} &= -\frac{2\,r_B\sin(\delta_B+(\gamma-2\beta_s))}{1+r_B^2}\,,
\end{alignedat}
\end{gathered}
\label{eq:truth}
\end{equation}
where $r_B$ and $\delta_B$ are the magnitude ratio and the strong phase difference between the suppressed and favoured \B decay amplitudes, respectively.

The formalism is extended to four-body \Bs final states such as \BsDsKpipi in Ref.~\cite{LHCb-PAPER-2020-030}.
After integration over the selected phase-space region, the expressions in \cref{eq:truth} use the amplitude ratio and effective strong phase for that region, with an additional coherence factor $0\leq\kappa_B\leq1$ multiplying the numerators of
$S_f$, $S_{\bar f}$, $A_f^{\Delta\Gamma}$ and $A_{\bar f}^{\Delta\Gamma}$.
The expression for $C_f$ is unchanged.
The factor $\kappa_B$ describes the dilution of the interference caused by the variation in the relative magnitude and strong phase difference of the decay amplitudes across the selected phase space.
These hadronic parameters are specific to the decay and differ from those of \BsDsK.

The same equations apply also to \BdDpi decays, with $\beta_s$ replaced by $-\beta$.
Another difference is that $\Delta\Gamma$ is negligible in the \Bd system, which removes sensitivity to $A^{\Delta\Gamma}_f$ and $A^{\Delta\Gamma}_{\bar f}$.
Finally, the amplitude ratio \ldmpi is expected to be approximately 1.6\%~\cite{LHCb-PAPER-2020-021}, which is much smaller than the analogous ratio in \BsDsK decays, $\ldsk \approx 34\%$.
The observables $C_f=-C_{\bar f}$ are consequently fixed to unity in the analysis of $B^0\to D^\mp\pi^\pm$ decays to maximise experimental sensitivity to \g.
Sensitivity to \g is reduced with respect to \BsDsK decays, since both remaining observables $S_f$ and $S_{\bar{f}}$ are proportional to \ldmpi.
In the simultaneous combination, with an external constraint on $\beta$, the constraints on \g from other channels allow \ldmpi to be determined from data, rather than being fixed to an estimate based on \suthreef symmetry as in Ref.~\cite{LHCb-PAPER-2018-009}.

\section{Inputs and method}
\label{sec:inputs}

The \lhcb measurements used as inputs to the combination are listed in \cref{tab:inputs_run123}.
All of them rely on data collected during Runs~1 and 2 of the \lhc, with the exception of a study of \BuDh decays with \DKShh, which uses data recorded in 2024~\cite{LHCb-PAPER-2026-010}.
This is the first \lhcb measurement of $\gamma$ using Run~3 data, and analyses the decay mode with the best sensitivity.
The measurement of \g using \BuDh decays with \DKShh in the Run~1 and Run~2 datasets~\cite{LHCb-PAPER-2020-019}
employs a binned Dalitz-plot analysis with external strong-phase inputs from the \besiii and CLEO collaborations.
A more recent measurement uses the same \lhcb dataset in a simultaneous, unbinned analysis of \lhcb and \besiii data,
achieving similar precision~\cite{LHCb-PAPER-2025-063,LHCb-PAPER-2025-064}.
The binned result is retained in this combination because it allows correlations arising from shared strong-phase inputs with other measurements using \DKShh decays to be accounted for.
\begin{table}[htbp]
  \caption{Measurements used in the combination.
    Those that are new, or that have changed, since the previous published combination~\cite{LHCb-PAPER-2021-033} are highlighted in the last column.
    LHCb Run~1 took place from 2011 to 2012, collecting proton--proton collision data at centre-of-mass energies, $\sqrt{s}$, of $7$ and $8\,\text{TeV}$, corresponding to integrated luminosities of $1$ and $2\,\text{fb}^{-1}$, respectively; Run~2 took place from 2015 to 2018 at $\sqrt{s}=13\,\text{TeV}$, corresponding to an integrated luminosity of $6\,\text{fb}^{-1}$; the Run~3 input listed in this table uses a partial dataset taken in 2024 at $\sqrt{s}=13.6\,\text{TeV}$, corresponding to an integrated luminosity of $5.8\,\text{fb}^{-1}$.
    In the table, ``PR'' (for ``partially reconstructed'') refers to measurements in which the neutral pion or photon from a $D^*\to D\pi^0/\gamma$ decay is not reconstructed, and ``FR'' to measurements in which the $D^*\to D\pi^0/\gamma$ candidate is fully reconstructed.
    }
  \label{tab:inputs_run123}
  \centering
  \renewcommand{\arraystretch}{1.0}
  \resizebox{\textwidth}{!}{
  \begin{tabular}{l l l l c}
    \toprule
        $\D$ decay & Observable(s) & Ref. & Dataset & Status since Ref.~\cite{LHCb-PAPER-2021-033} \\
    \midrule
    \Dzhh      & $\Delta A_{\CP}$                                        & \cite{LHCb-PAPER-2019-006,LHCb-PAPER-2015-055,LHCb-PAPER-2014-013} & \runOneTwo & {\it As before} \\
    \DzKK      & $A_{\CP}(\Kp\Km)$                                       & \cite{LHCb-PAPER-2022-024,LHCb-PAPER-2014-013,LHCb-PAPER-2016-035} & \runOneTwo    & {\bf Updated}       \\
    \Dzhh      & $y_{\CP}-\ycpRS$                                        & \cite{LHCb-PAPER-2018-038,LHCb-PAPER-2021-041} & \runOneTwo   & {\bf Updated}       \\
    \Dzhh      & \DYKK, \DYpipi                                          & \cite{LHCb-PAPER-2014-069,LHCb-PAPER-2016-063,LHCb-PAPER-2019-032,LHCb-PAPER-2020-045} & \runOneTwo & {\bf Updated} \\
    \DzKpi & $\RKpi$, $\AKpitilde$, $\cKpi^{(\prime)}$, $\DcKpitilde^{(\prime)}$
                                                                         & \cite{LHCb-PAPER-2017-046,LHCb-PAPER-2024-008} & \runOneTwo & {\bf Updated} \\
    \DzKpi & $R^{\pm}$, $(x'^\pm)^2$, $y'^\pm$              & \cite{LHCb-PAPER-2016-033,LHCb-PAPER-2024-044} & \runOneTwo & {\bf Updated} \\
    \DzKpipipi & $r$, $\kappa y'$, $(\xD^2+\yD^2)/4$                     & \cite{LHCb-PAPER-2015-057} & Run 1    & {\it As before} \\
    \DzKpipipi & $r_i^\pm$, $(\kappa y'^\pm)_i$, $x^{2\pm}+y^{2\pm}$     & \cite{LHCb-PAPER-2025-029} & Run 2    & {\bf New}       \\
    \DzKSpipi  & \xD, \yD                                                & \cite{LHCb-PAPER-2015-042} & 2011     & {\it As before} \\
    \DzKSpipi  & $x_{\CP}$, $y_{\CP}$, $\Delta x$, $\Delta y$            & \cite{LHCb-PAPER-2019-001,LHCb-PAPER-2021-009,LHCb-PAPER-2022-020} & \runOneTwo    & {\bf Updated}   \\
    \Dzpipipiz & \DYpppeff                                               & \cite{LHCb-PAPER-2024-003} & Run 2    & {\bf New}       \\
    \addlinespace[0.6ex]
    \midrule
    $\B$ decay & $\D$ decay & Ref. & Dataset & Status since Ref.~\cite{LHCb-PAPER-2021-033} \\
    \midrule
    \BuDh      & \Dhhpr                 & \cite{LHCb-PAPER-2020-036} & \runOneTwo & {\it As before} \\
    \BuDh      & \Dhhpipi               & \cite{LHCb-PAPER-2022-037,LHCb-PAPER-2025-019}
                                                                     & \runOneTwo & {\bf New}       \\
    \BuDh      & \Dhhprpiz              & \cite{LHCb-PAPER-2021-036} & \runOneTwo & {\bf Updated}   \\
    \BuDh      & \DKpipipi              & \cite{LHCb-PAPER-2022-017} & \runOneTwo & {\bf Updated}   \\
    \BuDh      & \DKShh                 & \cite{LHCb-PAPER-2020-019,LHCb-PAPER-2026-010} & \runOneTwo,\ 2024 & {\bf Updated} \\
    \BuDh      & \DKSKpi                & \cite{LHCb-PAPER-2019-044} & \runOneTwo & {\it As before} \\
    \BuDsth  (PR)  & \Dhhpr             & \cite{LHCb-PAPER-2020-036} & \runOneTwo & {\it As before} \\
    \BuDsth  (PR)  & \DKShh             & \cite{LHCb-PAPER-2023-029} & \runOneTwo & {\bf New}       \\
    \BuDsth  (FR)  & \DKShh             & \cite{LHCb-PAPER-2023-012} & \runOneTwo & {\bf New}       \\
    \BuDKst    & \Dhhpr                 & \cite{LHCb-PAPER-2024-023} & \runOneTwo & {\bf Updated}   \\
    \BuDKst    & \Dhpipipi              & \cite{LHCb-PAPER-2024-023} & \runOneTwo & {\bf Updated}   \\
    \BuDKst    & \DKShh                 & \cite{LHCb-PAPER-2024-023} & \runOneTwo & {\bf New}       \\
    \BuDhpipi  & \Dhhpr                 & \cite{LHCb-PAPER-2015-020} & Run 1    & {\it As before} \\
    \BdDKst    & \Dhhpr                 & \cite{LHCb-PAPER-2023-040} & \runOneTwo & {\bf Updated}   \\
    \BdDKst    & \Dhpipipi              & \cite{LHCb-PAPER-2023-040} & \runOneTwo & {\bf Updated}   \\
    \BdDKst    & \DKShh                 & \cite{LHCb-PAPER-2023-009} & \runOneTwo & {\bf Updated}   \\
    \BdDpi     & \DKpipi                & \cite{LHCb-PAPER-2018-009} & Run 1    & {\it As before} \\
    \BsDsK     & \Dshhh                 & \cite{LHCb-PAPER-2017-047,LHCb-PAPER-2024-020} & \runOneTwo & {\bf Updated} \\
    \BsDsKpipi & \Dshhh                 & \cite{LHCb-PAPER-2020-030} & \runOneTwo & {\it As before} \\
    \bottomrule
  \end{tabular}}
\end{table}

The external constraints, which provide auxiliary information on the hadronic parameters and coherence factors of multibody \B and \D decays,
and on the mixing phases $\beta$ and \phis, are summarised in \cref{tab:inputs_aux}.
\begin{table}[t]
  \caption{Auxiliary inputs used in the combination.
    Those that are new, or that have changed, since the previous published combination~\cite{LHCb-PAPER-2021-033} are highlighted in the last column.
    In the sixth row the same notation as Ref.~\cite{BESIII:2022qkh} is adopted,
    which uses $A_{K\pi}$ to denote the asymmetry between $D\to K^+\pi^-$ decay rates in events tagged by $C\!P$-even and $C\!P$-odd $D$ decays,
    not to be confused with the observable in \cref{eq:ws-to-rs}.
  }
  \label{tab:inputs_aux}
  \centering
  \renewcommand{\arraystretch}{1.3}
  \resizebox{\textwidth}{!}{
  \begin{tabular}{l l l l c}
    \toprule
    Decay & Parameters & Source & Ref. & Status since Ref.~\cite{LHCb-PAPER-2021-033} \\
    \midrule
    \BuDKst              & \kbdkst                                                  & LHCb   & \cite{LHCb-PAPER-2017-030} & {\it As before} \\
    \BdDKst              & \kbdkstz                                                 & LHCb   & \cite{LHCb-PAPER-2015-059} & {\it As before} \\
    \BdDpi               & $\beta$                                                  & HFLAV  & \cite{HFLAV:2022esi:zenodo.19371140} & {\bf Updated} \\
    \BsDsK\!\!$(\pi\pi)$ & \phis                                                    & HFLAV  & \cite{HFLAV:2024ctg}       & {\bf Updated}   \\
    \DKpi                & $\cos\ddkpi$, $\sin\ddkpi$, $(\rdkpi)^2$, $\xD^2$, \yD   & CLEO & \cite{CLEO:2012fel}        & {\it As before} \\
    \DKpi                & $A_{K\pi}$, $A_{K\pi}^{\pi\pi\piz}$, $\rdkpi\cos\ddkpi$, $\rdkpi\sin\ddkpi$
                                                                                    & BESIII & \cite{BESIII:2022qkh,BESIII:2025pod,BESIII:2025xed} & {\bf New} \\
    \Dhhpiz              & \Fppp, \Fkkp                                             & CLEO \& BESIII & \cite{Malde:2015mha,BESIII:2024nnf}      & {\bf Updated}       \\
    \Dpipipipi           & \Fpppp                                                   & CLEO \& BESIII & \cite{Malde:2015mha,BESIII:2022wqs}      & {\bf Updated} \\
    \DKKpipi             & \Fkkpp                                                   & BESIII & \cite{BESIII:2025wqu}      & {\bf New}       \\
    \DKpipiz             & \rdkpp, \ddkpp, \kdkpp                                   & BESIII & \cite{BESIII:2026oav}      & {\bf Updated}   \\
    \DKpipipi            & \ddkppp, \kdkppp                                         & CLEO \& BESIII & \cite{Evans:2019wza,BESIII:2026oav}      & {\bf Updated}   \\
    \DKSKpi              & \rdkskpi, \ddkskpi, \kdkskpi                             & CLEO & \cite{Insler:2012pm}       & {\it As before} \\
    \DKSKpi              & \rdkskpi                                                 & LHCb   & \cite{LHCb-PAPER-2015-026} & {\it As before} \\
    \DKSpipi             & $c_i, s_i$                                               & BESIII & \cite{BESIII:2025nsp}      & {\bf Updated}   \\
    \bottomrule
  \end{tabular}}
\end{table}
In particular, the results of the \lhcb measurements of \DzKSpipi decays relying on the bin-flip method~\cite{LHCb-PAPER-2019-001,LHCb-PAPER-2021-009,LHCb-PAPER-2022-020}
have been updated to employ the determination of hadronic parameters without U-spin constraints in Ref.~\cite{BESIII:2025nsp},
which is also employed, with a different division of the Dalitz plot into disjoint regions, by the first \lhcb measurement of \g with 2024 data~\cite{LHCb-PAPER-2026-010}. 
The results are reported in \cref{app:derived-inputs}, where the uncertainties on the hadronic parameters contribute up to 70\% of the total uncertainties. 
Conversely, Ref.~\cite{LHCb-PAPER-2015-042} has significantly larger statistical uncertainties, thus updating external inputs has negligible impact on the total precision.
\lhcb measurements of \g based on the Run~1 and~2 samples~\cite{LHCb-PAPER-2020-019,LHCb-PAPER-2023-029,LHCb-PAPER-2023-012,LHCb-PAPER-2024-023,LHCb-PAPER-2023-009}, where the external inputs also make a small contribution to the total uncertainty, have not been reinterpreted and employ the previous determinations of the hadronic parameters from Refs.~\cite{CLEO:2010iul,BESIII:2020khq}.
Correlations among the measurements above are taken into account only between the charm results in Refs.~\cite{LHCb-PAPER-2021-009,LHCb-PAPER-2022-020}
and among the measurements of \g relying on the Run~1 and 2 samples~\cite{LHCb-PAPER-2020-019,LHCb-PAPER-2023-029,LHCb-PAPER-2023-012,LHCb-PAPER-2024-023,LHCb-PAPER-2023-009}.
Residual correlations have negligible impact on the combination since the uncertainty from the knowledge of hadronic parameters is subleading in beauty measurements.

\lhcb measurements with \DKpipipi decays~\cite{LHCb-PAPER-2022-017,LHCb-PAPER-2025-029} in the regions of phase space defined in Ref.~\cite{Evans:2019wza}
are combined with updated constraints on the \D-decay hadronic parameters from the \besiii~\cite{BESIII:2026oav} and CLEO~\cite{Evans:2019wza} experiments in a dedicated combiner.
This approach is the same as that adopted for other multibody charm decays such as $\DKShh$ and $\DKKpipi$, where the dependence on the relevant charm hadronic parameters is also marginalised in independent combinations. 
For this particular case, this approach is also adopted due to the highly non-Gaussian nature of the constraints on the charm hadronic parameters. 
The direct inclusion of these constraints with the full likelihood would imply the inclusion of more than 300 additional observables, mostly associated with the observed yields of various decays at the \besiii and CLEO experiments, that are relevant only to the determination of the \DKpipipi hadronic parameters. 
The output of the standalone combination is a multidimensional Gaussian constraint related to the parameters of interest,
\g, \rbdk, \dbdk, \rbdpi, \dbdpi, and the charm mixing and \CP-violation parameters,
marginalised over the $\DKpipipi$ hadronic parameters.
The quality of this fit, whose results are detailed in \cref{app:derived-inputs}, is good, with $\chisq/\text{ndf} = 25.04/23$ for the \lhcb observables.
The validity of the Gaussian approximation is cross-checked against the use of the full likelihood, and negligible difference with the baseline treatment is found.

As discussed in \cref{sec:charm-cpv,sec:charm-observables-cpv-decay}, \CP symmetry in CF and DCS charm decays and universality of charm-mixing weak phases are assumed.
The latter assumption corresponds to setting \AcpDhh to zero in \cref{eq:DeltaY}, such that $\DYhh = - \xth\sin\phiM$, and maximises the sensitivity of the combination to \AcpDhh under the assumption that $\cot\ddhhQCD$ is of the order of unity.
The same assumption also applies to the study of \Dzpipipiz decays.

The effect of \CP violation in \Dz mixing on the measurements of \B decays is neglected,
as are any differences in the weak-phase difference of the two decay amplitudes corresponding to $\decay{\bquark}{\cquark\uquarkbar q}$ and $\decay{\bquark}{\uquark\cquarkbar q}$ transitions, between $q = s,d$ and among all possible final states of the intermediate \D meson.
All these effects have an impact on $\gamma$ smaller than $0.1\degrees$.
As mentioned in \cref{sec:timedepg}, it is assumed that $-2\betas=\phis$.
Regeneration of \KS mesons and \CP violation in \Kz mixing affect all charm and beauty measurements with a neutral kaon in the final state~\cite{Stahl:2014xlz,Bjorn:2019kov,Kagan:2020vri,LHCb-PAPER-2026-011}.
However, these effects are corrected for directly in the \lhcb measurements of charm mixing, and the small effects on beauty measurements are assessed as systematic uncertainties.

The combination follows a frequentist approach implemented in the \textsc{GammaCombo} software package~\cite{GammaCombo-2026}.
A \chisq function is defined from the global likelihood as \mbox{$\chisq(\vec{\alpha}) \equiv - 2 \log(\mathcal{L}(\vec{\alpha}))$}, where $\mathcal{L}(\vec{\alpha})$ is the product of the likelihoods of all individual measurements and $\vec{\alpha}$ is the vector of theory parameters to be determined.
Each measurement is assumed to be distributed according to a multivariate Gaussian distribution and,
for measurements with asymmetric uncertainties, the uncertainty is taken to be the arithmetic average of the upper and lower uncertainties.
The parameters $\vec{\alpha}$ are constrained to lie in their physically allowed phase space.
In particular, the parameters \rd, \rb, \xth and \yth are forced to be non-negative, the coherence parameters are constrained to lie in the interval $[0,1]$, and the angle \g in the interval $[0\degrees, 180\degrees]$ (to lift the degeneracy with \db).
The best-fit point, $\vec{\alpha}_\text{min}$, is defined by the minimum of the \chisq function over this phase space.

Confidence regions for a subset $\vec{\beta}$ of the $\vec{\alpha}$ parameters are computed using two methods.
The first is the profile-likelihood \textsc{Prob} treatment of nuisance parameters~\cite{Cowan:2010js}, which is based on the function $\Delta\chisq(\vec{\beta}) \equiv \min_{\vec{\gamma}}(\chisq(\vec{\alpha})) - \chisq(\vec{\alpha}_\text{min})$,
where $\vec{\gamma}$ is the set of all $\vec{\alpha}$ parameters other than $\vec{\beta}$.
Confidence intervals are calculated assuming that this function is distributed according to a \chisq distribution with a number of degrees of freedom equal to the size of $\vec{\beta}$.
However, this method is in general prone to undercoverage when statistical sensitivity is low, or parameters are near the boundaries of allowed values.

The second method is the $\hat{\mu}$ or \textsc{Plugin} treatment of nuisance parameters, which provides better coverage at the cost of increased computational requirements.
Here, a set of pseudoexperiments is generated, in each of which a new instance of all measurements included in the combination is generated using $\mathcal{L}(\vec{\alpha}_\text{min})$ as the probability density function.
Then, the $\vec{\alpha}$ parameters are fit to the pseudoexperiment obtaining a new global minimum $\vec{\alpha}^{\prime}_\text{min}$ and, finally, the $\Delta{\chisq}^{\prime}(\vec{\beta})$ function is calculated by minimising the $\chisq$ of the pseudoexperiment with fixed $\vec{\beta}$ parameters, where the prime indicates that the difference is taken with respect to the \chisq calculated at the minimum of the pseudoexperiment, $\vec{\alpha}^{\prime}_\text{min}$.
The \omcl for each point $\vec{\beta}$ is then defined as the fraction of pseudoexperiments for which \mbox{$\Delta{\chisq}^{\prime}(\vec{\beta}) > \Delta{\chisq}(\vec{\beta})$}.
The coverage of this method is not guaranteed for the full parameter space, since the parameters $\vec{\gamma}$ are fixed in the likelihood while generating pseudoexperiments, but it is verified for the best-fit point~\cite{Bodhisattva:2009uba}.

All numerical results quoted in \cref{sec:results} are obtained with the \textsc{Plugin} method.
Since the one-dimensional confidence intervals obtained with the two methods differ by less than 2\% on average, all the \omcl profiles and two-dimensional confidence regions shown in the figures are obtained using the \textsc{Prob} method.

\section{Results}
\label{sec:results}

The combination uses a total of \ninputsRunThree input observables to determine \nparamsRunThree
free parameters, giving a best-fit $\chisq$ of \chisqminRunThree that corresponds to a p-value of $\goodnessOfFitRunThree$.
The resulting confidence intervals for the parameters of greatest interest are given in \cref{tab:results_run123}.
The correlations between the fitted parameters and the
contribution of each input measurement to the $\chisq$ are given in \cref{app:additional-results}.

\begin{table}[!p]
  \begin{center}
    \caption{
      Confidence intervals and best-fit values for each of the parameters of interest,
      computed with the toy-based \textsc{Plugin} method~\cite{LHCb-PAPER-2013-020}.
      Entries marked with an asterisk show where the scan has hit a physical boundary.
    }
    \label{tab:results_run123}
    \setlength{\tabcolsep}{5pt}
    \renewcommand{\arraystretch}{1.2}
    \resizebox{0.9\textwidth}{!}{
    \begin{tabular}{lr|rr|rr}
    \hline
    \multirow{2}{*}{Quantity}& \multirow{2}{*}{Value} & \multicolumn{2}{c|}{68.3\% CL} & \multicolumn{2}{c}{95.4\% CL} \\
                             &                        & Uncertainty & Interval          & Uncertainty & Interval        \\
    \hline
$\g [^\circ]$               & $63.2$      & $^{+2.3}_{-2.2}$          & $[61.0,65.5]$             & $\pm 4.5$                 & $[58.7,67.7]$              \\
$\rdk [\%]$                 & $9.58$      & $^{+0.17}_{-0.18}$        & $[9.40,9.75]$             & $\pm 0.35$                & $[9.23,9.93]$              \\
$\ddk [^\circ]$             & $123.4$     & $^{+2.4}_{-2.6}$          & $[120.8,125.8]$           & $^{+4.8}_{-5.2}$          & $[118.2,128.2]$            \\
$\rdpi [\%]$                & $0.49$      & $^{+0.05}_{-0.04}$        & $[0.44,0.53]$             & $^{+0.09}_{-0.08}$        & $[0.40,0.58]$              \\
$\ddpi [^\circ]$            & $280$       & $^{+9}_{-10}$             & $[270,289]$               & $^{+17}_{-19}$            & $[261,297]$                \\
$\rdstk [\%]$               & $10.7$      & $\pm 1.0$                 & $[9.7,11.6]$              & $^{+1.9}_{-2.0}$          & $[8.7,12.6]$               \\
$\ddstk [^\circ]$           & $310$       & $^{+6}_{-7}$              & $[304,317]$               & $^{+12}_{-15}$            & $[296,323]$                \\
$\rdstpi [\%]$              & $0.73$      & $^{+0.39}_{-0.31}$        & $[0.42,1.12]$             & $^{+0.83}_{-0.63}$        & $[0.10,1.56]$              \\
$\ddstpi [^\circ]$          & $37$        & $^{+39}_{-20}$            & $[17,76]$                 & $^{+90}_{-35}$            & $[2,127]$                  \\
$\rdkst [\%]$               & $10.5$      & $^{+0.9}_{-1.0}$          & $[9.6,11.4]$              & $^{+1.7}_{-1.9}$          & $[8.6,12.2]$               \\
$\ddkst [^\circ]$           & $50$        & $^{+14}_{-12}$            & $[38,64]$                 & $^{+29}_{-23}$            & $[27,79]$                  \\
$\rdkstz [\%]$              & $23.2$      & $\pm 1.5$                 & $[21.7,24.7]$             & $^{+2.9}_{-3.2}$          & $[20.0,26.1]$              \\
$\ddkstz [^\circ]$          & $192$       & $\pm 6$                   & $[186,198]$               & $^{+13}_{-12}$            & $[180,205]$                \\
$\ldsk [\%]$                & $34$        & $\pm 4$                   & $[30,37]$                 & $\pm 7$                   & $[27,41]$                  \\
$\ddsk [^\circ]$            & $350$       & $\pm 6$                   & $[344,355]$               & $^{+11}_{-12}$            & $[338,361]$                \\
$\ldskpipi [\%]$            & $46$        & $\pm 8$                   & $[38,54]$                 & $^{+16}_{-17}$            & $[29,62]$                  \\
$\ddskpipi [^\circ]$        & $347$       & $\pm 12$                  & $[335,359]$               & $\pm 24$                  & $[323,371]$                \\
$\ldmpi [\%]$               & $3.0$       & $^{+1.4}_{-1.2}$          & $[1.8,4.4]$               & $^{+3.3}_{-2.6}$          & $[0.4,6.3]$                \\
$\ddmpi [^\circ]$           & $32$        & $^{+25}_{-39}$            & $[-7,57]$                 & $^{+43}_{-81}$            & $[-49,75]$                 \\
$\rdkpipi [\%]$             & $8.0$       & $^{+2.7}_{-3.5}$          & $[4.5,10.7]$              & $^{+4.9}_{-8.0}$          & $[0.0,12.9]^*$             \\
$\rdpipipi [\%]$            & $6.1$       & $^{+2.2}_{-2.5}$          & $[3.6,8.3]$               & $^{+4.9}_{-6.1}$          & $[0.0,11.0]^*$             \\
$\rDkpi [\%]$               & $5.864$     & $\pm 0.010$               & $[5.855,5.874]$           & $^{+0.020}_{-0.019}$      & $[5.845,5.884]$            \\
$\dDkpi [^\circ]$           & $192.9$     & $^{+2.3}_{-2.4}$          & $[190.5,195.2]$           & $^{+4.7}_{-4.8}$          & $[188.1,197.6]$            \\
$\AcpDKK [10^{-3}]$         & $0.9$       & $\pm 0.5$                 & $[0.4,1.4]$               & $\pm 1.0$                 & $[-0.1,1.9]$               \\
$\AcpDpipi [10^{-3}]$       & $2.4$       & $\pm 0.6$                 & $[1.9,3.0]$               & $^{+1.2}_{-1.1}$          & $[1.3,3.6]$                \\
\hline
$\xth [10^{-3}]$            & $4.2$       & $^{+0.4}_{-0.5}$          & $[3.7,4.6]$               & $\pm 0.9$                 & $[3.3,5.1]$                \\
$\yth [10^{-3}]$            & $6.18$      & $^{+0.17}_{-0.18}$        & $[6.00,6.35]$             & $^{+0.35}_{-0.36}$        & $[5.82,6.53]$              \\
$\phiM [\text{mrad}]$             & $4$         & $^{+11}_{-10}$            & $[-6,15]$                 & $\pm 22$                  & $[-18,26]$                 \\
$\phiG [\text{mrad}]$             & $19$        & $^{+23}_{-24}$            & $[-5,42]$                 & $^{+47}_{-48}$            & $[-29,66]$                 \\
\hline
$\xD [10^{-3}]$             & $4.2$       & $^{+0.4}_{-0.5}$          & $[3.8,4.6]$               & $\pm 0.9$                 & $[3.3,5.1]$                \\
$\yD [10^{-3}]$             & $6.18$      & $^{+0.17}_{-0.18}$        & $[6.00,6.35]$             & $^{+0.35}_{-0.36}$        & $[5.82,6.53]$              \\
$\qopD - 1$                 & $-0.007$    & $^{+0.013}_{-0.012}$      & $[-0.019,0.006]$          & $^{+0.025}_{-0.024}$      & $[-0.031,0.018]$           \\
$\phiD [\text{mrad}]$             & $-14$       & $\pm 16$                  & $[-30,2]$                 & $^{+31}_{-33}$            & $[-47,17]$                 \\
\hline
    \end{tabular}
    }
  \end{center}
\end{table}

\begin{figure}[!htb]
  \centering
  \includegraphics[width=0.495\textwidth]{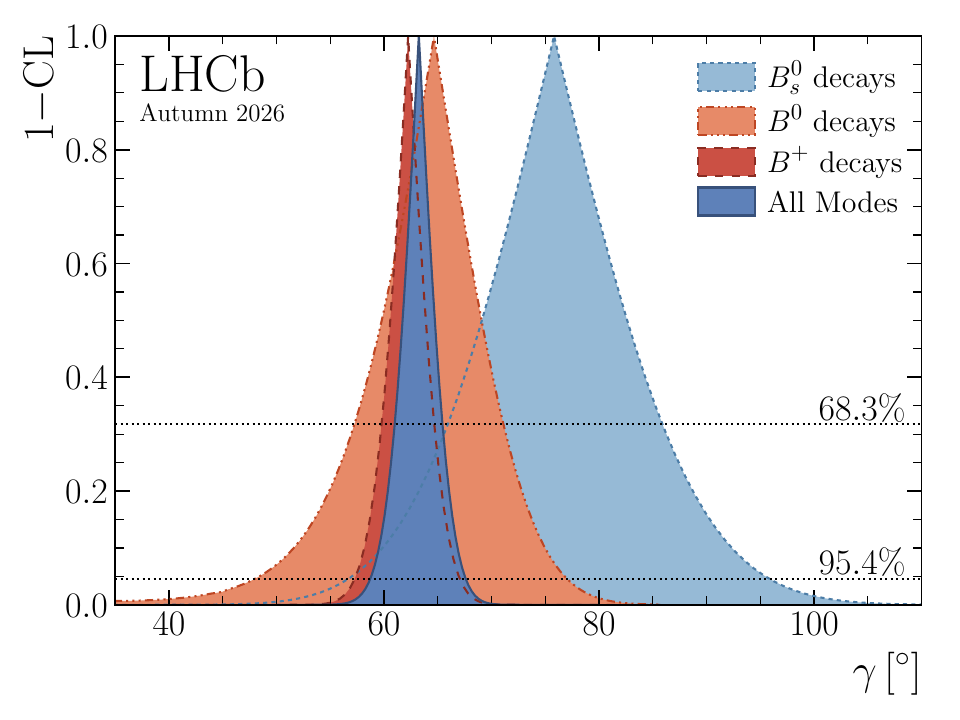}
  \includegraphics[width=0.495\textwidth]{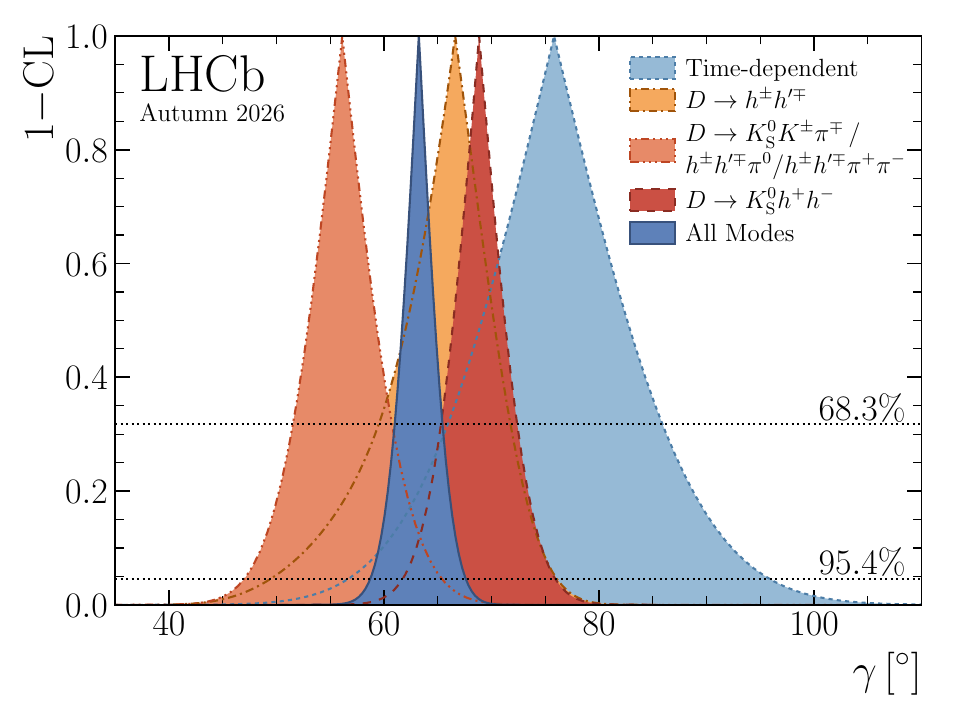}
  \includegraphics[width=0.495\textwidth]{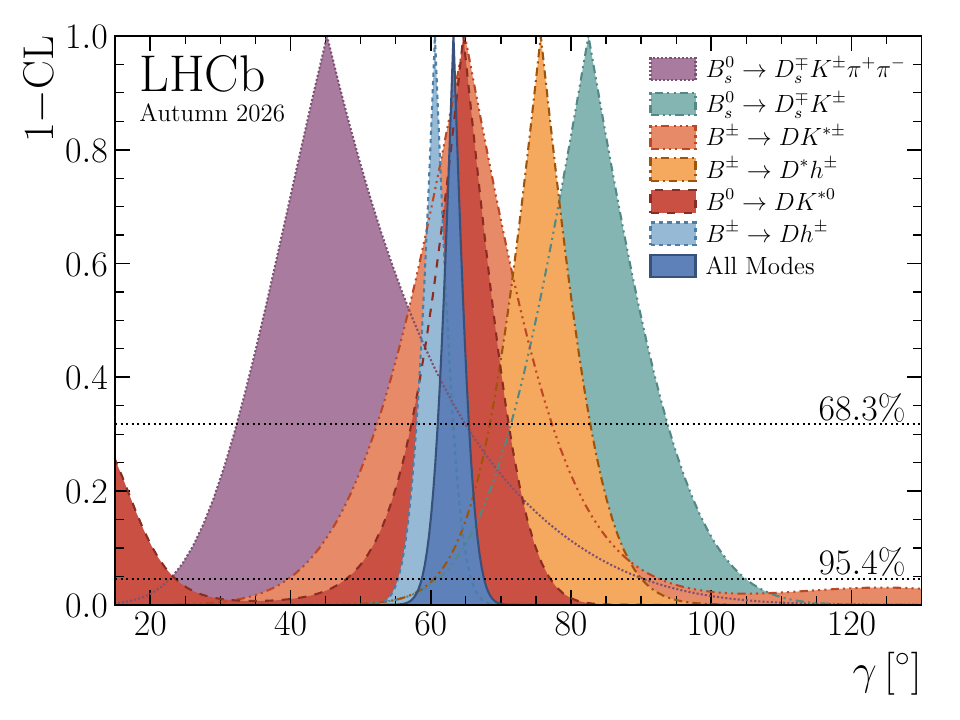}
  \includegraphics[width=0.495\textwidth]{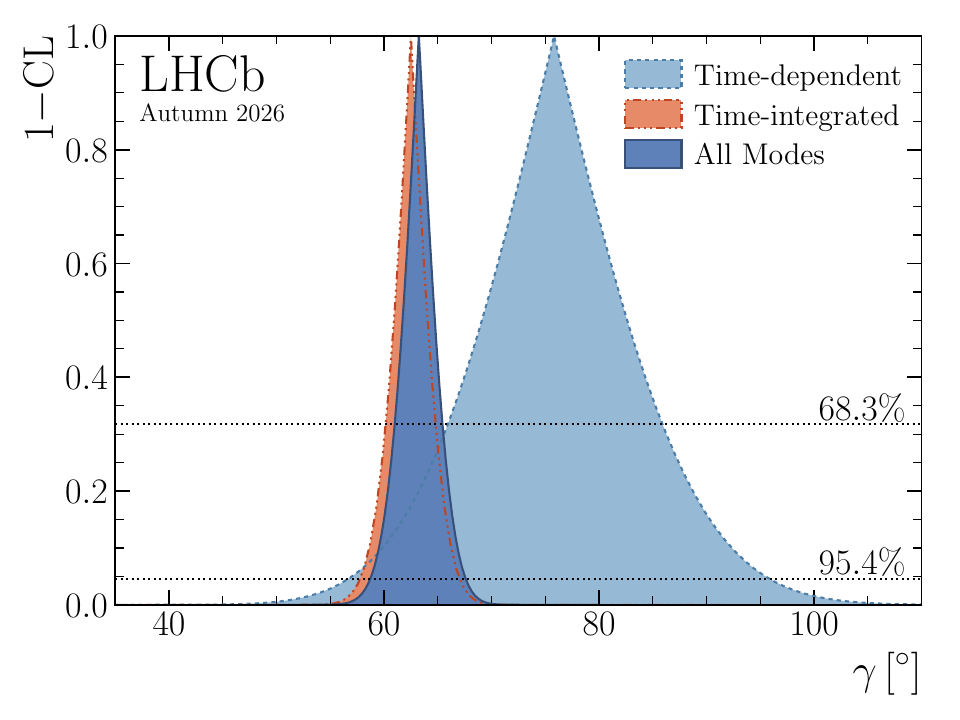}
  \caption{One-dimensional profile-likelihood scans of $1-\mathrm{CL}$ as a function of the CKM angle~$\gamma$. 
    In each panel the total combination
    is shown in dark blue with a solid line. The inputs are split 
    (top left) by initial $B$ state, 
    (top right) by $D$ final state,
    (bottom left) by $B$ decay mode, and
    (bottom right) into time-dependent and time-integrated measurements.}
  \label{fig:scan_run123}
\end{figure}

The obtained value of \g, \resultGRunThree, is compatible with and more precise than the previous
\lhcb combination, \gammaPrevious~\cite{LHCb-PAPER-2021-033}.
The p-value, or equivalently one minus the confidence level (\omcl), as a function of \g is shown
in \cref{fig:scan_run123} for the total combination and for subcombinations in which the input
observables are split by initial \B meson, \D final state, \B decay mode, or time dependence of the measurement.
All subcombinations include the same charm inputs as the total combination, so that they differ only in their beauty inputs.
The corresponding confidence intervals for \g are given in \cref{tab:splits_run123}.
The subcombinations agree with one another and with the total combination.
The largest deviation from the total, of less than two standard deviations, is that of the \BuDsth decays, \resultGBuDsthRunThree.
The time-dependent measurements give \resultGTDRunThree, above but consistent with the time-integrated value of \resultGTIRunThree.

\begin{table}[!tb]
  \centering
  \caption{Confidence intervals and best-fit values for $\gamma$ of the subcombinations, by category.}
  \label{tab:splits_run123}
  \renewcommand{\arraystretch}{1.2}
  \resizebox{\textwidth}{!}{
  \begin{tabular}{lc|rr|rr}
  \hline
  \multirow{2}{*}{Subcombination} & \multirow{2}{*}{Value $[\degrees]$} & \multicolumn{2}{c|}{68.3\% CL} & \multicolumn{2}{c}{95.4\% CL} \\
                                   &                                     & Uncertainty & Interval         & Uncertainty & Interval        \\
  \hline
\textit{Initial $B$ meson} & & & & & \\
$B_s^0$ decays              & $76$        & $\pm 10$                  & $[66,86]$                 & $^{+21}_{-20}$            & $[56,97]$                  \\
$B^0$ decays                & $65$        & $^{+6}_{-7}$              & $[57,71]$                 & $^{+12}_{-18}$            & $[47,77]$                  \\
$B^+$ decays                & $62.2$      & $\pm 2.5$                 & $[59.7,64.7]$             & $\pm 5.0$                 & $[57.2,67.2]$              \\
\hline
\textit{$D$ final state} & & & & & \\
Time-dependent              & $76$        & $\pm 10$                  & $[66,86]$                 & $^{+21}_{-20}$            & $[56,97]$                  \\
$D\to h^\pm h^{\prime\mp}$  & $67$        & $^{+6}_{-7}$              & $[59,72]$                 & $^{+10}_{-18}$            & $[49,77]$                  \\
\begin{tabular}[c]{@{}l@{}}$D\to K_S^0 K^\pm \pi^\mp\, /$\\$h^\pm h^{\prime\mp}\pi^0 / h^\pm h^{\prime\mp}\pi^+\pi^-$\end{tabular}& $56$        & $\pm 5$                   & $[52,61]$                 & $\pm 9$                   & $[47,66]$                  \\
$D\to K_S^0 h^+ h^-$        & $69$        & $\pm 4$                   & $[65,72]$                 & $\pm 7$                   & $[62,76]$                  \\
\hline
\textit{$B$ decay mode} & & & & & \\
$B_s^0 \to D_s^\mp K^\pm \pi^+ \pi^-$& $45$        & $^{+22}_{-13}$            & $[32,67]$                 & $^{+48}_{-22}$            & $[23,93]$                  \\
$B_s^0 \to D_s^\mp K^\pm$   & $82$        & $^{+12}_{-11}$            & $[71,94]$                 & $^{+23}_{-21}$            & $[61,105]$                 \\
$B^{\pm} \to D K^{*\pm}$    & $65$        & $^{+13}_{-12}$            & $[53,78]$                 & $^{+28}_{-25}$            & $[40,93]$                  \\
$B^{\pm} \to D^{*} h^{\pm}$ & $76$        & $^{+7}_{-8}$              & $[68,83]$                 & $^{+14}_{-16}$            & $[60,90]$                  \\
$B^{0} \to D K^{*0}$        & $65$        & $^{+7}_{-8}$              & $[57,71]$                 & $^{+12}_{-18}$            & $[47,77]$                  \\
$B^{\pm} \to D h^{\pm}$     & $60.6$      & $\pm 2.6$                 & $[58.0,63.2]$             & $^{+5.2}_{-5.3}$          & $[55.3,65.8]$              \\
\hline
\textit{Time dependence} & & & & & \\
Time-dependent              & $76$        & $\pm 10$                  & $[66,86]$                 & $^{+21}_{-20}$            & $[56,97]$                  \\
Time-integrated             & $62.5$      & $^{+2.3}_{-2.4}$          & $[60.1,64.8]$             & $^{+4.6}_{-4.8}$          & $[57.7,67.1]$              \\
\hline

  \end{tabular}
  }
\end{table}

% Run3 1D scans

The time-dependent measurements are based on \BsDsK, \BsDsKpipi and \BdDpi decays.
The \BdDpi decay contributes two observables: the mixing-induced asymmetries $S_f$ and $S_{\bar f}$ defined in \cref{sec:timedepg}.
These observables depend on the weak phase $\g + 2\beta$, on the ratio of the suppressed to favoured
decay amplitudes, \ldmpi, and on their strong-phase difference, \ddmpi.
Since these two observables depend on three unknowns,
and since \g is constrained by the other modes and $\beta$ by the HFLAV average,
this input effectively yields measurements of \ldmpi and \ddmpi.
The result for \ldmpi, \resultLdmpiRunThree, agrees with the prediction based on \suthreef symmetry, $0.0163 \pm 0.0034$~\cite{LHCb-PAPER-2020-021}.

The sensitivity of the time-dependent subcombination to \g therefore comes from the \Bs modes,
which constrain the combination $\g - 2\betas$ through \cref{eq:truth}.
The angle \g is determined under the assumption $-2\betas = \phis$, which neglects
penguin-loop and BSM contributions to the $\Bs \to \jpsi\phi$ decay.
Without this assumption, the combination gives $\g - 2\betas = \resultGphisRunThree$.

\begin{figure}[tb]
    \centering
    \includegraphics[width=0.45\linewidth]{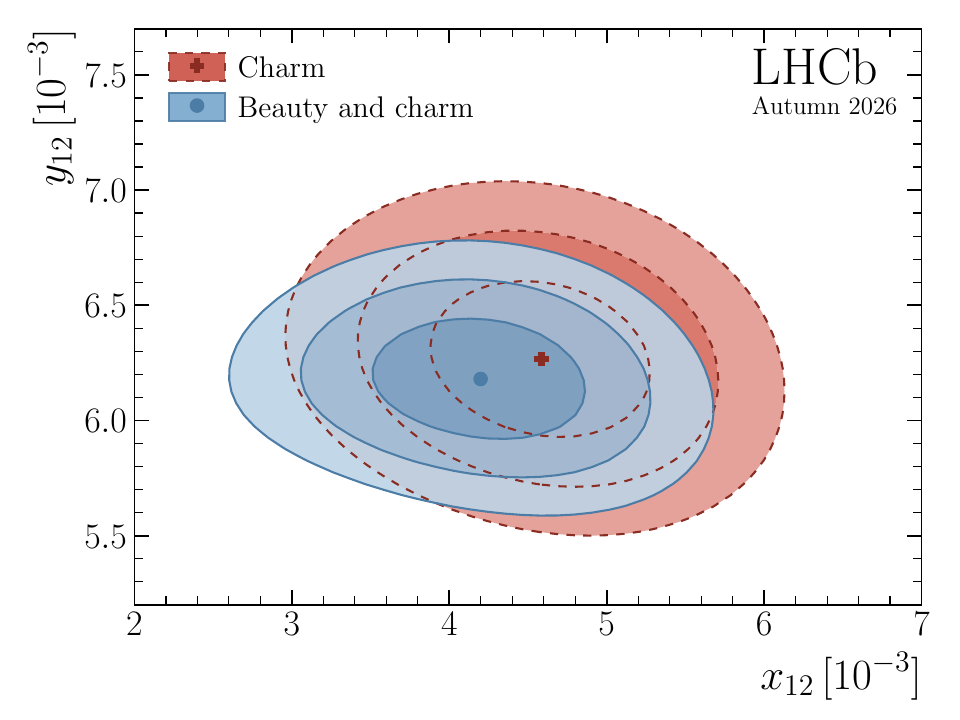}
    \includegraphics[width=0.45\linewidth]{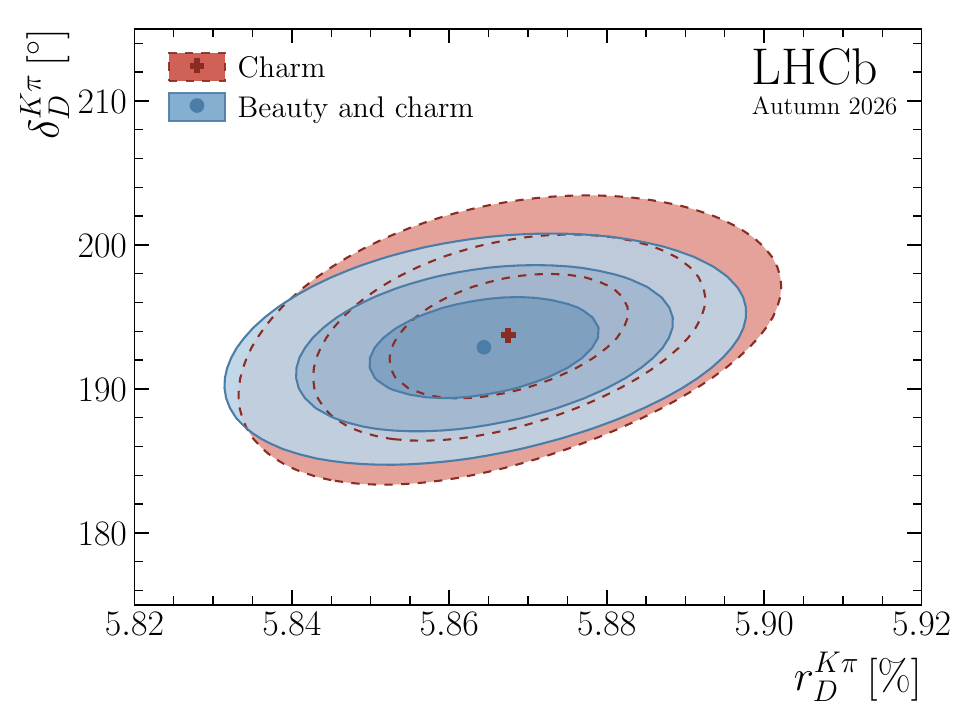}
    \caption{
        Two-dimensional profile-likelihood contours for (left) the charm-mixing parameters in the theoretical parametrisation and (right) the hadronic parameters of $D^0\to K^\pm\pi^\mp$ decays.
        The blue and red contours show the determinations with and without including the beauty inputs in \cref{tab:inputs_run123}, respectively.
        The mixing parameters in the phenomenological parametrisation are indistinguishable from those in the theoretical parametrisation.
        Contours are displayed at the 1, 2, $3\,\sigma$ confidence levels, encompassing $68.3\%$, $95.4\%$, $99.7\%$ of the distributions, respectively.
    }
    \label{fig:charm-mixing}
\end{figure}

The charm-mixing parameters are determined to be $\xth = \resultXtwelveRunThree$ and $\yth = \resultYtwelveRunThree$ in the theoretical parametrisation, and the phenomenological parameters \xD and \yD take the same values at the quoted precision.
The value of \xD is larger than that found in previous determinations~\cite{LHCb-PAPER-2021-033}.
This is mostly due to the \DKSpipi hadronic parameters recently measured by the \besiii collaboration~\cite{BESIII:2025nsp},
which are smaller in absolute value than the previous determinations~\cite{CLEO:2010iul,BESIII:2020khq}
and have been used to update the \lhcb measurements of \DzKSpipi decays~\cite{LHCb-PAPER-2019-001,LHCb-PAPER-2021-009,LHCb-PAPER-2022-020}.
Two-dimensional profile-likelihood contours for the mixing parameters are shown in \cref{fig:charm-mixing}.
Including the beauty inputs significantly improves the precision on \yth with respect to a charm-only combination,
since they contribute to constraining the \DKpi and \DKpipipi hadronic parameters.
The phase \dDkpi is determined to be \resultDdKpiRunThree.
It deviates significantly from the U-spin-symmetry limit, $\dDkpi = 180\degrees$,
and provides a test bench for models of rescattering in \Dz decays~\cite{Chau:1993ec,Buccella:1994nf,Browder:1995ay,Falk:1999ts,Gao:2006nb,Buccella:2019kpn}.

\begin{figure}
  \centering
  \includegraphics[width=0.45\linewidth]{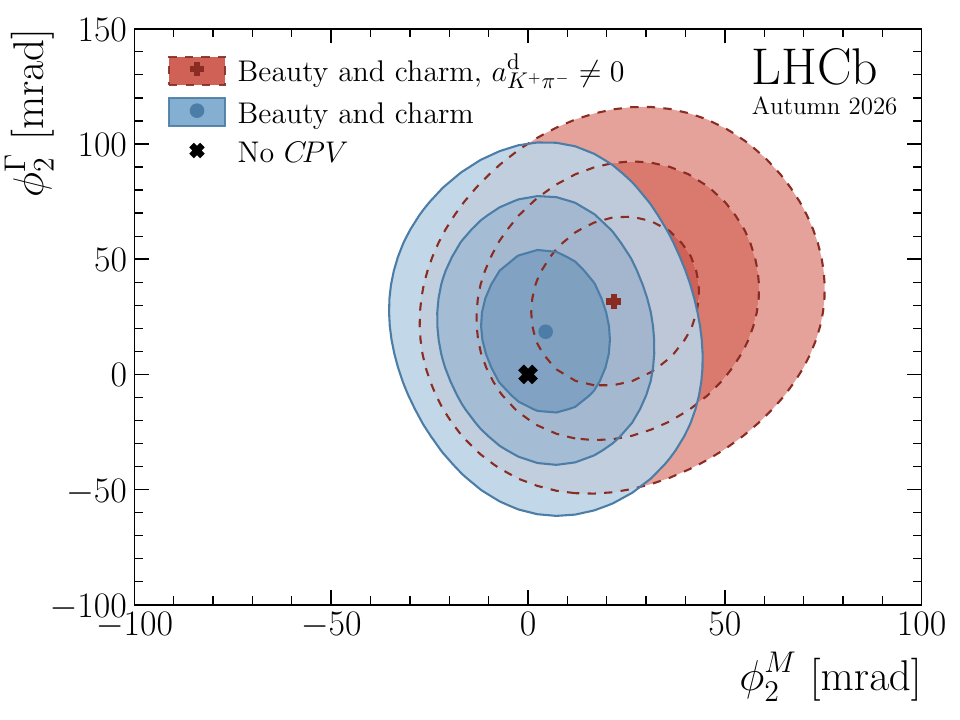}
  \includegraphics[width=0.45\linewidth]{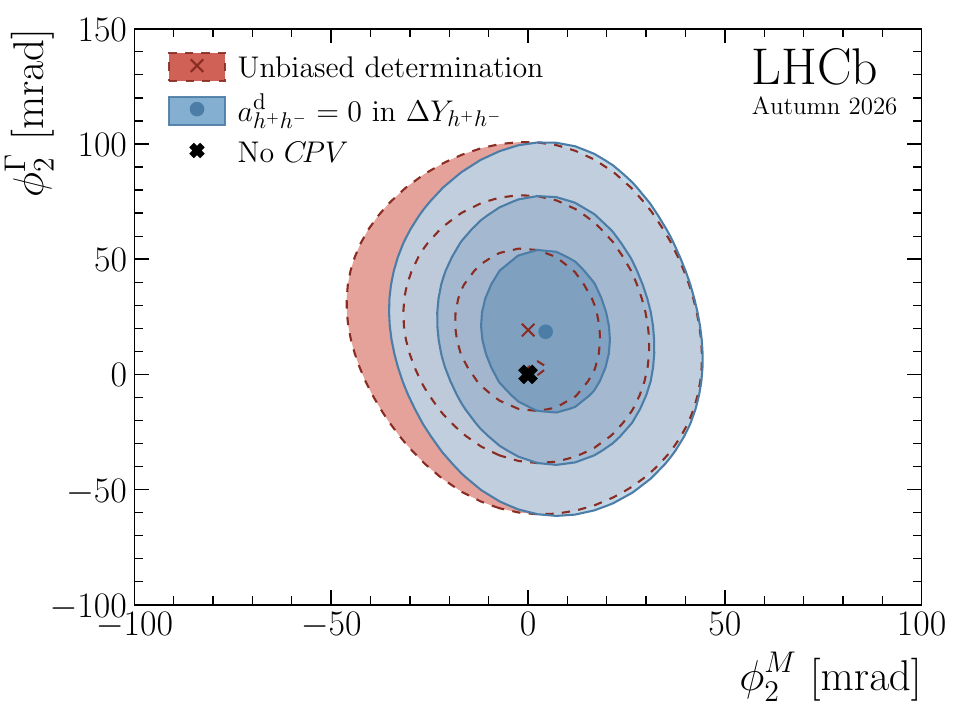}
  \includegraphics[width=0.45\linewidth]{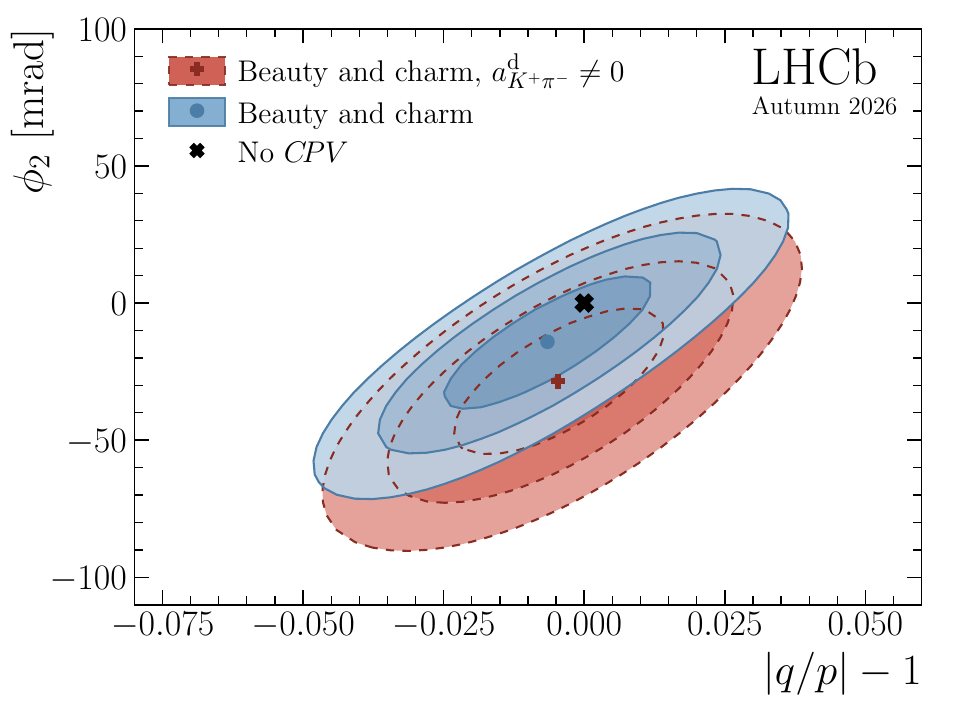}
  \includegraphics[width=0.45\linewidth]{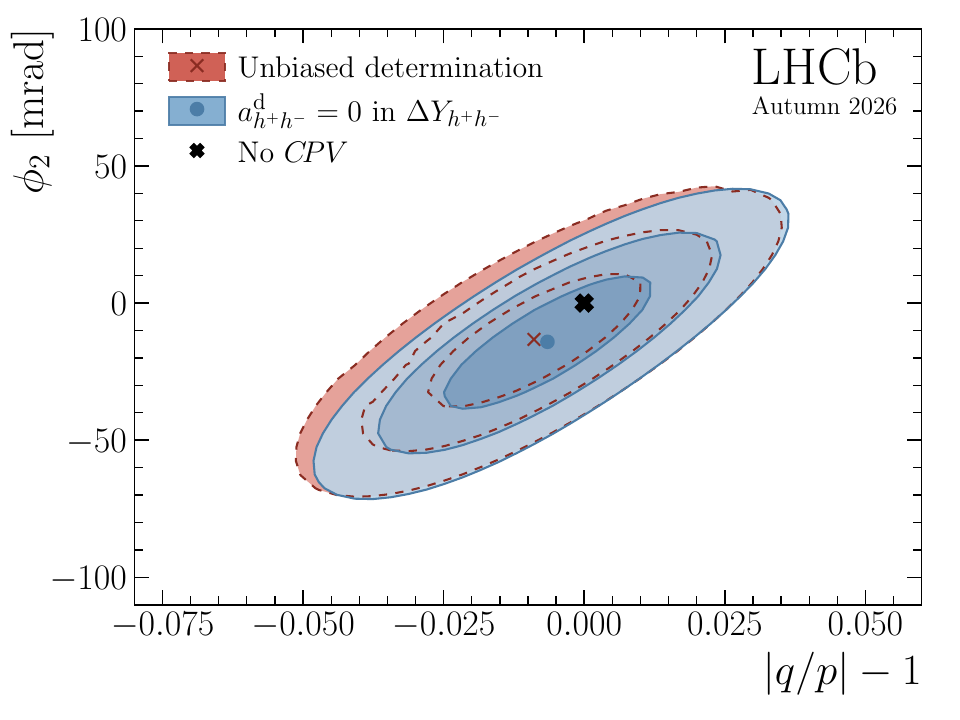}
  \includegraphics[width=0.45\linewidth]{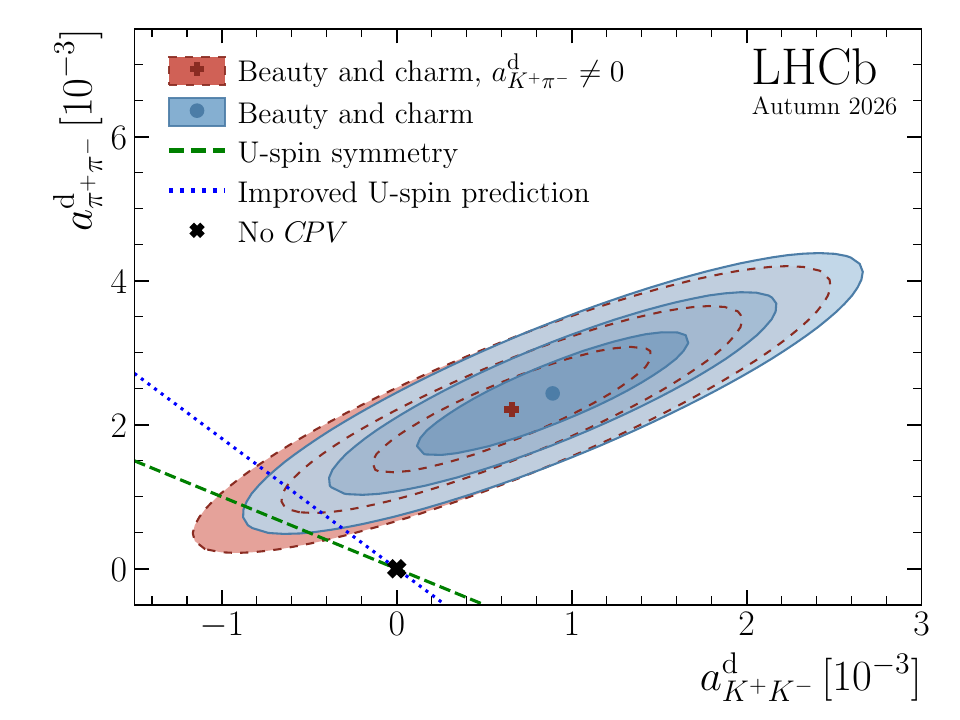}
  \includegraphics[width=0.45\linewidth]{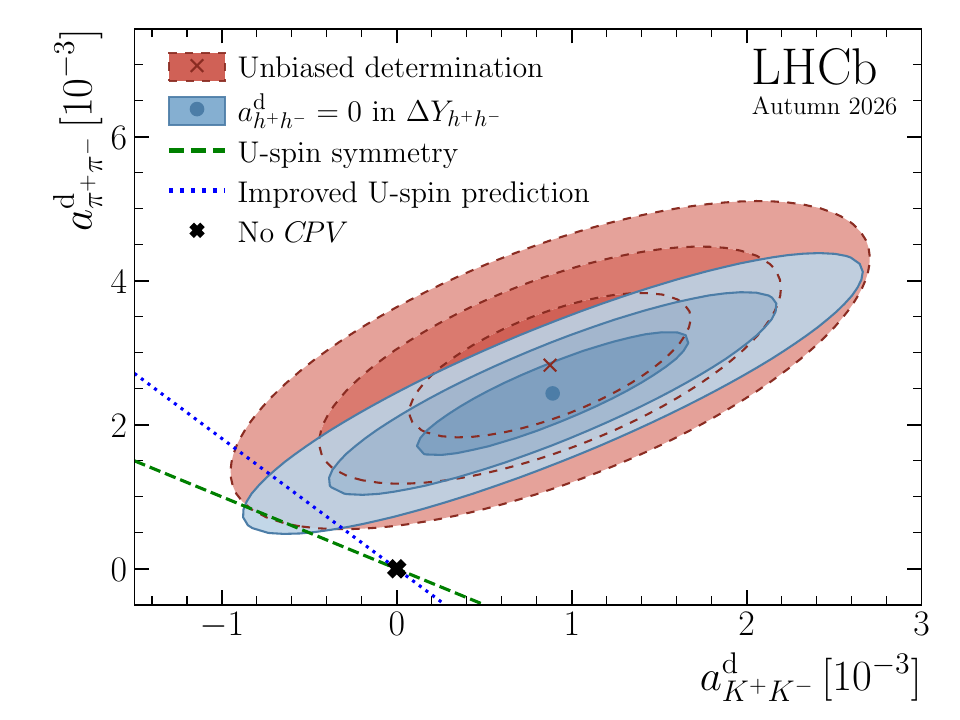}
  \caption{
    Two-dimensional profile-likelihood contours for the parameters quantifying
    (top) $C\!P$ violation in $D^0$ mixing in the theoretical parametrisation,
    (middle) the same in the phenomenological parametrisation, and
    (bottom) the $C\!P$ asymmetry in $D^0\to K^+K^-$ and $D^0\to \pi^+\pi^-$ decays.
    The left plots compare the results of (blue) the baseline combination not allowing for $C\!P$ violation in DCS decays and (red)
        the combination allowing for $C\!P$ violation in DCS $D^0\to K^+\pi^-$ decays.
    The right plots compare (blue) the baseline combination and (red) the combination allowing for the final-state dependence of the $\Delta Y_{K^+K^-}$ and $\Delta Y_{\pi^+\pi^-}$ parameters.
    The origin of the plots, which corresponds to $C\!P$ conservation, is shown for reference as are, for the bottom plots,
    the lines corresponding to exact and improved U-spin limits in \cref{eq:u-spin,eq:u-spin-improved}.
    Contours are displayed at the 1, 2, $3\,\sigma$ confidence levels, encompassing $68.3\%$, $95.4\%$, $99.7\%$ of the distributions, respectively.
  }
  \label{fig:charm-cpv}
\end{figure}

The dispersive and absorptive phases describing \CP violation in
charm mixing are determined to be \mbox{$\phiM = \resultPhiMRunThree$} and \mbox{$\phiG = \resultPhiGRunThree$},
corresponding to \mbox{$\qopD - 1 = \resultQPRunThree$} and \mbox{$\phiD = \resultPhiRunThree$} in the phenomenological parametrisation,
while the \CP asymmetries in \DzKK and \Dzpipi decays are \mbox{$\AcpDKK = \resultAcpDKKRunThree$} and
\mbox{$\AcpDpipi = \resultAcpDpipiRunThree$}.
The corresponding profile-likelihood contours are shown in the left plots of \cref{fig:charm-cpv},
which compare the baseline combination with an alternative configuration where \CP violation is allowed in the most precisely measured DCS decay, \DzKpPim.
Releasing the assumption that \mbox{$\AcpDKpi = 0$} significantly degrades the sensitivity to \phiM, due to the correlation between \AKpi and \DcKpi in \cref{eq:ws-to-rs}.
The values of \AcpDKK and \AcpDpipi are also affected, since the most precise measurement of the WS-to-RS ratio in \DzKpi decays determines the observable $\AKpit \approx \AcpDKpi - 2\,\AcpDKK$~\cite{LHCb-PAPER-2024-008}.
The numerical results of this alternative combination are reported in \cref{app:alternative-fits}, and are compatible with the baseline.

\subsection{Unbiased determination of \texorpdfstring{\boldmath \CP}{CP} violation in \texorpdfstring{\boldmath \Dzhh}{D0 -> h+ h-} decays}
An alternative combination is performed that drops the assumption of universality of charm-mixing weak phases and allows for final-state dependence of the \DYhh parameters through $\AcpDhh \neq 0$ in \cref{eq:DeltaY},
to obtain unbiased determinations of \AcpDKK and \AcpDpipi.
The time-dependent measurement of \Dzpipipiz decays in Ref.~\cite{LHCb-PAPER-2024-003} is removed from the inputs of this combination,
because no absolute measurements of \CP violation in this decay have been performed at \lhcb,
and the variation of the analogue of the strong phase \ddhhQCD across the Dalitz plot cannot be accounted for in a model-independent way.

The results are reported in \cref{tab:results-run123-dy-fsc} of \cref{app:alternative-fits} and are compared with those of the baseline combination in the right plots of \cref{fig:charm-cpv}.
Since the expression of \DYhh in \cref{eq:DeltaY} depends on the parameters $\cot\ddhhQCD$ (with $h = K, \pi$),
which do not appear in the expression of any other observable included in the combination,
the measurements of \DYhh and \AcpDhh nearly decouple from the rest of the combination.
Only a subleading correlation with the phases \phiM and \phiG remains, induced by the measurement of the observable $\AKpit \approx \AcpDKpi - 2\,\AcpDKK$ in the analysis of the WS-to-RS ratio of \DzKpi decays~\cite{LHCb-PAPER-2024-008}.
As the \DYhh measurements no longer constrain \phiM, the precision on this phase is marginally reduced, giving \mbox{$\phiM = \resultPhiMUnbiased$}.
The constraints set by the $\Delta A_{\CP}$ measurement in the (\AcpDKK, \AcpDpipi) plane are also weakened,
since in this case the uncertainty on the contribution of \DYhh to the time-integrated asymmetry of \cref{eq:acp-dz-to-hh} is larger than the uncertainty on the asymmetry itself.
This set of assumptions results in $\AcpDKK = \resultAcpDKKUnbiased$ and $\AcpDpipi = \resultAcpDpipiUnbiased$, with a correlation of $65\%$.
While this alternative combination could in principle be used to determine the values of $\cot\ddhhQCD$ and \rQCD separately, by reparametrising \AcpDhh, this possibility is not explored due to the low achievable precision.

\section{Conclusions}
\label{sec:conclusions}

The determination of the angle \g of the Unitarity Triangle is one of the key goals of the \lhcb collaboration,
with the original letter of intent~\cite{CERN-LHCC-1995-005} targeting a sensitivity of 
around 14\degrees for each inverse femtobarn of data accumulated at nominal beam energy,
under optimistic assumptions for the reconstruction efficiency and tagging power of the experiment.
With most of the foreseen sensitivity now harvested from the $9\invfb$ data set acquired between 2011 and 2018,
the uncertainty on \g has nearly reached 2\degrees, 
greatly expanding on the originally conceived programme of measurements, achieving the sensitivity set forth in the roadmap of the experiment \cite{LHCb-PUB-2009-029-with-link}, and significantly surpassing projections made with a more realistic understanding of the detector performance~\cite{LHCb-TDR-012}.
Moreover, the first measurement of \g with the upgraded \lhcb detector has been performed~\cite{LHCb-PAPER-2026-010},
with the sensitivity of a single year of data taking comparable to that achieved with the same method using the previous six years of data taking.
Including this measurement, the combination yields $\g = \resultGRunThree$.
Together, these results represent a stringent test of the SM picture of \CP violation in the quark sector,
with the precision of the direct measurement of \g now surpassing that of the angle $\alpha = \left(84.1^{\,+\,3.7}_{\,-\,3.0}\right)\degrees$~\cite{HFLAV:2022esi:zenodo.19371140} and comparable to that of the indirect determination from CKM unitarity constraints,
 \mbox{$\g = \gCKMfitter$}~\cite{CKMfitter2015} in a frequentist framework
and \mbox{$\g = \gUTfit$}~\cite{UTfit-UT} in a Bayesian one.
The result is three times more precise than the direct determination by the Belle and Belle~II experiments, \mbox{$\gamma = (78.6^{\,+\,7.2}_{\,-\,7.3})\degrees$}~\cite{Belle:2024knt}, with which it agrees at the level of two standard deviations.
The current generation of the LHCb experiment~\cite{LHCb-DP-2022-002} is expected to probe the CKM paradigm with subdegree precision in the near future.

Mixing and \CP violation in the charm sector are also crucial tests of the CKM picture, as they are in general sensitive to higher energy scales than beauty hadrons, and they provide more stringent bounds on BSM interactions affecting predominantly up-type quarks.
The first observation of \CP violation in charm decays~\cite{LHCb-PAPER-2019-006}, for which an updated interpretation is given in the present article, remains one of the most notable discoveries of the LHCb experiment.
By combining information from the beauty and charm sectors, unprecedented precision on the charm-mixing parameters has now been achieved,
with relative uncertainties of approximately $10\%$ and $3\%$ for \xth and \yth, respectively.
The two phases responsible for possible \CP violation in mixing remain consistent with zero,
with a precision of around $10\mrad$ and $20\mrad$ for \phiM and \phiG, respectively.
It can thus be anticipated that, with the unprecedented samples of charm decays collected by the current generation of the \lhcb experiment and by future upgrades~\cite{LHCb-PII-Physics},
sensitivity to the full range of SM predictions~\cite{Kagan:2020vri} can be achieved.

% Comment this in for paper drafts; do not include this in analysis note, conference and figure reports
\input{LHCb/acknowledgements}

% Appendices
\clearpage
\section*{Appendices}

\appendix
\crefalias{section}{appendix}

\section{Charm observables in the phenomenological parametrisation}
\label{app:phenomenological}
This appendix gives the expressions for the observables of \cref{sec:charm-observables} in the phenomenological parametrisation.
The observables of the WS-to-RS ratio in \DzKpi decays are conveniently reparametrised as $c_{K\pi}^\pm \equiv \cKpi \pm \DcKpi$ and $c_{K\pi}^{\prime\pm} \equiv \cKpipr \pm \DcKpipr$, with
\begin{equation}
\begin{aligned}
    c_{K\pi}^\pm         &\approx -\left\lvert \frac{q}{p} \right\rvert^{\pm 1} (\yD \cos(\ddkpi \pm \phiD) - \xD \sin(\ddkpi \pm \phiD) ),\\
    c_{K\pi}^{\prime\pm} &\approx \dfrac{1}{4} \left(\xD^2 + \yD^2 \right)\left\lvert \frac{q}{p} \right\rvert^{\pm 2}.
\end{aligned}
\end{equation}
The bin-flip observables in \cref{eq:binflip} are equal to
\begin{equation}
\begin{aligned}
    x_{\CP}  &= \frac{1}{2}\Big(  \xD\cos\phiD\Big(\Big\lvert\frac{q}{p}\Big\rvert + \Big\lvert\frac{p}{q}\Big\rvert\Big)
                                     + \yD\sin\phiD\Big(\Big\lvert\frac{q}{p}\Big\rvert - \Big\lvert\frac{p}{q}\Big\rvert\Big)\Big), \\
    y_{\CP}  &= \frac{1}{2}\Big(  \yD\cos\phiD\Big(\Big\lvert\frac{q}{p}\Big\rvert + \Big\lvert\frac{p}{q}\Big\rvert\Big)
                                     - \xD\sin\phiD\Big(\Big\lvert\frac{q}{p}\Big\rvert - \Big\lvert\frac{p}{q}\Big\rvert\Big)\Big), \\
    \Delta x &= \frac{1}{2}\Big(  \xD\cos\phiD\Big(\Big\lvert\frac{q}{p}\Big\rvert - \Big\lvert\frac{p}{q}\Big\rvert\Big)
                                     + \yD\sin\phiD\Big(\Big\lvert\frac{q}{p}\Big\rvert + \Big\lvert\frac{p}{q}\Big\rvert\Big)\Big), \\
    \Delta y &= \frac{1}{2}\Big(  \yD\cos\phiD\Big(\Big\lvert\frac{q}{p}\Big\rvert - \Big\lvert\frac{p}{q}\Big\rvert\Big)
                                     - \xD\sin\phiD\Big(\Big\lvert\frac{q}{p}\Big\rvert + \Big\lvert\frac{p}{q}\Big\rvert\Big)\Big),
\end{aligned}
\end{equation}
while the \ycpRS parameter in \cref{eq:ycp-rs}, equal to $\rdkpi(x_\CP\sin\ddkpi + y_\CP\cos\ddkpi)$, is expressed as
\begin{equation}
\begin{aligned}
    \ycpRS = \frac{1}{2}\rdkpi \bigg(& (\yD\cos\ddkpi + \xD\sin\ddkpi)\cos\phiD\left(\left\lvert\frac{q}{p}\right\rvert + \left\lvert\frac{p}{q}\right\rvert\right) \\
                                  & + (\yD\sin\ddkpi - \xD\cos\ddkpi)\sin\phiD\left(\left\lvert\frac{q}{p}\right\rvert - \left\lvert\frac{p}{q}\right\rvert\right)\bigg).
\end{aligned}
\end{equation}
Finally, the \DYhh parameter in \cref{eq:DeltaY} is equal to
\begin{equation}
\begin{aligned}
    \DYhh \approx \frac{1}{2}\bigg(&  \xD(\sin\phiD + \cos\phiD\,\AcpDhh\cot\ddhhQCD)\left(\left\lvert\frac{q}{p}\right\rvert + \left\lvert\frac{p}{q}\right\rvert\right) \\
                                  & - \yD\cos\phiD                                 \left(\left\lvert\frac{q}{p}\right\rvert - \left\lvert\frac{p}{q}\right\rvert - 2\,\AcpDhh \right)\bigg).
\end{aligned}
\end{equation}

\section{Reinterpretation of published LHCb results}
\label{app:derived-inputs}

This appendix reports the inputs to the combination that are derived from published \lhcb results,
reinterpreting them with updated external inputs.

The bin-flip measurements of \DzKSpipi decays using the Run~1~\cite{LHCb-PAPER-2019-001} and Run~2~\cite{LHCb-PAPER-2021-009,LHCb-PAPER-2022-020} samples are reinterpreted using the latest determination of the strong-phase parameters $(c_i,s_i)$ without U-spin constraints by the \besiii collaboration~\cite{BESIII:2025nsp}.
The updated results and their correlations are reported in \cref{tab:binflip-updated-r1,tab:binflip-updated-r2}, respectively.
\begin{table}[tb]
    \centering
    \caption{
        (Left) Results of the bin-flip measurement of $D^0\to K^0_{\mathrm{S}}\pi^+\pi^-$ decays using Run~1 data~\cite{LHCb-PAPER-2019-001},
        updated to use the determination of the hadronic parameters without U-spin constraints by the BESIII collaboration in Table~4 of Ref.~\cite{BESIII:2025nsp},
        and (right) the corresponding correlation matrix.
    }
    \begin{tabular}{CS[table-format=-1.1]@{$\,\pm\,$}S[table-format=1.1]}
    \toprule
    \text{Observable}& \multicolumn{2}{c}{Value $[10^{-3}]$} \\
    \midrule
    x_\CP    &       3.8 & 2.0 \\
    y_\CP    &       6.4 & 3.2 \\
    \Delta x &      -0.6 & 0.9 \\
    \Delta y &       0.5 & 1.5 \\
    \bottomrule
    \end{tabular}
    \hspace{1cm}
    \begin{tabular}{CS[table-format=-1.3]S[table-format=-1.3]S[table-format=-1.3]S[table-format=-1.3]}
    \toprule
             & {$x_\CP$} & {$y_\CP$} & {$\Delta x$} & {$\Delta y$} \\
    \midrule
    x_\CP    & 1.      & -0.031  & -0.006     &  0.002 \\
    y_\CP    &         &  1.     &  0.014     & -0.002 \\
    \Delta x &         &         &  1.        & -0.044 \\
    \Delta y &         &         &            &  1.    \\
    \bottomrule
    \end{tabular}
    \label{tab:binflip-updated-r1}
\end{table}
\begin{table}[tb]
    \centering
    \caption{
        (Left) Results of the bin-flip measurements of $D^0\to K^0_{\mathrm{S}}\pi^+\pi^-$ decays using Run~2 data~\cite{LHCb-PAPER-2021-009,LHCb-PAPER-2022-020},
        updated to use the determination of the hadronic parameters without U-spin constraints by the BESIII collaboration in Table~4 of Ref.~\cite{BESIII:2025nsp},
        and (right) the corresponding correlation matrix.}
    \begin{tabular}{CS[table-format=-1.2]@{$\,\pm\,$}S[table-format=1.2]}
    \toprule
    \text{Observable}& \multicolumn{2}{c}{Value $[10^{-3}]$} \\
    \midrule
    x_\CP    &       4.90 & 0.66 \\
    y_\CP    &       5.12 & 1.18 \\
    \Delta x &      -0.33 & 0.22 \\
    \Delta y &       0.28 & 0.35 \\
    \bottomrule
    \end{tabular}
    \hspace{1cm}
    \begin{tabular}{CS[table-format=-1.3]S[table-format=-1.3]S[table-format=-1.3]S[table-format=-1.3]}
    \toprule
             & {$x_\CP$} & {$y_\CP$} & {$\Delta x$} & {$\Delta y$} \\
    \midrule
    x_\CP    & 1.      & 0.042   & -0.083     & -0.007 \\
    y_\CP    &         & 1.      & -0.004     & -0.072 \\
    \Delta x &         &         &  1.        &  0.012 \\
    \Delta y &         &         &            &  1.    \\
    \bottomrule
    \end{tabular}
    \label{tab:binflip-updated-r2}
\end{table}

The results of the \DKpipipi standalone combination described in \cref{sec:inputs} are reported in \cref{tab:k3pi-beauty-charm}.
In this combination, the charm-mixing parameters are parametrised as
$\rmix \equiv \sqrt{\xD^2 + \yD^2}$
and
$\pmix \equiv \arctan(\xD/\yD)$,
such that
$\xD = \rmix\sin \pmix$ and
$\yD = \rmix\cos \pmix$,
while \CP violation in mixing is described by the parameters $\Delta x$ and $\Delta y$ defined in \cref{eq:binflip}.
\begin{table}[tb]
    \centering
    \caption{
        (Top) Results of the standalone $D\to K^\pm\pi^\mp\pi^+\pi^-$ combination including both beauty and charm measurements,
        and (bottom) the corresponding correlation matrix.
    }
    \begin{tabular}{CC}
    \toprule
    \text{Observable}& \text{Value} \\
    \midrule
    \gamma          & (58    \pm  8)\degrees          \\
    \rdk            & (9.6   \pm  0.4)\times10^{-2} \\
    \ddk            & (123   \pm 11)\degrees          \\
    \rdpi           & (5.6   \pm  1.3)\times10^{-3} \\
    \ddpi           & (268   \pm 16)\degrees          \\
    \rmix           & (7.1   \pm  0.5)\times10^{-3} \\
    \pmix           & (42    \pm 16)\degrees          \\
    \Delta x        & (-0.02 \pm  0.27)\times10^{-3} \\
    \Delta y        & (-0.01 \pm  0.21)\times10^{-3} \\
    \bottomrule
    \end{tabular}
    \par\vspace{2ex}
    \begin{tabular}{CS[table-format=-1.2]S[table-format=-1.2]S[table-format=-1.2]S[table-format=-1.2]S[table-format=-1.2]S[table-format=-1.2]S[table-format=-1.2]S[table-format=-1.2]S[table-format=-1.2]}
    \toprule
                  & {$\gamma$} & {\rdk} & {\ddk} & {\rdpi} & {\ddpi} & {\rmix} & {\pmix} & {$\Delta x$} & {$\Delta y$} \\
    \midrule
    \gamma        & 1.       & -0.21    &  0.24         & -0.17      & -0.05           &  0.21          & -0.27          &  0.05      & -0.05 \\
    \rdk          &          &  1.      & -0.52         &  0.10      & -0.02           &  0.03          &  0.18          & -0.05      &  0.05 \\
    \ddk          &          &          &  1.           & -0.10      &  0.13           & -0.14          &  0.18          & -0.01      &  0.00 \\
    \rdpi         &          &          &               &  1.        & -0.19           &  0.08          &  0.13          & -0.03      &  0.02 \\
    \ddpi         &          &          &               &            &  1.             & -0.16          & -0.05          &  0.04      & -0.03 \\
    \rmix         &          &          &               &            &                 &  1.            &  0.40          & -0.07      &  0.12 \\
    \pmix         &          &          &               &            &                 &                &  1.            & -0.16      &  0.13 \\
    \Delta x      &          &          &               &            &                 &                &                &  1.        & -0.63 \\
    \Delta y      &          &          &               &            &                 &                &                &            &  1.   \\
    \bottomrule
    \end{tabular}
    \label{tab:k3pi-beauty-charm}
\end{table}
An alternative combination including only the charm measurements is also performed, with $\chisqndf = 7.65/12$.
Its results are reported in \cref{tab:k3pi-charm}.
\begin{table}[tb]
    \centering
    \caption{
        (Left) Results of the standalone $D\to K^\pm\pi^\mp\pi^+\pi^-$ combination including only charm measurements,
        and (right) the corresponding correlation matrix.
    }
    \begin{tabular}{CC}
    \toprule
    \text{Observable}& \text{Value} \\
    \midrule
    \rmix        & (7.5   \pm  0.7)\times10^{-3}  \\
    \pmix        & (48    \pm 16)\degrees           \\
    \Delta x     & (0.01  \pm  0.29)\times10^{-3} \\
    \Delta y     & (-0.02 \pm  0.19)\times10^{-3} \\
    \bottomrule
    \end{tabular}
    \hspace{1cm}
    \begin{tabular}{CS[table-format=-1.2]S[table-format=-1.2]S[table-format=-1.2]S[table-format=-1.2]}
    \toprule
                 & {\rmix}        & {\pmix}        & {$\Delta x$} & {$\Delta y$} \\
    \midrule
    \rmix        & 1.             & -0.52          & -0.09      &  0.11 \\
    \pmix        &                &  1.            &  0.14      & -0.10 \\
    \Delta x     &                &                &  1.        & -0.59 \\
    \Delta y     &                &                &            &  1.   \\
    \bottomrule
    \end{tabular}
    \label{tab:k3pi-charm}
\end{table}

\clearpage

\section{Additional combination results}
\label{app:additional-results}

\begin{figure}[!ht]
  \centering
  \includegraphics[width=\textwidth]{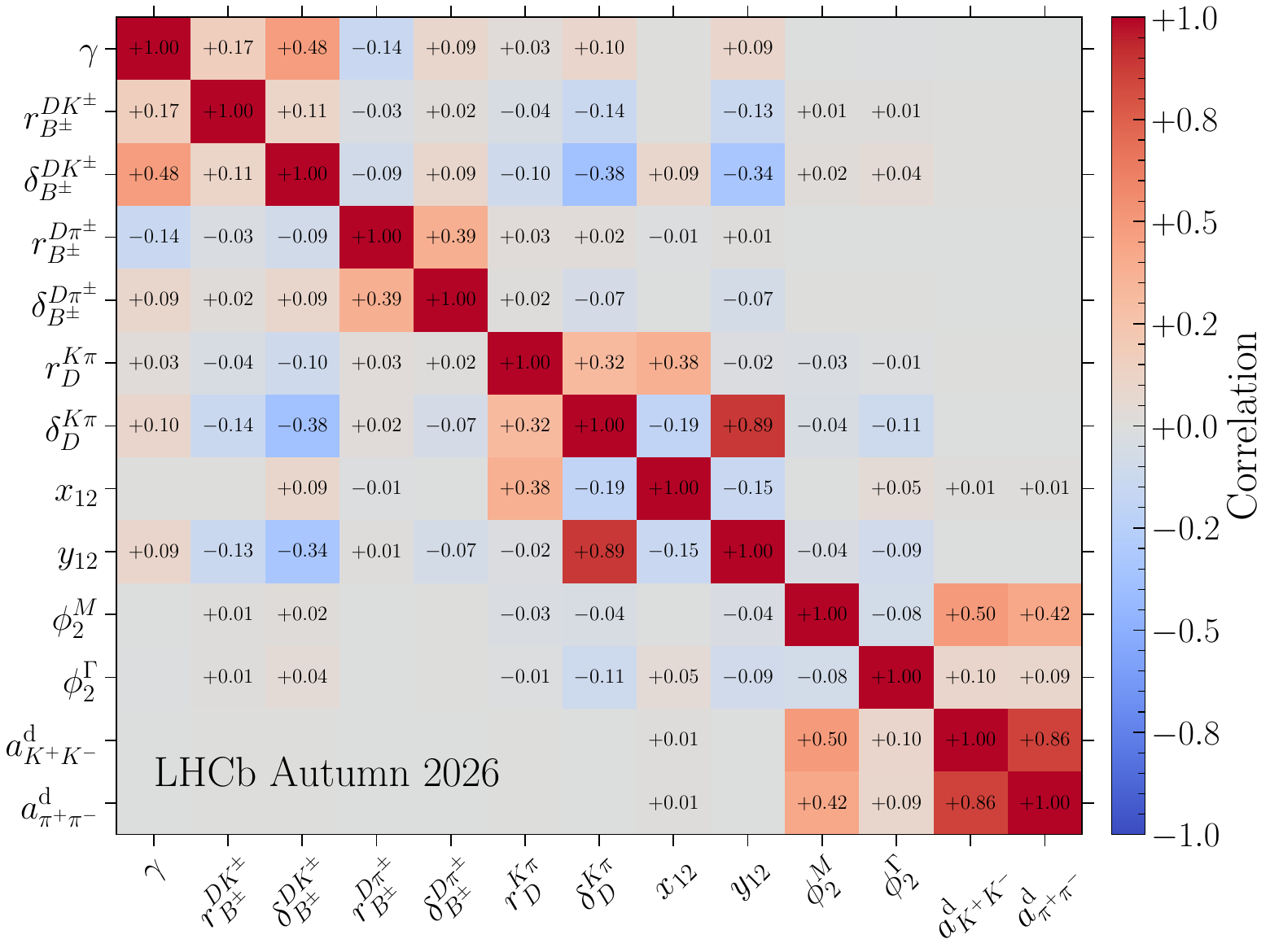}
  \caption{Reduced correlation matrix for the parameters of
    greatest interest, as given in \cref{tab:corrs_small_run123}. Values smaller than 0.01 in absolute value are not written.}
  \label{fig:corrs_small_run123}
\end{figure}

\begin{figure}[!ht]
  \centering
  \includegraphics[width=\textwidth]{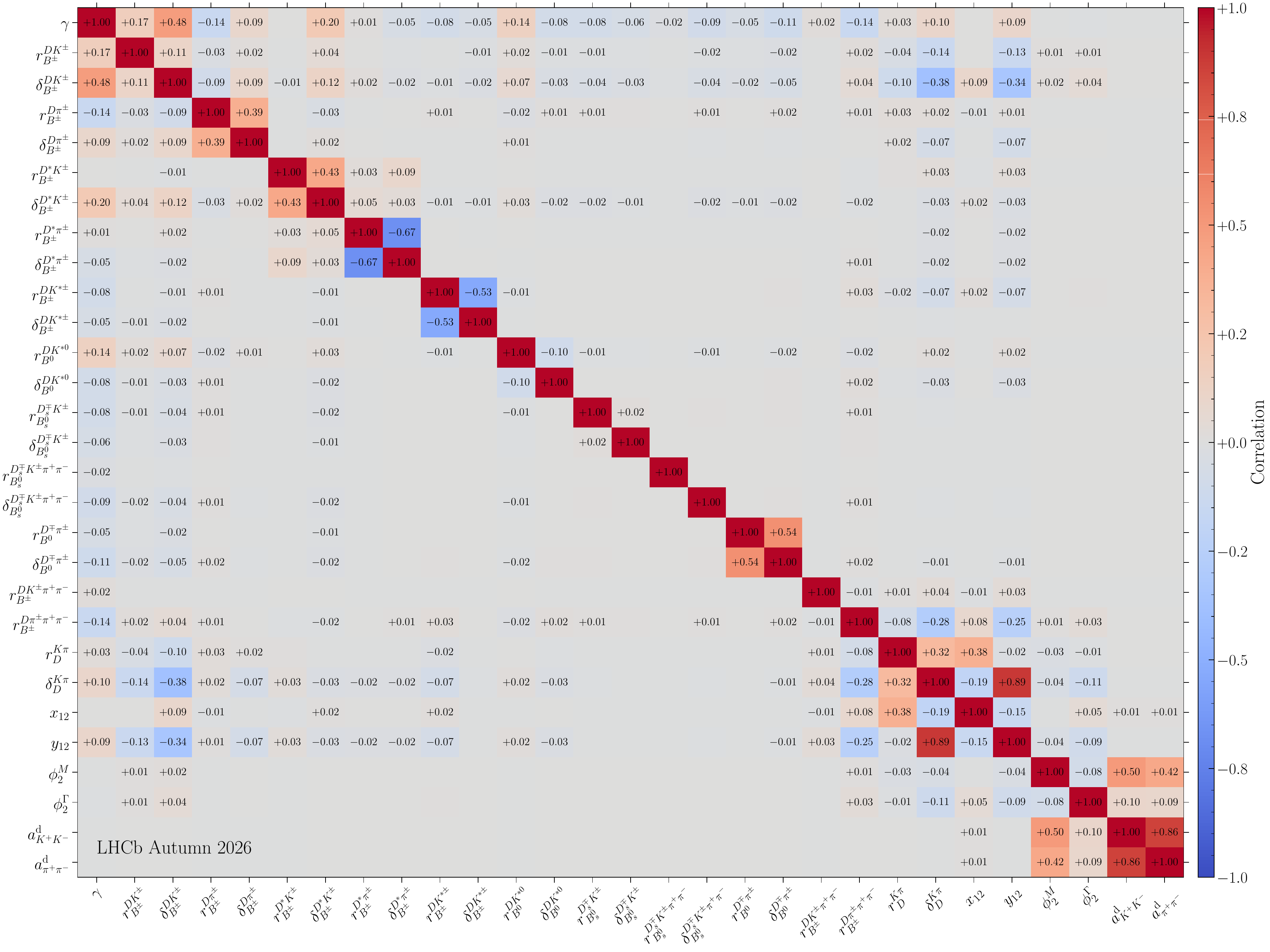}
  \caption{
    Correlation matrix of the combination, as given in \cref{tab:corrs_run123}.
    Values smaller than 0.01 in absolute value are not written.
  }
  \label{fig:corrs_run123}
\end{figure}

The correlation matrix for the parameters of greatest interest presented in
\cref{tab:results_run123} is given in \cref{tab:corrs_small_run123} and visualised in
\cref{fig:corrs_small_run123}, with
the charm-mixing parameters expressed in the theoretical parametrisation.
The correlations between \g and the charm parameters in the phenomenological parametrisation are given in \cref{tab:corrs_pheno_run123},
while the full correlation matrix is given in \cref{tab:corrs_run123} and visualised in
\cref{fig:corrs_run123}.
Finally, the contribution of each input measurement to the global \chisq is listed in \cref{tab:chi2_contrib_run123}.

\begin{table}[!ht]
  \centering
  \caption{Reduced correlation matrix for the parameters of greatest interest.
  Values smaller than 0.01 in absolute value are replaced with a ``-'' symbol.}
  \label{tab:corrs_small_run123}
  \renewcommand{\arraystretch}{1.2}
  \setlength{\tabcolsep}{5pt}
  \resizebox{\textwidth}{!}{
\begin{tabular}{C*{13}{S[table-format=-1.2]}}
\hline
            &        {\g} &      {\rdk} &      {\ddk} &     {\rdpi} &     {\ddpi} &    {\rDkpi} &    {\dDkpi} &      {\xth} &      {\yth} &     {\phiM} &     {\phiG} &   {\AcpDKK} & {\AcpDpipi} \\
\hline
\g          &           1 &        0.17 &        0.48 &       -0.14 &        0.09 &        0.03 &        0.10 &         {-} &        0.09 &         {-} &         {-} &         {-} &         {-} \\
\rdk        &             &           1 &        0.11 &       -0.03 &        0.02 &       -0.04 &       -0.14 &         {-} &       -0.13 &        0.01 &        0.01 &         {-} &         {-} \\
\ddk        &             &             &           1 &       -0.09 &        0.09 &       -0.10 &       -0.38 &        0.09 &       -0.34 &        0.02 &        0.04 &         {-} &         {-} \\
\rdpi       &             &             &             &           1 &        0.39 &        0.03 &        0.02 &       -0.01 &        0.01 &         {-} &         {-} &         {-} &         {-} \\
\ddpi       &             &             &             &             &           1 &        0.02 &       -0.07 &         {-} &       -0.07 &         {-} &         {-} &         {-} &         {-} \\
\rDkpi      &             &             &             &             &             &           1 &        0.32 &        0.38 &       -0.02 &       -0.03 &       -0.01 &         {-} &         {-} \\
\dDkpi      &             &             &             &             &             &             &           1 &       -0.19 &        0.89 &       -0.04 &       -0.11 &         {-} &         {-} \\
\xth        &             &             &             &             &             &             &             &           1 &       -0.15 &         {-} &        0.05 &        0.01 &        0.01 \\
\yth        &             &             &             &             &             &             &             &             &           1 &       -0.04 &       -0.09 &         {-} &         {-} \\
\phiM       &             &             &             &             &             &             &             &             &             &           1 &       -0.08 &        0.50 &        0.42 \\
\phiG       &             &             &             &             &             &             &             &             &             &             &           1 &        0.10 &        0.09 \\
\AcpDKK     &             &             &             &             &             &             &             &             &             &             &             &           1 &        0.86 \\
\AcpDpipi   &             &             &             &             &             &             &             &             &             &             &             &             &           1 \\
\hline
\end{tabular}
  }
\end{table}

\begin{table}[!ht]
  \centering
  \caption{Reduced correlation matrix for $\gamma$ and the charm
    parameters, with the charm-mixing parameters expressed in the phenomenological parametrisation.
    Values smaller than 0.01 in absolute value are replaced with a ``-'' symbol.}
  \label{tab:corrs_pheno_run123}
  \renewcommand{\arraystretch}{1.2}
  \setlength{\tabcolsep}{5pt}
  \resizebox{0.75\textwidth}{!}{
\begin{tabular}{C*{9}{S[table-format=-1.2]}}
\hline
            &        {\g} &    {\rDkpi} &    {\dDkpi} &       {\xD} &       {\yD} &     {\qopD} &     {\phiD} &   {\AcpDKK} & {\AcpDpipi} \\
\hline
\g          &           1 &        0.03 &        0.10 &         {-} &        0.09 &         {-} &         {-} &         {-} &         {-} \\
\rDkpi      &             &           1 &        0.32 &        0.38 &       -0.02 &       -0.01 &        0.03 &         {-} &         {-} \\
\dDkpi      &             &             &           1 &       -0.19 &        0.89 &        0.09 &        0.10 &         {-} &         {-} \\
\xD         &             &             &             &           1 &       -0.15 &       -0.06 &         {-} &        0.01 &        0.01 \\
\yD         &             &             &             &             &           1 &        0.08 &        0.09 &         {-} &         {-} \\
\qopD       &             &             &             &             &             &           1 &        0.81 &        0.11 &        0.09 \\
\phiD       &             &             &             &             &             &             &           1 &       -0.20 &       -0.17 \\
\AcpDKK     &             &             &             &             &             &             &             &           1 &        0.86 \\
\AcpDpipi   &             &             &             &             &             &             &             &             &           1 \\
\hline
\end{tabular}
  }
\end{table}

\begin{landscape}
\begin{table}[!tb]
  \centering
  \caption{Correlation matrix of the combination. Values smaller than 0.01 in absolute value are
    replaced with a ``-'' symbol.}
  \label{tab:corrs_run123}
  \renewcommand{\arraystretch}{1.2}
  \setlength{\tabcolsep}{5pt}
  % the rotated page is \textheight wide, not \textwidth
  \makebox[\textwidth][c]{\resizebox{1.5\textheight}{!}{
\begin{tabular}{C*{29}{S[table-format=-1.2]}}
\hline
            &        {\g} &      {\rdk} &      {\ddk} &     {\rdpi} &     {\ddpi} &    {\rdstk} &    {\ddstk} &   {\rdstpi} &   {\ddstpi} &    {\rdkst} &    {\ddkst} &   {\rdkstz} &   {\ddkstz} &     {\ldsk} &     {\ddsk} & {\ldskpipi} & {\ddskpipi} &    {\ldmpi} &    {\ddmpi} &  {\rdkpipi} & {\rdpipipi} &    {\rDkpi} &    {\dDkpi} &      {\xth} &      {\yth} &     {\phiM} &     {\phiG} &   {\AcpDKK} & {\AcpDpipi} \\
\hline
\g          &           1 &        0.17 &        0.48 &       -0.14 &        0.09 &         {-} &        0.20 &        0.01 &       -0.05 &       -0.08 &       -0.05 &        0.14 &       -0.08 &       -0.08 &       -0.06 &       -0.02 &       -0.09 &       -0.05 &       -0.11 &        0.02 &       -0.14 &        0.03 &        0.10 &         {-} &        0.09 &         {-} &         {-} &         {-} &         {-} \\
\rdk        &             &           1 &        0.11 &       -0.03 &        0.02 &         {-} &        0.04 &         {-} &         {-} &         {-} &       -0.01 &        0.02 &       -0.01 &       -0.01 &         {-} &         {-} &       -0.02 &         {-} &       -0.02 &         {-} &        0.02 &       -0.04 &       -0.14 &         {-} &       -0.13 &        0.01 &        0.01 &         {-} &         {-} \\
\ddk        &             &             &           1 &       -0.09 &        0.09 &       -0.01 &        0.12 &        0.02 &       -0.02 &       -0.01 &       -0.02 &        0.07 &       -0.03 &       -0.04 &       -0.03 &         {-} &       -0.04 &       -0.02 &       -0.05 &         {-} &        0.04 &       -0.10 &       -0.38 &        0.09 &       -0.34 &        0.02 &        0.04 &         {-} &         {-} \\
\rdpi       &             &             &             &           1 &        0.39 &         {-} &       -0.03 &         {-} &         {-} &        0.01 &         {-} &       -0.02 &        0.01 &        0.01 &         {-} &         {-} &        0.01 &         {-} &        0.02 &         {-} &        0.01 &        0.03 &        0.02 &       -0.01 &        0.01 &         {-} &         {-} &         {-} &         {-} \\
\ddpi       &             &             &             &             &           1 &         {-} &        0.02 &         {-} &         {-} &         {-} &         {-} &        0.01 &         {-} &         {-} &         {-} &         {-} &         {-} &         {-} &         {-} &         {-} &         {-} &        0.02 &       -0.07 &         {-} &       -0.07 &         {-} &         {-} &         {-} &         {-} \\
\rdstk      &             &             &             &             &             &           1 &        0.43 &        0.03 &        0.09 &         {-} &         {-} &         {-} &         {-} &         {-} &         {-} &         {-} &         {-} &         {-} &         {-} &         {-} &         {-} &         {-} &        0.03 &         {-} &        0.03 &         {-} &         {-} &         {-} &         {-} \\
\ddstk      &             &             &             &             &             &             &           1 &        0.05 &        0.03 &       -0.01 &       -0.01 &        0.03 &       -0.02 &       -0.02 &       -0.01 &         {-} &       -0.02 &       -0.01 &       -0.02 &         {-} &       -0.02 &         {-} &       -0.03 &        0.02 &       -0.03 &         {-} &         {-} &         {-} &         {-} \\
\rdstpi     &             &             &             &             &             &             &             &           1 &       -0.67 &         {-} &         {-} &         {-} &         {-} &         {-} &         {-} &         {-} &         {-} &         {-} &         {-} &         {-} &         {-} &         {-} &       -0.02 &         {-} &       -0.02 &         {-} &         {-} &         {-} &         {-} \\
\ddstpi     &             &             &             &             &             &             &             &             &           1 &         {-} &         {-} &         {-} &         {-} &         {-} &         {-} &         {-} &         {-} &         {-} &         {-} &         {-} &        0.01 &         {-} &       -0.02 &         {-} &       -0.02 &         {-} &         {-} &         {-} &         {-} \\
\rdkst      &             &             &             &             &             &             &             &             &             &           1 &       -0.53 &       -0.01 &         {-} &         {-} &         {-} &         {-} &         {-} &         {-} &         {-} &         {-} &        0.03 &       -0.02 &       -0.07 &        0.02 &       -0.07 &         {-} &         {-} &         {-} &         {-} \\
\ddkst      &             &             &             &             &             &             &             &             &             &             &           1 &         {-} &         {-} &         {-} &         {-} &         {-} &         {-} &         {-} &         {-} &         {-} &         {-} &         {-} &         {-} &         {-} &         {-} &         {-} &         {-} &         {-} &         {-} \\
\rdkstz     &             &             &             &             &             &             &             &             &             &             &             &           1 &       -0.10 &       -0.01 &         {-} &         {-} &       -0.01 &         {-} &       -0.02 &         {-} &       -0.02 &         {-} &        0.02 &         {-} &        0.02 &         {-} &         {-} &         {-} &         {-} \\
\ddkstz     &             &             &             &             &             &             &             &             &             &             &             &             &           1 &         {-} &         {-} &         {-} &         {-} &         {-} &         {-} &         {-} &        0.02 &         {-} &       -0.03 &         {-} &       -0.03 &         {-} &         {-} &         {-} &         {-} \\
\ldsk       &             &             &             &             &             &             &             &             &             &             &             &             &             &           1 &        0.02 &         {-} &         {-} &         {-} &         {-} &         {-} &        0.01 &         {-} &         {-} &         {-} &         {-} &         {-} &         {-} &         {-} &         {-} \\
\ddsk       &             &             &             &             &             &             &             &             &             &             &             &             &             &             &           1 &         {-} &         {-} &         {-} &         {-} &         {-} &         {-} &         {-} &         {-} &         {-} &         {-} &         {-} &         {-} &         {-} &         {-} \\
\ldskpipi   &             &             &             &             &             &             &             &             &             &             &             &             &             &             &             &           1 &         {-} &         {-} &         {-} &         {-} &         {-} &         {-} &         {-} &         {-} &         {-} &         {-} &         {-} &         {-} &         {-} \\
\ddskpipi   &             &             &             &             &             &             &             &             &             &             &             &             &             &             &             &             &           1 &         {-} &         {-} &         {-} &        0.01 &         {-} &         {-} &         {-} &         {-} &         {-} &         {-} &         {-} &         {-} \\
\ldmpi      &             &             &             &             &             &             &             &             &             &             &             &             &             &             &             &             &             &           1 &        0.54 &         {-} &         {-} &         {-} &         {-} &         {-} &         {-} &         {-} &         {-} &         {-} &         {-} \\
\ddmpi      &             &             &             &             &             &             &             &             &             &             &             &             &             &             &             &             &             &             &           1 &         {-} &        0.02 &         {-} &       -0.01 &         {-} &       -0.01 &         {-} &         {-} &         {-} &         {-} \\
\rdkpipi    &             &             &             &             &             &             &             &             &             &             &             &             &             &             &             &             &             &             &             &           1 &       -0.01 &        0.01 &        0.04 &       -0.01 &        0.03 &         {-} &         {-} &         {-} &         {-} \\
\rdpipipi   &             &             &             &             &             &             &             &             &             &             &             &             &             &             &             &             &             &             &             &             &           1 &       -0.08 &       -0.28 &        0.08 &       -0.25 &        0.01 &        0.03 &         {-} &         {-} \\
\rDkpi      &             &             &             &             &             &             &             &             &             &             &             &             &             &             &             &             &             &             &             &             &             &           1 &        0.32 &        0.38 &       -0.02 &       -0.03 &       -0.01 &         {-} &         {-} \\
\dDkpi      &             &             &             &             &             &             &             &             &             &             &             &             &             &             &             &             &             &             &             &             &             &             &           1 &       -0.19 &        0.89 &       -0.04 &       -0.11 &         {-} &         {-} \\
\xth        &             &             &             &             &             &             &             &             &             &             &             &             &             &             &             &             &             &             &             &             &             &             &             &           1 &       -0.15 &         {-} &        0.05 &        0.01 &        0.01 \\
\yth        &             &             &             &             &             &             &             &             &             &             &             &             &             &             &             &             &             &             &             &             &             &             &             &             &           1 &       -0.04 &       -0.09 &         {-} &         {-} \\
\phiM       &             &             &             &             &             &             &             &             &             &             &             &             &             &             &             &             &             &             &             &             &             &             &             &             &             &           1 &       -0.08 &        0.50 &        0.42 \\
\phiG       &             &             &             &             &             &             &             &             &             &             &             &             &             &             &             &             &             &             &             &             &             &             &             &             &             &             &           1 &        0.10 &        0.09 \\
\AcpDKK     &             &             &             &             &             &             &             &             &             &             &             &             &             &             &             &             &             &             &             &             &             &             &             &             &             &             &             &           1 &        0.86 \\
\AcpDpipi   &             &             &             &             &             &             &             &             &             &             &             &             &             &             &             &             &             &             &             &             &             &             &             &             &             &             &             &             &           1 \\
\hline
\end{tabular}
  }}
\end{table}
\end{landscape}

\begin{table}[!ht]
  \caption{
    Contribution of each input measurement to the total $\chi^2$ of the combination, and number of observables it provides.
    The fit contains 52 free parameters.
  }
  \label{tab:chi2_contrib_run123}
  \centering
  \renewcommand{\arraystretch}{1.1}
  \resizebox{0.6\textwidth}{!}{
\begin{tabular}{l|lrc}
\toprule
& Measurement & $\chi^2$ & No.~of obs.\\
\midrule
\multirow{15}{*}{\rotatebox{90}{Beauty sector}}
	& $B^{\pm}\to Dh^{\pm},\, D\to h^{\pm}h^{\prime\mp}$ & $ 5.20$ & $  8$ \\
	& $B^{\pm}\to Dh^{\pm},\, D\to h^+h^-\pi^+\pi^-$ & $ 3.70$ & $  6$ \\
	& $B^{\pm}\to Dh^{\pm},\, D\to K^+K^-\pi^+\pi^-\ \mathrm{(binned)}$ & $ 4.92$ & $  5$ \\
	& $B^{\pm}\to Dh^{\pm},\, D\to h^{\pm}h^{\prime\mp}\pi^0$ & $ 6.65$ & $ 11$ \\
	& $B\to D^{(*)}h^{(*)},\, D\to K^0_S h^+h^-$ & $25.29$ & $ 26$ \\
	& $B^{\pm}\to Dh^{\pm},\, D\to K^0_S h^+h^-\ (2024)$ & $ 7.18$ & $  6$ \\
	& $B^{\pm}\to Dh^{\pm},\, D\to K^0_S K^{\pm}\pi^{\mp}$ & $ 7.78$ & $  7$ \\
	& $B^{\pm}\to D^{*}h^{\pm},\, D\to h^{\pm}h^{\prime\mp}$ & $ 9.55$ & $ 16$ \\
	& $B^{\pm}\to DK^{*\pm},\, D\to h^{\pm}h^{\prime\mp}(\pi^+\pi^-)$ & $28.53$ & $ 12$ \\
	& $B^{\pm}\to Dh^{\pm}\pi^+\pi^-,\, D\to h^{\pm}h^{\prime\mp}$ & $ 1.26$ & $ 11$ \\
	& $B^{0}\to DK^{*0},\, D\to h^{\pm}h^{\prime\mp}(\pi^+\pi^-)$ & $15.51$ & $ 12$ \\
	& $B^{0}\to D^{\mp}\pi^{\pm}$ & $ 0.00$ & $  2$ \\
	& $B^{0}_{s}\to D^{\mp}_{s}K^{\pm}\ \mathrm{Run\ 1}$ & $ 5.78$ & $  5$ \\
	& $B^{0}_{s}\to D^{\mp}_{s}K^{\pm}\ \mathrm{Run\ 2}$ & $ 1.29$ & $  5$ \\
	& $B^{0}_{s}\to D^{\mp}_{s}K^{\pm}\pi^+\pi^-$ & $ 2.70$ & $  5$ \\
\midrule
	& $D\to K^{\mp}\pi^{\pm}\pi^+\pi^-\ \mathrm{combination}$ & $ 2.59$ & $  9$ \\
\midrule
\multirow{11}{*}{\rotatebox{90}{Charm sector}}
	& $A_{C\!P}(h^+h^-)$ & $ 7.72$ & $  8$ \\
	& $y_{C\!P}-y_{C\!P}^{K\pi}\ \mathrm{Run\ 1}$ & $ 0.32$ & $  1$ \\
	& $y_{C\!P}-y_{C\!P}^{K\pi}\ \mathrm{Run\ 2}$ & $ 1.64$ & $  1$ \\
	& $\Delta Y$ & $ 2.01$ & $  2$ \\
	& $D^0\to K^{\pm}\pi^{\mp}\ \mathrm{(single\ tag)}$ & $ 5.02$ & $  9$ \\
	& $D^0\to K^{\pm}\pi^{\mp}\ \mathrm{(double\ tag)}$ & $ 4.85$ & $  6$ \\
	& $D^0\to K^{\pm}\pi^{\mp}\pi^+\pi^-\ (x^2+y^2)/4,\ \mathrm{Run\ 1}$ & $ 3.58$ & $  1$ \\
	& $D^0\to K^0_S\pi^+\pi^-\ (x,\, y)$ & $ 5.52$ & $  2$ \\
	& $D^0\to K^0_S\pi^+\pi^-\ \mathrm{bin\text{-}flip}\ \mathrm{Run\ 1}$ & $ 0.46$ & $  4$ \\
	& $D^0\to K^0_S\pi^+\pi^-\ \mathrm{bin\text{-}flip}\ \mathrm{Run\ 2}$ & $ 3.31$ & $  4$ \\
	& $\Delta Y^{\mathrm{eff}}\ (\pi^+\pi^-\pi^0)$ & $ 0.03$ & $  1$ \\
\midrule
\multirow{14}{*}{\rotatebox{90}{External constraints}}
	& $\kappa_B^{DK^{*\pm}}$ & $ 0.07$ & $  1$ \\
	& $\kappa_B^{DK^{*0}}$ & $ 0.00$ & $  1$ \\
	& $\beta$ & $ 0.00$ & $  1$ \\
	& $\phi_s$ & $ 0.01$ & $  1$ \\
	& $D^0\to K^{\pm}\pi^{\mp}\ \mathrm{(CLEO)}$ & $10.72$ & $  5$ \\
	& $D^0\to K^{\pm}\pi^{\mp}\ \mathrm{(BESIII)}$ & $ 5.30$ & $  7$ \\
	& $F^{+}_{hh\pi^0}\ \mathrm{(CLEO)}$ & $ 5.96$ & $  2$ \\
	& $F^{+}_{hh\pi^0}\ \mathrm{(BESIII)}$ & $ 0.52$ & $  2$ \\
	& $F^{+}_{4\pi}\ \mathrm{(CLEO)}$ & $ 1.10$ & $  1$ \\
	& $F^{+}_{4\pi}\ \mathrm{(BESIII)}$ & $ 0.37$ & $  1$ \\
	& $F^{+}_{KK\pi\pi}$ & $ 0.04$ & $  1$ \\
	& $D\to K^{\mp}\pi^{\pm}\pi^+\pi^-,\, K^{\mp}\pi^{\pm}\pi^0\ \mathrm{hadronic\ params}$ & $ 0.22$ & $  6$ \\
	& $D\to K^0_S K^{\pm}\pi^{\mp}\ \mathrm{hadronic\ params}$ & $ 3.79$ & $  3$ \\
	& $r_D^{K^0_S K\pi}$ & $ 0.60$ & $  1$ \\
\midrule
& \textbf{Total} & $191.06$ & $218$ \\
\bottomrule
\end{tabular}
  }
\end{table}

\begin{figure}[!htb]
  \centering
  \includegraphics[width=0.48\textwidth]{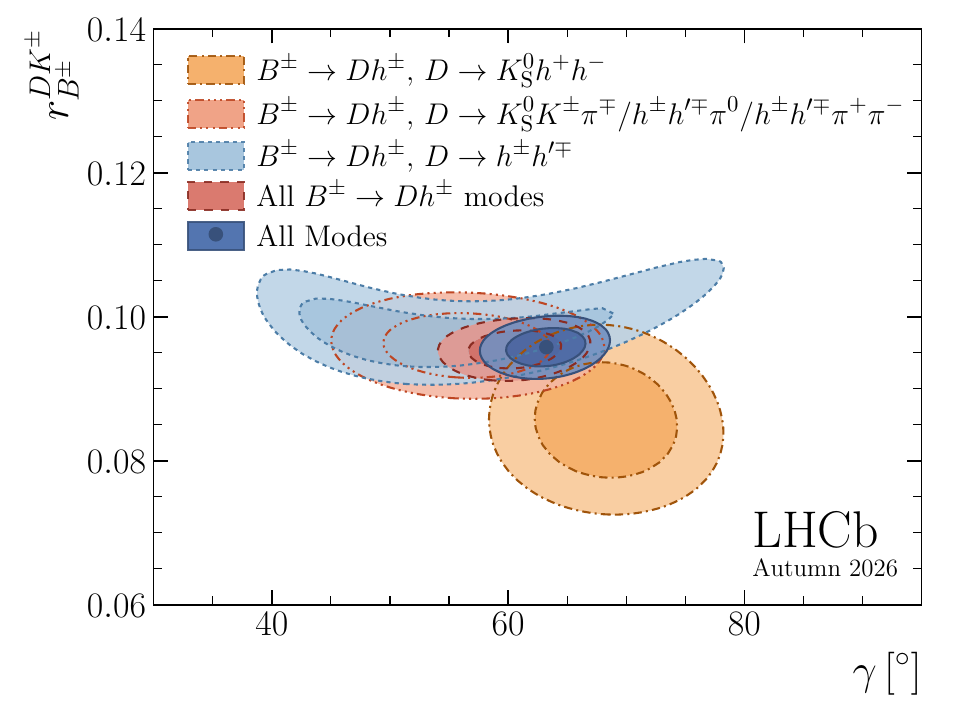}
  \includegraphics[width=0.48\textwidth]{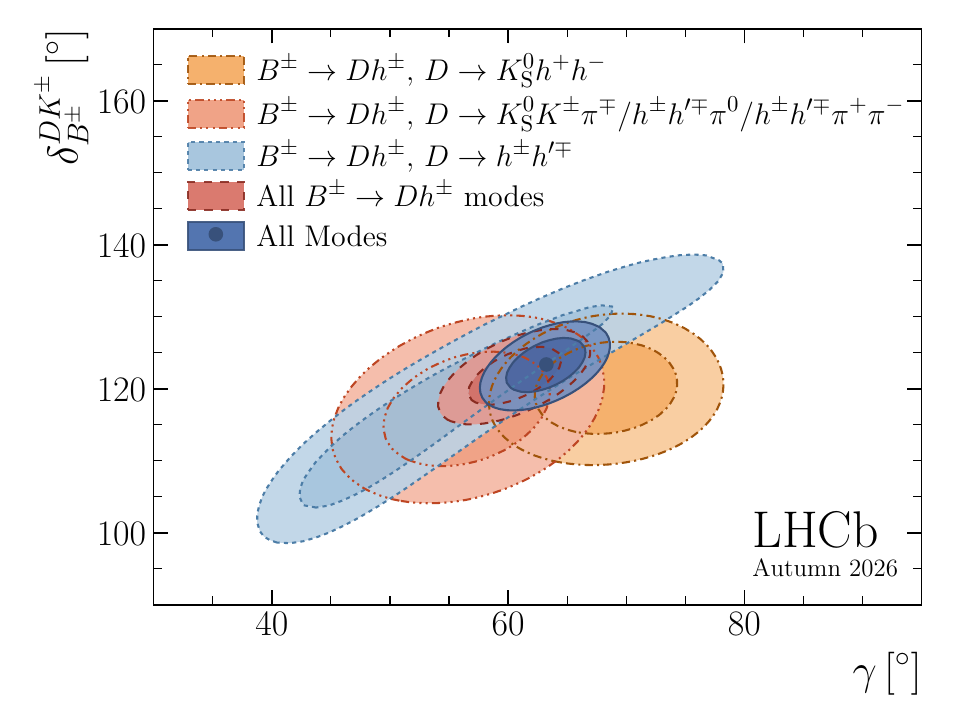}
  \includegraphics[width=0.48\textwidth]{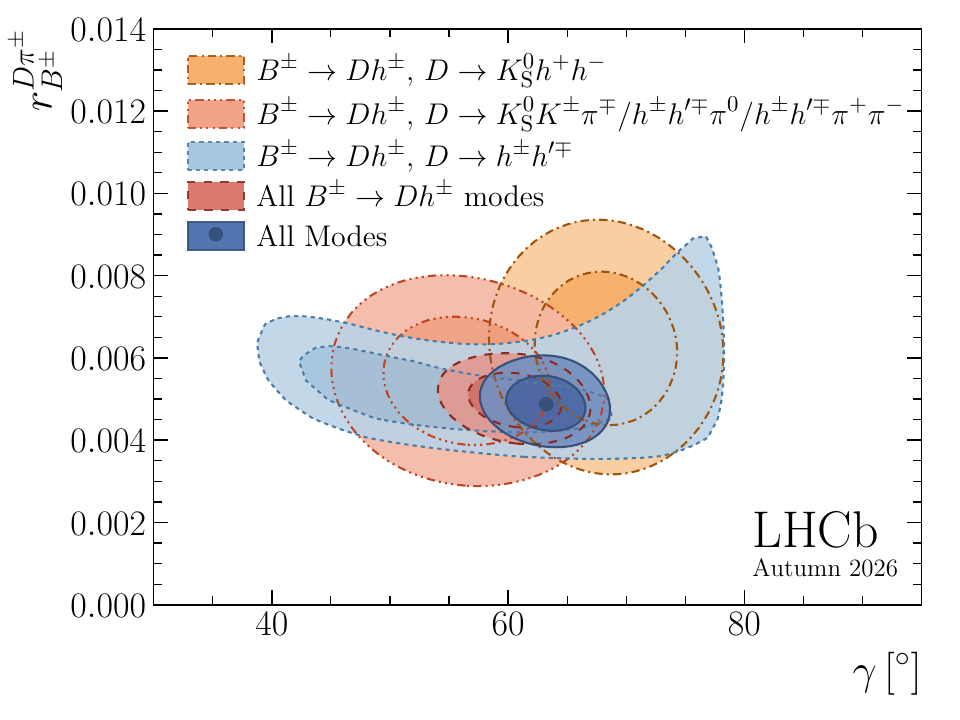}
  \includegraphics[width=0.48\textwidth]{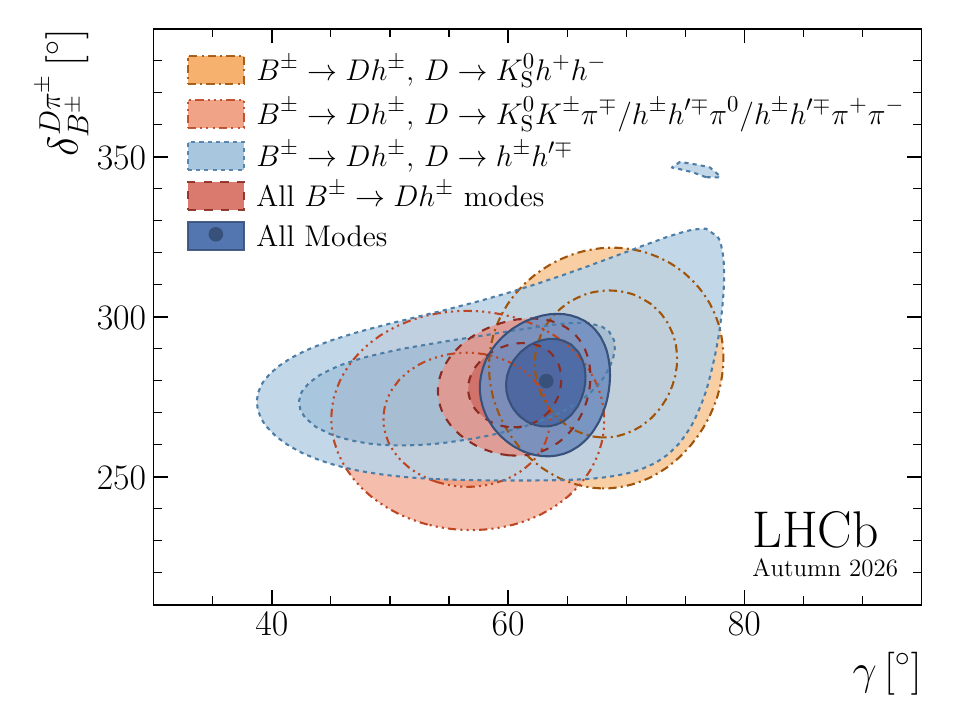}
  \caption{Profile-likelihood contours in the planes of $\gamma$ against the hadronic parameters of
    $B^-\to Dh^-$ decays, showing the breakdown of sensitivity among subcombinations of decay modes.
    Contours are displayed at the $1\sigma$ and $2\sigma$ confidence levels, encompassing $68.3\%$ and $95.4\%$ of the distributions, respectively.}
  \label{fig:res:breakdown_2d_dh1_run123}
\end{figure}

\begin{figure}[!htb]
  \centering
  \includegraphics[width=0.48\textwidth]{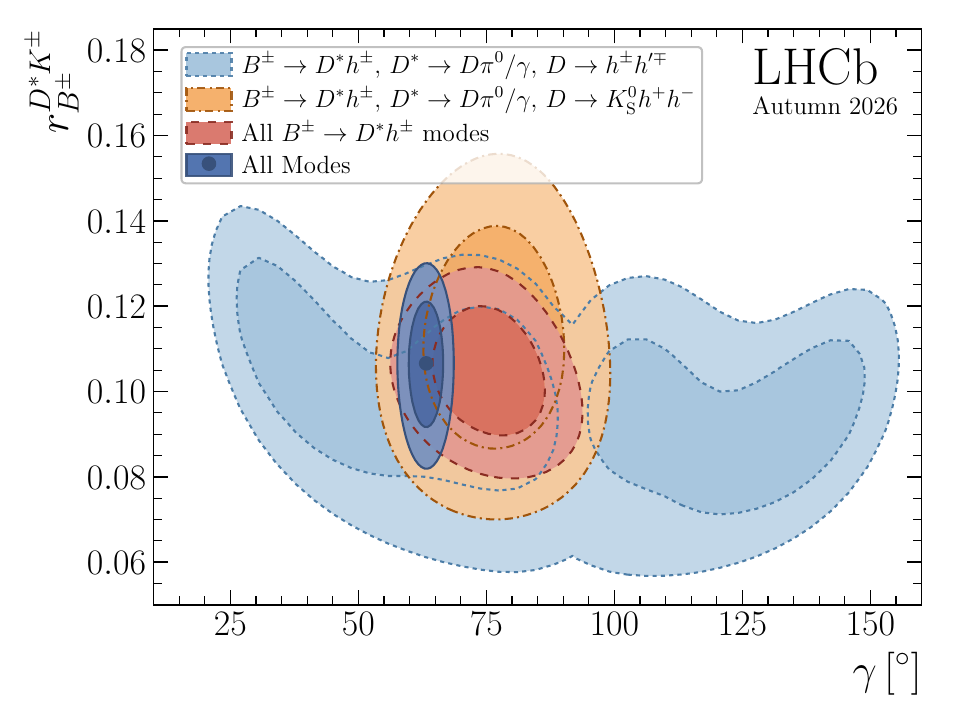}
  \includegraphics[width=0.48\textwidth]{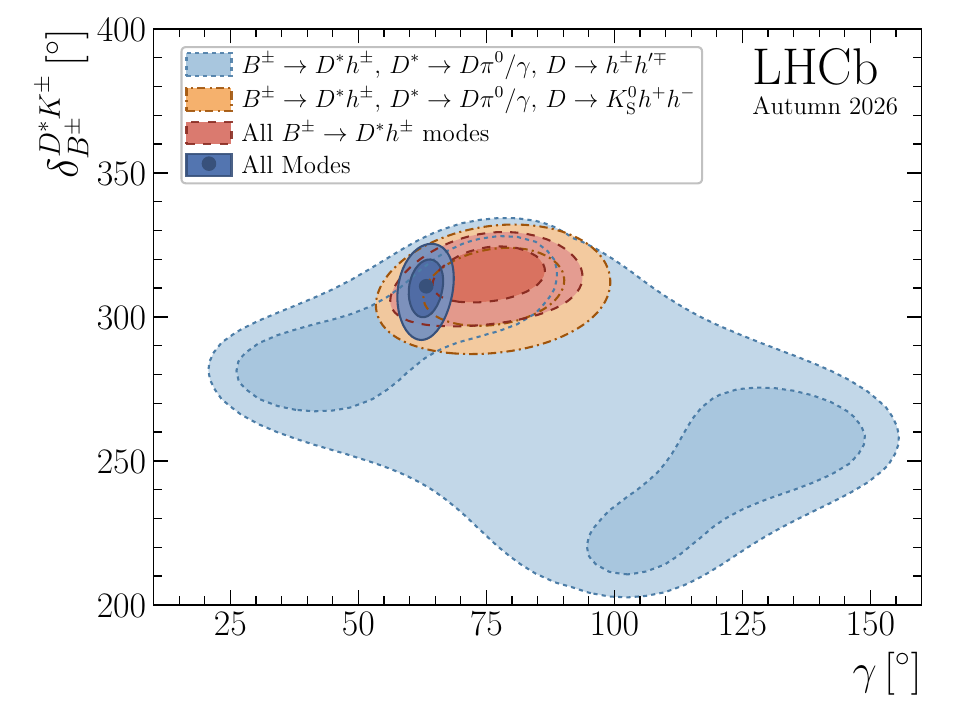}
  \includegraphics[width=0.48\textwidth]{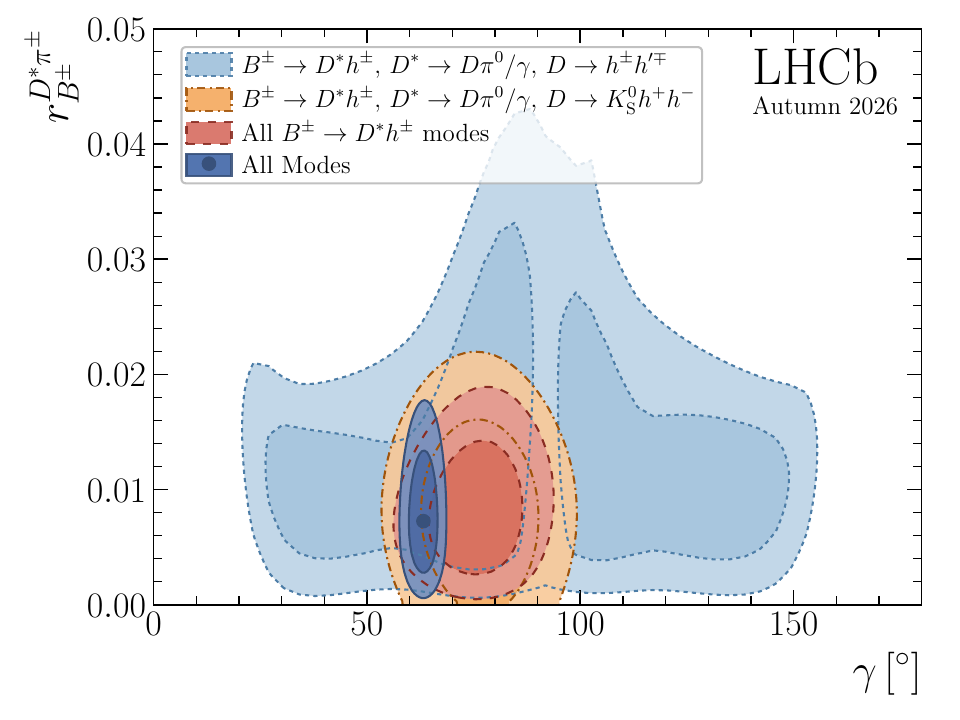}
  \includegraphics[width=0.48\textwidth]{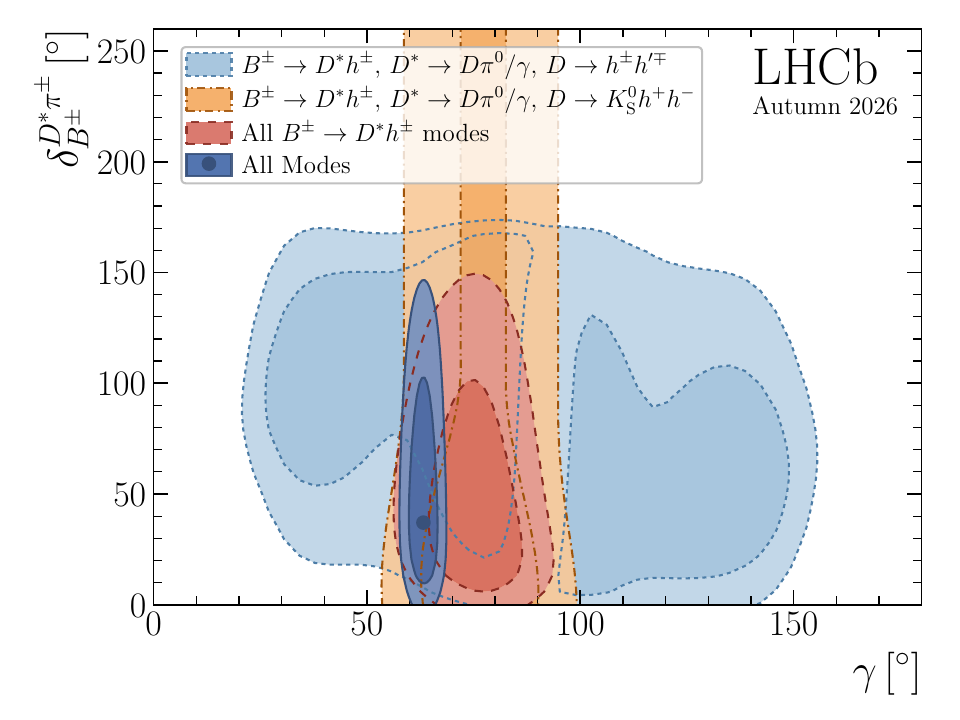}
  \caption{Profile-likelihood contours in the planes of $\gamma$ against the hadronic parameters of
    $B^-\to D^*h^-$ decays, showing the breakdown of sensitivity among subcombinations of decay modes.
    Contours are displayed at the $1\sigma$ and $2\sigma$ confidence levels, encompassing $68.3\%$ and $95.4\%$ of the distributions, respectively.}
  \label{fig:res:breakdown_2d_dh2_run123}
\end{figure}

\begin{figure}[!htb]
  \centering
  \includegraphics[width=0.48\textwidth]{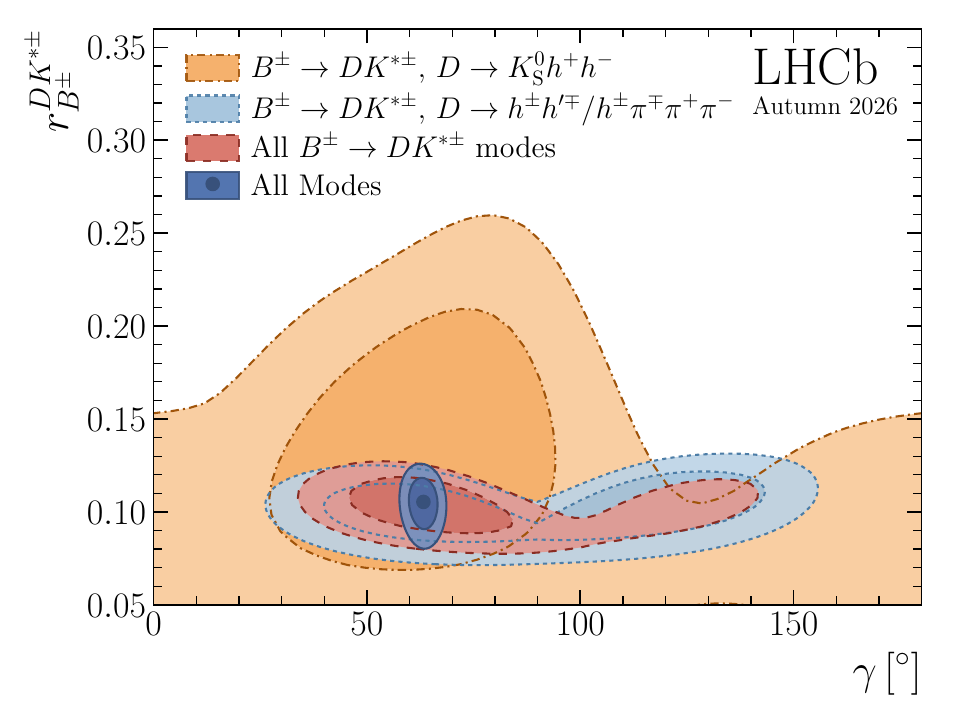}
  \includegraphics[width=0.48\textwidth]{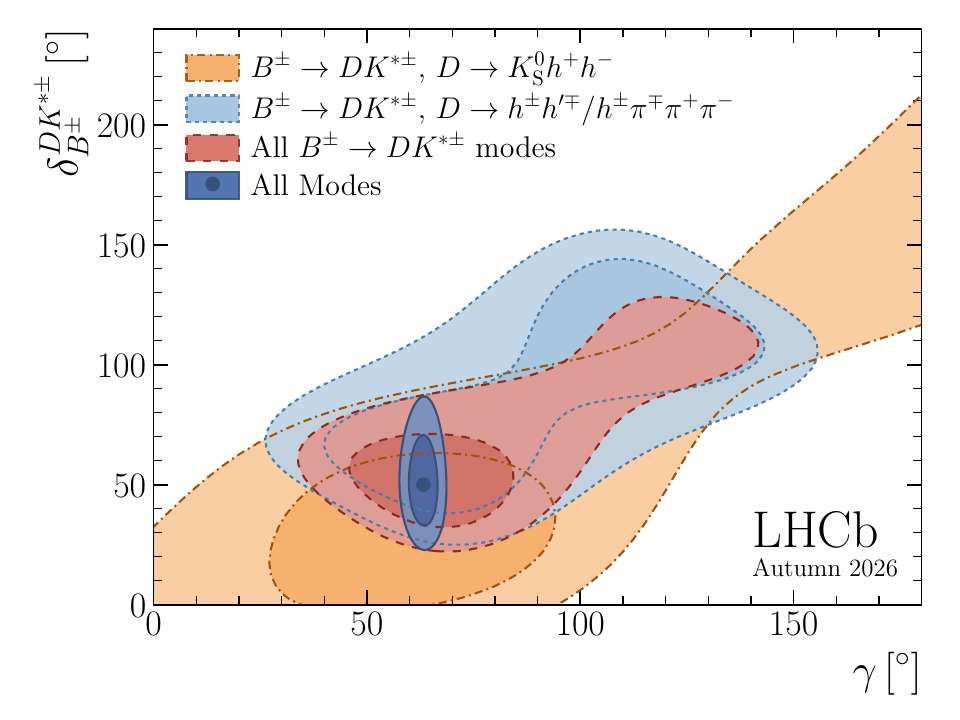}
  \includegraphics[width=0.48\textwidth]{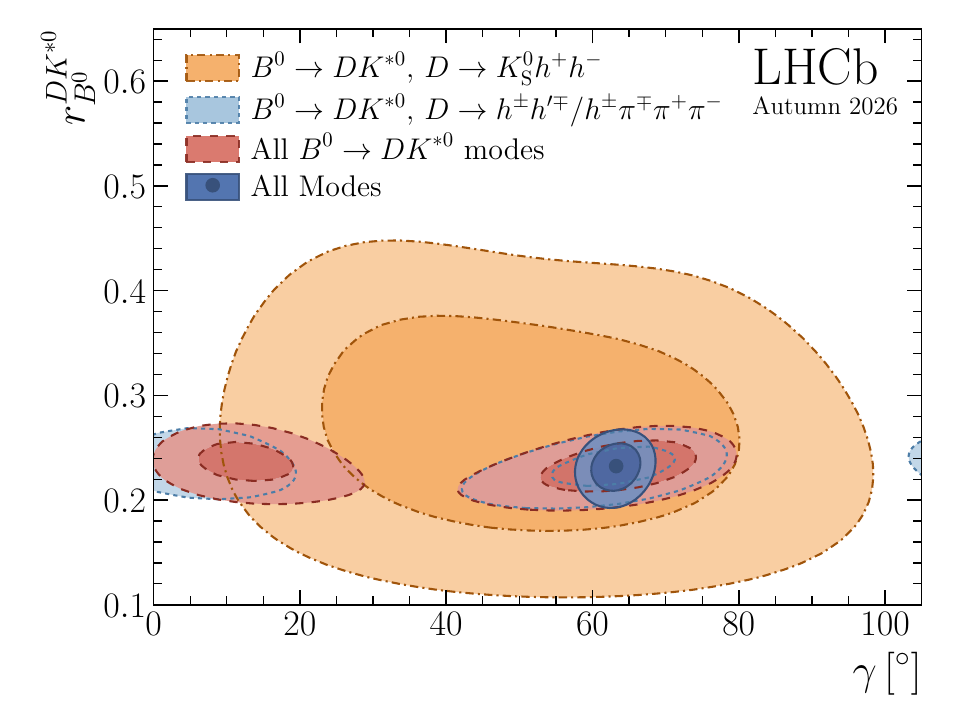}
  \includegraphics[width=0.48\textwidth]{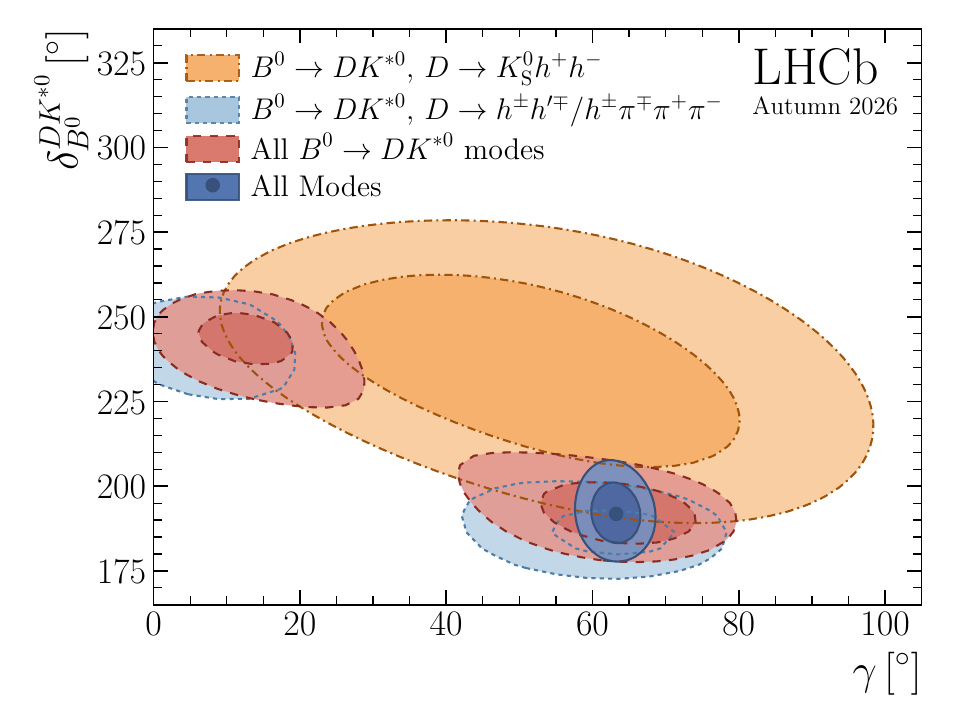}
  \caption{Profile-likelihood contours in the planes of $\gamma$ against the hadronic parameters of
    $B^-\to DK^{*-}$ and $B^0\to DK^{*0}$ decays, showing the breakdown of sensitivity among subcombinations of decay modes.
    Contours are displayed at the $1\sigma$ and $2\sigma$ confidence levels, encompassing $68.3\%$ and $95.4\%$ of the distributions, respectively.}
  \label{fig:res:breakdown_2d_dh3_run123}
\end{figure}

\begin{figure}
    \centering
    \includegraphics[width=0.45\linewidth]{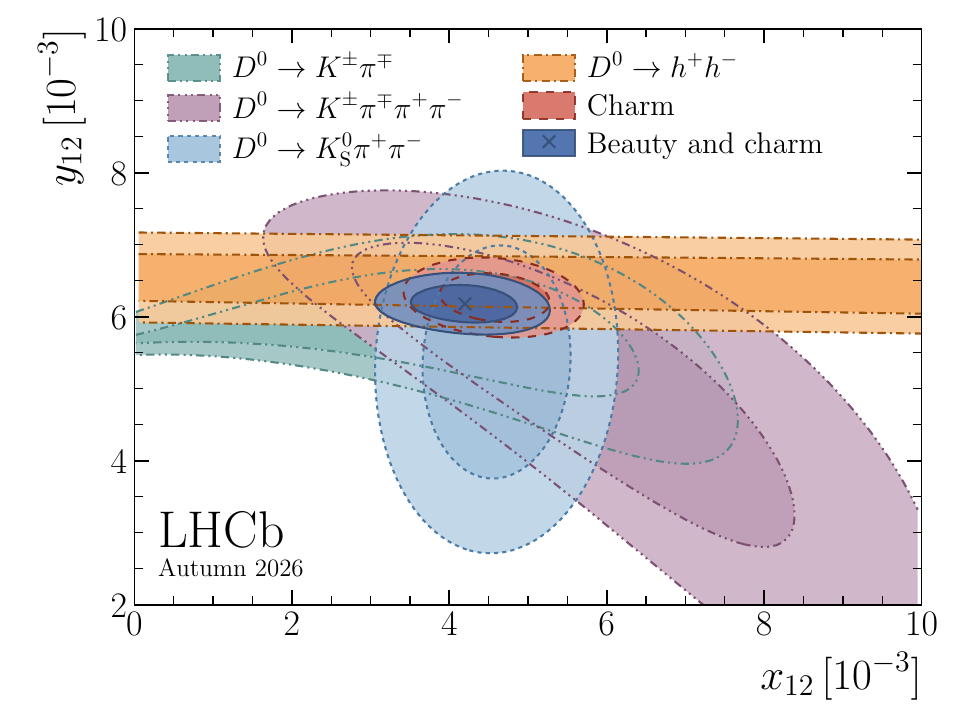}
    \includegraphics[width=0.45\linewidth]{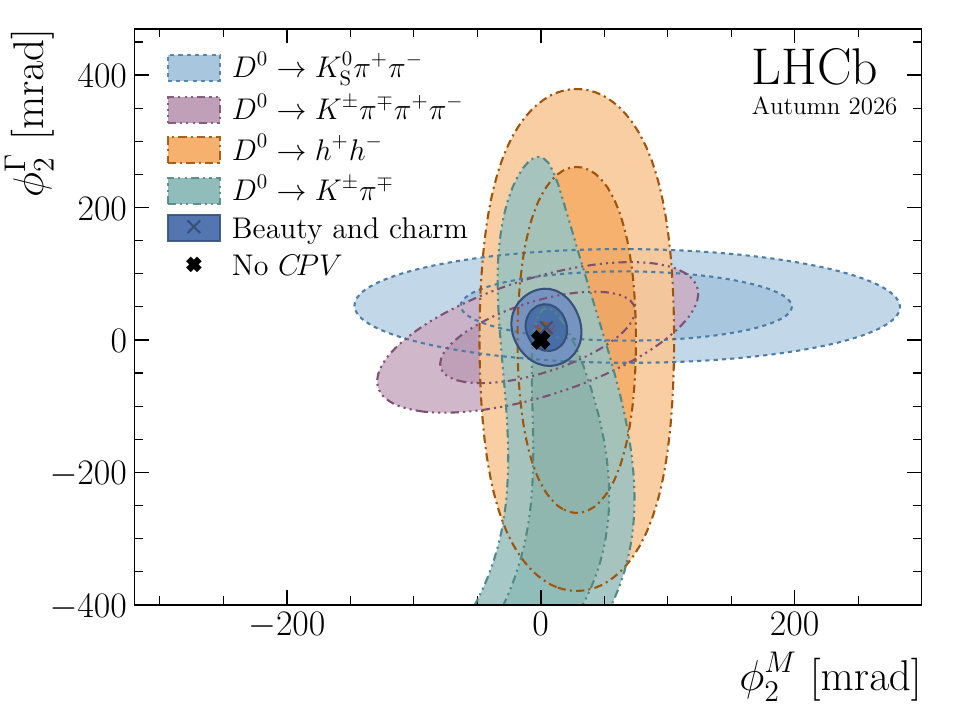}
    \includegraphics[width=0.45\linewidth]{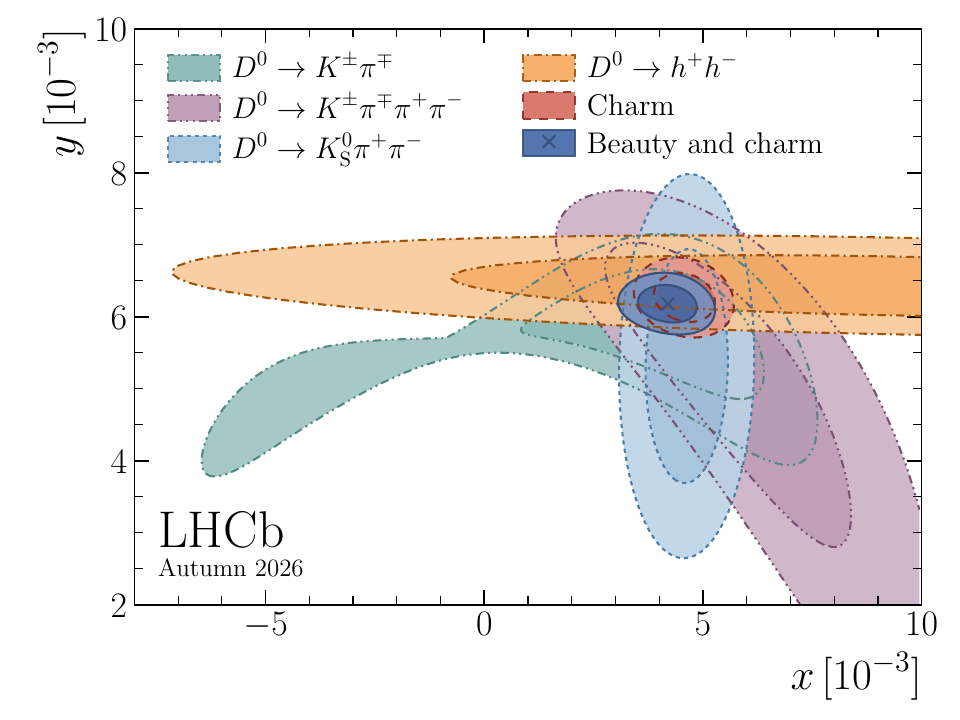}
    \includegraphics[width=0.45\linewidth]{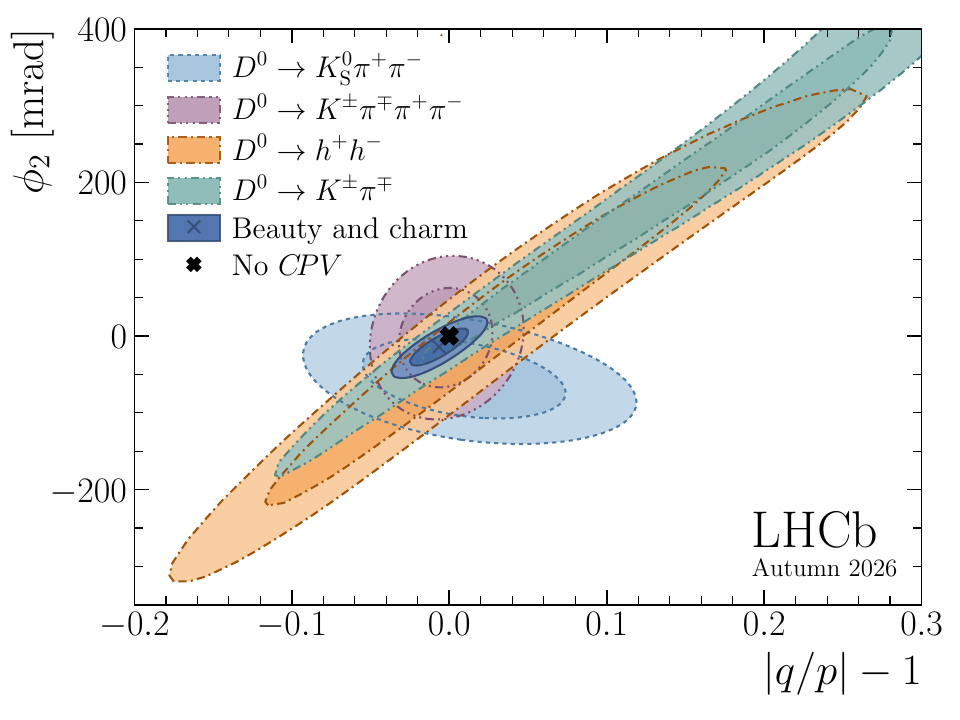}
    \caption{
        Profile-likelihood contours for the charm-mixing parameters, showing the breakdown of sensitivity among different decay modes.
        The plots correspond to the parameters quantifying (left) $D^0$ mixing and (right) $C\!P$ violation in $D^0$ mixing in the (top) theoretical and (bottom) phenomenological parametrisation.
        The origin of the right-hand plots, which corresponds to $C\!P$ conservation in $D^0$ mixing, is shown for reference.
        The $C\!P$-violation (mixing) parameters are fixed to their best-fit values in the left (right) plots.
        The parameters $r_{D}^{K\pi}$ and $\delta_{D}^{K\pi}$ are additionally fixed to their best-fit values in all contours of $D^0\to h^+h^-$ decays.
        Contours are displayed at the $1\sigma$ and $2\sigma$ confidence levels, encompassing $68.3\%$ and $95.4\%$ of the distributions, respectively.
    }
    \label{fig:charm-mix-breakdowns}
\end{figure}

\begin{figure}
    \centering
    \includegraphics[width=0.45\linewidth]{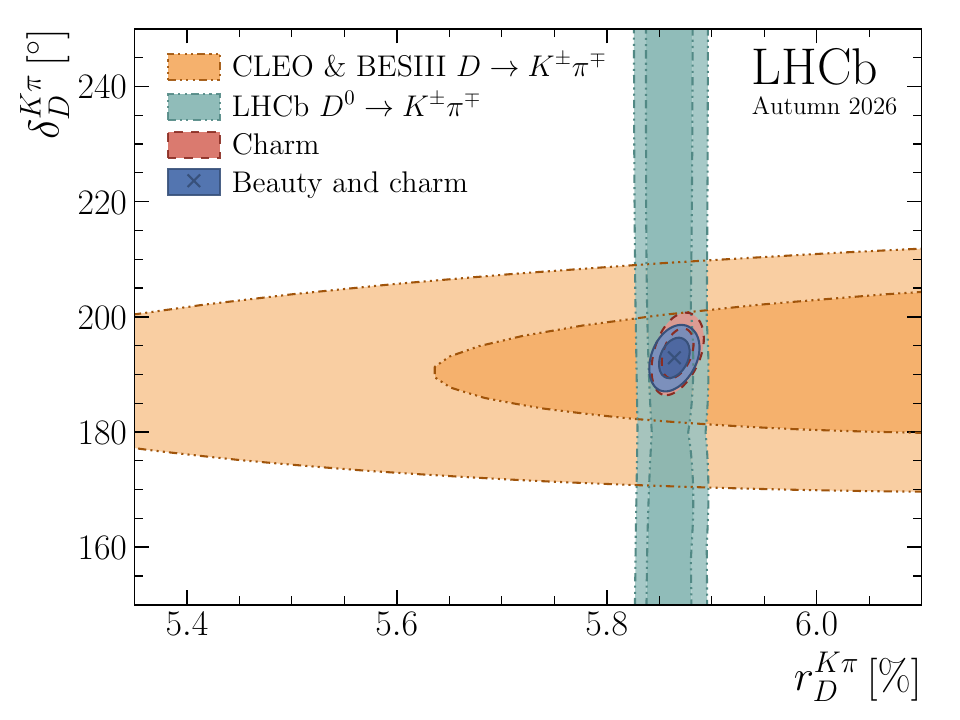}
    \caption{
        Profile-likelihood contours for the $D\to K^+\pi^-$ hadronic parameters, showing the breakdown of sensitivity among different experiments and combinations.
        Contours are displayed at the $1\sigma$ and $2\sigma$ confidence levels, encompassing $68.3\%$ and $95.4\%$ of the distributions, respectively.
    }
    \label{fig:charm-kpi-breakdown}
\end{figure}

The sensitivity of different measurements to \g and to the hadronic parameters of \BuDh, \BuDsth, \BuDKst and \BdDKst decays
is shown by the two-dimensional contours in \cref{fig:res:breakdown_2d_dh1_run123,fig:res:breakdown_2d_dh2_run123,fig:res:breakdown_2d_dh3_run123}.
The sensitivity to the charm-mixing parameters of the measurements of different \Dz decay channels,
and that to the \DKpi hadronic parameters \rdkpi and \ddkpi of the CLEO, \besiii and \lhcb measurements,
are compared in \cref{fig:charm-mix-breakdowns,fig:charm-kpi-breakdown}, respectively.


\clearpage

\section{Results of combinations with alternative theory assumptions}
\label{app:alternative-fits}
The results of the alternative combinations allowing for \CP violation in \DzKpPim decays,
and dropping the hypothesis of universality of charm-mixing weak phases, are presented in
\cref{tab:results-run123-cpv-d0-to-kpi,tab:results-run123-dy-fsc}, respectively.
Correlations among parameters are omitted since they differ from those of the baseline combination only marginally.

\begin{table}[!ht]
  \begin{center}
    \caption{Confidence intervals and best-fit values for each of the parameters of interest for the fit allowing for $C\!P$ violation in $D^0\to K^+\pi^-$ decays, computed with the toy-based \textsc{Plugin} method~\cite{LHCb-PAPER-2013-020}. Entries marked with an asterisk show where the scan has hit a physical boundary.}
    \label{tab:results-run123-cpv-d0-to-kpi}
    \setlength{\tabcolsep}{5pt}
    \renewcommand{\arraystretch}{1.2}
    \resizebox{0.9\textwidth}{!}{
    \begin{tabular}{lc|rr|rr}
    \hline
    \multirow{2}{*}{Quantity}& \multirow{2}{*}{Value} & \multicolumn{2}{c|}{68.3\% CL} & \multicolumn{2}{c}{95.4\% CL} \\
                             &                        & Uncertainty & Interval          & Uncertainty & Interval        \\
    \hline
$\g [^\circ]$               & $63.2$      & $^{+2.3}_{-2.2}$          & $[61.0,65.5]$             & $^{+4.4}_{-4.5}$          & $[58.7,67.6]$              \\
$\rdk [\%]$                 & $9.58$      & $^{+0.17}_{-0.18}$        & $[9.40,9.75]$             & $^{+0.35}_{-0.36}$        & $[9.22,9.93]$              \\
$\ddk [^\circ]$             & $123.3$     & $\pm 2.5$                 & $[120.8,125.8]$           & $^{+4.9}_{-5.0}$          & $[118.3,128.2]$            \\
$\rdpi [\%]$                & $0.49$      & $\pm 0.04$                & $[0.44,0.53]$             & $^{+0.09}_{-0.08}$        & $[0.40,0.58]$              \\
$\ddpi [^\circ]$            & $280$       & $\pm 9$                   & $[270,289]$               & $^{+17}_{-19}$            & $[261,297]$                \\
$\rdstk [\%]$               & $10.7$      & $\pm 1.0$                 & $[9.7,11.6]$              & $^{+1.9}_{-2.1}$          & $[8.6,12.6]$               \\
$\ddstk [^\circ]$           & $310$       & $^{+6}_{-7}$              & $[303,317]$               & $^{+12}_{-15}$            & $[296,323]$                \\
$\rdstpi [\%]$              & $0.73$      & $^{+0.39}_{-0.31}$        & $[0.42,1.12]$             & $^{+0.83}_{-0.62}$        & $[0.11,1.56]$              \\
$\ddstpi [^\circ]$          & $37$        & $^{+39}_{-20}$            & $[17,76]$                 & $^{+92}_{-34}$            & $[3,129]$                  \\
$\rdkst [\%]$               & $10.5$      & $^{+0.9}_{-1.0}$          & $[9.6,11.4]$              & $^{+1.8}_{-1.9}$          & $[8.6,12.3]$               \\
$\ddkst [^\circ]$           & $50$        & $^{+14}_{-12}$            & $[38,64]$                 & $^{+30}_{-23}$            & $[27,80]$                  \\
$\rdkstz [\%]$              & $23.2$      & $\pm 1.5$                 & $[21.7,24.7]$             & $^{+2.9}_{-3.2}$          & $[20.0,26.1]$              \\
$\ddkstz [^\circ]$          & $192$       & $\pm 6$                   & $[186,198]$               & $\pm 12$                  & $[180,204]$                \\
$\ldsk [\%]$                & $34$        & $\pm 4$                   & $[30,37]$                 & $\pm 7$                   & $[27,41]$                  \\
$\ddsk [^\circ]$            & $350$       & $\pm 6$                   & $[344,355]$               & $^{+11}_{-12}$            & $[338,361]$                \\
$\ldskpipi [\%]$            & $46$        & $\pm 8$                   & $[38,54]$                 & $^{+16}_{-17}$            & $[29,62]$                  \\
$\ddskpipi [^\circ]$        & $347$       & $\pm 12$                  & $[335,359]$               & $\pm 24$                  & $[323,371]$                \\
$\ldmpi [\%]$               & $3.0$       & $^{+1.5}_{-1.2}$          & $[1.8,4.5]$               & $^{+3.3}_{-2.6}$          & $[0.4,6.3]$                \\
$\ddmpi [^\circ]$           & $32$        & $^{+25}_{-38}$            & $[-6,57]$                 & $^{+44}_{-82}$            & $[-50,76]$                 \\
$\rdkpipi [\%]$             & $8.0$       & $^{+2.8}_{-3.5}$          & $[4.5,10.8]$              & $^{+4.9}_{-8.0}$          & $[0.0,12.9]^*$             \\
$\rdpipipi [\%]$            & $6.1$       & $^{+2.2}_{-2.5}$          & $[3.6,8.3]$               & $^{+4.7}_{-6.1}$          & $[0.0,10.8]^*$             \\
$\rDkpi [\%]$               & $5.865$     & $\pm 0.010$               & $[5.855,5.874]$           & $^{+0.019}_{-0.020}$      & $[5.845,5.884]$            \\
$\dDkpi [^\circ]$           & $192.9$     & $^{+2.4}_{-2.3}$          & $[190.6,195.3]$           & $^{+4.6}_{-4.9}$          & $[188.0,197.5]$            \\
$\AcpDKK [10^{-3}]$         & $0.7$       & $\pm 0.5$                 & $[0.1,1.2]$               & $^{+1.0}_{-1.1}$          & $[-0.4,1.7]$               \\
$\AcpDpipi [10^{-3}]$       & $2.2$       & $\pm 0.6$                 & $[1.6,2.8]$               & $\pm 1.2$                 & $[1.0,3.4]$                \\
$\AD [\%]$                  & $-0.49$     & $^{+0.26}_{-0.25}$        & $[-0.74,-0.23]$           & $\pm 0.52$                & $[-1.01,0.03]$             \\
\hline
$\xth [10^{-3}]$            & $4.2$       & $^{+0.4}_{-0.5}$          & $[3.8,4.6]$               & $\pm 0.9$                 & $[3.3,5.1]$                \\
$\yth [10^{-3}]$            & $6.19$      & $^{+0.17}_{-0.18}$        & $[6.01,6.36]$             & $\pm 0.34$                & $[5.85,6.53]$              \\
$\phiM [\mrad]$             & $22$        & $^{+15}_{-14}$            & $[8,37]$                  & $^{+30}_{-29}$            & $[-7,52]$                  \\
$\phiG [\mrad]$             & $32$        & $^{+24}_{-25}$            & $[7,56]$                  & $^{+48}_{-49}$            & $[-17,80]$                 \\
\hline
$\xD [10^{-3}]$             & $4.2$       & $\pm 0.5$                 & $[3.8,4.7]$               & $\pm 0.9$                 & $[3.3,5.1]$                \\
$\yD [10^{-3}]$             & $6.19$      & $^{+0.17}_{-0.18}$        & $[6.01,6.36]$             & $^{+0.35}_{-0.33}$        & $[5.86,6.54]$              \\
$\qopD - 1$                 & $-0.004$    & $^{+0.012}_{-0.013}$      & $[-0.017,0.008]$          & $^{+0.024}_{-0.025}$      & $[-0.029,0.020]$           \\
$\phiD [\mrad]$             & $-28$       & $^{+17}_{-19}$            & $[-47,-10]$               & $^{+35}_{-37}$            & $[-65,7]$                  \\
\hline
    \end{tabular}
    }
  \end{center}
\end{table}

\begin{table}[!ht]
  \begin{center}
    \caption{Confidence intervals and best-fit values for each of the parameters of interest for the fit dropping the hypothesis of universality of charm-mixing weak phases, computed with the toy-based \textsc{Plugin} method~\cite{LHCb-PAPER-2013-020}. Entries marked with an asterisk show where the scan has hit a physical boundary.}
    \label{tab:results-run123-dy-fsc}
    \setlength{\tabcolsep}{5pt}
    \renewcommand{\arraystretch}{1.2}
    \resizebox{0.9\textwidth}{!}{
    \begin{tabular}{lc|rr|rr}
    \hline
    \multirow{2}{*}{Quantity}& \multirow{2}{*}{Value} & \multicolumn{2}{c|}{68.3\% CL} & \multicolumn{2}{c}{95.4\% CL} \\
                             &                        & Uncertainty & Interval          & Uncertainty & Interval        \\
    \hline
$\g [^\circ]$               & $63.2$      & $^{+2.3}_{-2.2}$          & $[61.0,65.5]$             & $^{+4.4}_{-4.5}$          & $[58.7,67.6]$              \\
$\rdk [\%]$                 & $9.58$      & $\pm 0.18$                & $[9.40,9.76]$             & $\pm 0.36$                & $[9.22,9.94]$              \\
$\ddk [^\circ]$             & $123.4$     & $^{+2.5}_{-2.6}$          & $[120.8,125.9]$           & $^{+4.8}_{-5.2}$          & $[118.2,128.2]$            \\
$\rdpi [\%]$                & $0.49$      & $^{+0.05}_{-0.04}$        & $[0.44,0.53]$             & $^{+0.09}_{-0.08}$        & $[0.40,0.58]$              \\
$\ddpi [^\circ]$            & $280$       & $^{+9}_{-10}$             & $[270,289]$               & $^{+18}_{-19}$            & $[261,298]$                \\
$\rdstk [\%]$               & $10.7$      & $\pm 1.0$                 & $[9.7,11.6]$              & $^{+1.8}_{-2.1}$          & $[8.6,12.5]$               \\
$\ddstk [^\circ]$           & $310$       & $^{+6}_{-7}$              & $[304,317]$               & $^{+12}_{-16}$            & $[295,323]$                \\
$\rdstpi [\%]$              & $0.73$      & $^{+0.39}_{-0.32}$        & $[0.41,1.12]$             & $^{+0.82}_{-0.62}$        & $[0.11,1.55]$              \\
$\ddstpi [^\circ]$          & $37$        & $^{+40}_{-20}$            & $[17,77]$                 & $^{+91}_{-35}$            & $[2,128]$                  \\
$\rdkst [\%]$               & $10.5$      & $^{+0.9}_{-1.0}$          & $[9.6,11.4]$              & $^{+1.8}_{-2.0}$          & $[8.5,12.3]$               \\
$\ddkst [^\circ]$           & $50$        & $^{+14}_{-12}$            & $[38,64]$                 & $^{+29}_{-23}$            & $[27,79]$                  \\
$\rdkstz [\%]$              & $23.2$      & $\pm 1.5$                 & $[21.7,24.7]$             & $^{+2.9}_{-3.2}$          & $[20.0,26.1]$              \\
$\ddkstz [^\circ]$          & $192$       & $\pm 6$                   & $[186,198]$               & $\pm 12$                  & $[180,204]$                \\
$\ldsk [\%]$                & $33.7$      & $^{+3.6}_{-3.5}$          & $[30.2,37.3]$             & $^{+7.3}_{-7.1}$          & $[26.6,41.0]$              \\
$\ddsk [^\circ]$            & $350$       & $\pm 6$                   & $[344,355]$               & $^{+11}_{-12}$            & $[338,361]$                \\
$\ldskpipi [\%]$            & $46$        & $\pm 8$                   & $[38,54]$                 & $^{+16}_{-17}$            & $[29,62]$                  \\
$\ddskpipi [^\circ]$        & $347$       & $\pm 12$                  & $[335,359]$               & $^{+24}_{-25}$            & $[322,371]$                \\
$\ldmpi [\%]$               & $3.0$       & $^{+1.5}_{-1.2}$          & $[1.8,4.5]$               & $^{+3.3}_{-2.6}$          & $[0.4,6.3]$                \\
$\ddmpi [^\circ]$           & $32$        & $^{+25}_{-38}$            & $[-6,57]$                 & $^{+44}_{-83}$            & $[-51,76]$                 \\
$\rdkpipi [\%]$             & $8.0$       & $^{+2.8}_{-3.5}$          & $[4.5,10.8]$              & $^{+4.9}_{-8.0}$          & $[0.0,12.9]^*$             \\
$\rdpipipi [\%]$            & $6.1$       & $^{+2.2}_{-2.5}$          & $[3.6,8.3]$               & $^{+4.7}_{-6.1}$          & $[0.0,10.8]^*$             \\
$\rDkpi [\%]$               & $5.864$     & $\pm 0.010$               & $[5.855,5.874]$           & $^{+0.020}_{-0.019}$      & $[5.845,5.884]$            \\
$\dDkpi [^\circ]$           & $192.9$     & $^{+2.3}_{-2.4}$          & $[190.5,195.2]$           & $^{+4.6}_{-4.9}$          & $[188.0,197.5]$            \\
$\AcpDKK [10^{-3}]$         & $0.9$       & $\pm 0.5$                 & $[0.3,1.4]$               & $\pm 1.1$                 & $[-0.2,2.0]$               \\
$\AcpDpipi [10^{-3}]$       & $2.8$       & $\pm 0.7$                 & $[2.1,3.5]$               & $\pm 1.3$                 & $[1.5,4.1]$                \\
\hline
$\xth [10^{-3}]$            & $4.2$       & $^{+0.4}_{-0.5}$          & $[3.7,4.6]$               & $\pm 0.9$                 & $[3.3,5.1]$                \\
$\yth [10^{-3}]$            & $6.19$      & $^{+0.17}_{-0.18}$        & $[6.01,6.36]$             & $^{+0.34}_{-0.36}$        & $[5.83,6.53]$              \\
$\phiM [\mrad]$             & $0$         & $\pm 12$                  & $[-12,12]$                & $^{+24}_{-25}$            & $[-25,24]$                 \\
$\phiG [\mrad]$             & $19$        & $^{+24}_{-23}$            & $[-4,43]$                 & $^{+48}_{-46}$            & $[-27,67]$                 \\
\hline
$\xD [10^{-3}]$             & $4.2$       & $^{+0.4}_{-0.5}$          & $[3.7,4.6]$               & $\pm 0.9$                 & $[3.2,5.1]$                \\
$\yD [10^{-3}]$             & $6.18$      & $^{+0.18}_{-0.17}$        & $[6.01,6.36]$             & $\pm 0.35$                & $[5.83,6.53]$              \\
$\qopD - 1$                 & $-0.009$    & $\pm 0.013$               & $[-0.022,0.004]$          & $^{+0.026}_{-0.025}$      & $[-0.034,0.017]$           \\
$\phiD [\mrad]$             & $-13$       & $^{+16}_{-17}$            & $[-30,3]$                 & $\pm 33$                  & $[-47,19]$                 \\
\hline
    \end{tabular}
    }
  \end{center}
\end{table}

%% file: LHCb/acknowledgements.tex
\section*{Acknowledgements}
%
% These Acknowledgements valid from 16/06/2026
%
\noindent We express our gratitude to our colleagues in the CERN
accelerator departments for the excellent performance of the LHC. We
thank the technical and administrative staff at the LHCb
institutes.
We acknowledge support from CERN and from the national agencies:
ARC (Australia);
CAPES, CNPq, FAPERJ and FINEP (Brazil); 
MOST and NSFC (China); 
CNRS/IN2P3 and CEA (France);  
BMFTR, DFG and MPG (Germany);
NKFIH (Hungary);              
INFN (Italy); 
NWO (Netherlands); 
MNiSW and NCN (Poland); 
MEC/IFA (Romania); 
MICIU and AEI (Spain);
SNSF and SER (Switzerland); 
NASU (Ukraine); 
STFC (United Kingdom); 
DOE NP and NSF (USA).
%%%%%%%%%%%%%%%%%%%%%%%%%%%%%%%%%%%%%%%%%%%%%
We acknowledge the computing resources that are provided by ARDC (Australia), 
CBPF (Brazil),
CERN, 
IHEP and LZU (China),
IN2P3 (France), 
KIT and DESY (Germany), 
INFN (Italy), 
SURF (Netherlands),
Polish WLCG (Poland),
IFIN-HH (Romania), % http://dx.doi.org/10.13039/100019931,"Institutul National de Cercetare-Dezvoltare pentru Fizica si Inginerie Nucleara 'Horia Hulubei'"
PIC (Spain), CSCS (Switzerland), 
GridPP (United Kingdom),
and NSF (USA).  % added Feb2026
%%%%%%%%%%%%%%%%%%%%%%%%%%%%%%%%%%%%%%%%%% 
We are indebted to the communities behind the multiple open-source
software packages on which we depend.
%%%%%%%%%%%%%%%%%%%%%%%%%%%%%%%%%%%%%%%%%%
Individual groups or members have received support from
RTP (Australia), % added 06/03/2026
FWO Odysseus grant G0ASD25N (Belgium), % added 20/4/2026
Key Research Program of Frontier Sciences of CAS, CAS PIFI, CAS CCEPP (China); 
Minciencias (Colombia);
EPLANET, Marie Sk\l{}odowska-Curie Actions, ERC and NextGenerationEU (European Union);
A*MIDEX, ANR, IPhU and Labex P2IO, and R\'{e}gion Auvergne-Rh\^{o}ne-Alpes (France);
Alexander-von-Humboldt Foundation (Germany);
ICSC (Italy); 
Severo Ochoa and Mar\'ia de Maeztu Units of Excellence, GVA, XuntaGal, GENCAT, InTalent-Inditex and Prog.~Atracci\'on Talento CM (Spain);
the Leverhulme Trust, the Royal Society and UKRI (United Kingdom).

%% file: LHCb/Authorship_LHCb-PAPER-2026-058.tex
% LHCb collaboration author list
% Data extracted on November 13th, 2026 at 2:27pm for paper reference LHCb-PAPER-2026-058
\centerline
{\large\bf LHCb collaboration}
\begin
{flushleft}
\small
R.~Aaij$^{41}$\lhcborcid{0000-0003-0533-1952},
M.~Abdelfatah$^{74}$,
A.S.W.~Abdelmotteleb$^{62}$\lhcborcid{0000-0001-7905-0542},
C.~Abellan~Beteta$^{56}$\lhcborcid{0009-0009-0869-6798},
F.~Abudin\'en$^{64}$\lhcborcid{0000-0002-6737-3528},
T.~Ackernley$^{66}$\lhcborcid{0000-0002-5951-3498},
A.A.~Adefisoye$^{74}$\lhcborcid{0000-0003-2448-1550},
B.~Adeva$^{52}$\lhcborcid{0000-0001-9756-3712},
M.~Adinolfi$^{60}$\lhcborcid{0000-0002-1326-1264},
P.~Adlarson$^{91,46}$\lhcborcid{0000-0001-6280-3851}
% ...

{\footnotesize \it

% ...
$^{41}$Nikhef National Institute for Subatomic Physics, Amsterdam, Netherlands\\
% ...
$^{46}$National Center for Nuclear Research (NCBJ), Warsaw, Poland\\
% ...
$^{52}$Instituto Galego de F{\'\i}sica de Altas Enerx{\'\i}as (IGFAE), Universidade de Santiago de Compostela, Santiago de Compostela, Spain\\
% ...
$^{56}$Physik-Institut, Universit{\"a}t Z{\"u}rich, Z{\"u}rich, Switzerland\\
% ...
$^{60}$H.H. Wills Physics Laboratory, University of Bristol, Bristol, United Kingdom\\
% ...
$^{62}$Department of Physics, University of Warwick, Coventry, United Kingdom\\
% ...
$^{64}$School of Physics and Astronomy, University of Edinburgh, Edinburgh, United Kingdom\\
$^{65}$School of Physics and Astronomy, University of Glasgow, Glasgow, United Kingdom\\
$^{66}$Oliver Lodge Laboratory, University of Liverpool, Liverpool, United Kingdom\\
% ...
$^{74}$Syracuse University, Syracuse, NY, United States\\
% ...
$^{91}$Department of Physics and Astronomy, Uppsala University, Uppsala, Sweden, associated to $^{65}$\\
% ...
}
\end{flushleft}